\documentclass[11pt,a4paper]{article}
\usepackage{amssymb} \usepackage{amsmath} \usepackage{graphicx}
\usepackage{epsfig,latexsym}
\usepackage{tabularx}

\def\bef{\begin{figure}}
\def\eef{\end{figure}}

\newcommand{\be}[1]{\begin{equation}\label{#1}}
\newcommand{\beq}{\begin{equation}}
\newcommand{\ee}{\end{equation}}
\newcommand{\beqn}[1]{\begin{eqnarray}\label{#1}}
\newcommand{\eeqn}{\end{eqnarray}}
\newcommand{\bd}{\begin{displaymath}}
\newcommand{\ed}{\end{displaymath}}

\def\lsim{\raise0.3ex\hbox{$\;<$\kern-0.75em\raise-1.1ex
e\hbox{$\sim\;$}}}  
\def\gsim{\raise0.3ex\hbox{$\;>$\kern-0.75em\raise-1.1ex
\hbox{$\sim\;$}}}
\def\simlt{\mathrel{\lower2.5pt\vbox{\lineskip=0pt\baselineskip=0pt
           \hbox{$<$}\hbox{$\sim$}}}}
\def\simgt{\mathrel{\lower2.5pt\vbox{\lineskip=0pt\baselineskip=0pt
           \hbox{$>$}\hbox{$\sim$}}}}
\def\unity{{\hbox{1\kern-.8mm l}}}

\renewcommand{\to}{\rightarrow}
\renewcommand{\vec}[1]{\mathbf{#1}}

\renewcommand{\to}{\rightarrow}

\def\simlt{\stackrel{<}{{}_\sim}}
\def\simgt{\stackrel{>}{{}_\sim}}

\def\ee{\end{equation}}

\newcommand{\ve}{\vec{v}_{\scriptscriptstyle E}}
\newcommand{\vdm}{\vec{v}_{\scriptscriptstyle DM}}
\newcommand{\vdmp}{\vec{v'}_{\scriptscriptstyle DM}}
\newcommand{\vdms}{\vec{v''}_{\scriptscriptstyle DM}}
\newcommand{\vhalo}{\vec{v}_{\scriptscriptstyle halo}}
\newcommand{\vzh}{v_{\scriptscriptstyle 0,halo}}
\newcommand{\vs}{\vec{v}_{\scriptscriptstyle S}}
\newcommand{\vr}{\vec{v}_{\scriptscriptstyle rev}(t)}
\newcommand{\vrel}{\vec{v}_{\scriptscriptstyle rel}(t)}

\renewcommand{\to}{\rightarrow}

\def\lsim{\raise0.3ex\hbox{$\;<$\kern-0.75em\raise-1.1ex
\hbox{$\sim\;$}}}
\def\gsim{\raise0.3ex\hbox{$\;>$\kern-0.75em\raise-1.1ex
\hbox{$\sim\;$}}}
\def\cal{\mathcal}
\renewcommand{\vec}[1]{ \boldsymbol{#1}}
\normalsize
\begin{document}

\baselineskip=0.65cm

\begin{center}
\Large
{\bf DAMA annual modulation and mirror Dark Matter}
\rm
\end{center}

\large

\begin{center}

R. Cerulli$^{a}$, P. Villar$^{b}$, F. Cappella$^{a}$, 
R. Bernabei$^{c}$, P. Belli$^{c}$, A. Incicchitti$^{d}$,
A. Addazi$^{e,a}$, Z. Berezhiani$^{e,a}$

\vspace{2mm}

$^{a}${\it INFN, Laboratori Nazionali del Gran Sasso, I-67010 Assergi AQ, Italy}\\
$^{b}${\it Laboratorio de F\'isica Nuclear y Astropart\'iculas, Universidad de Zaragoza, \\
C/ Pedro Cerbuna 12, 50009 Zaragoza and \\
Laboratorio Subterr\'aneo de Canfranc, Paseo de los Ayerbe s.n., 22880 Canfranc Estaci\'on,
Huesca, SPAIN\\
$^{c}${\it Dipartimento di Fisica, Universit\`a di Roma ``Tor Vergata'' and \\
INFN -- Tor Vergata, I-00133 Rome, Italy}\\
$^{d}${\it Dipartimento di Fisica, Universit\`a di Roma ``La Sapienza'' and \\
INFN -- Roma, I-00185 Rome, Italy}\\
$^{e}${\it Dipartimento di Scienze Fisiche e Chimiche,
 Universit\`a di L'Aquila, I-67100 Coppito AQ, Italy}\\
}

\vspace{1mm}

\end{center}
	
\normalsize

\begin{abstract}
The DAMA experiment using ultra low background NaI(Tl) crystal scintillators has measured 
an annual modulation effect in the keV region 
which satisfies all the peculiarities of an effect induced by Dark Matter particles.
In this paper we analyze this annual modulation effect in terms of mirror Dark Matter, 
an exact duplicate of ordinary matter from parallel hidden sector,   
which chemical composition is dominated by mirror helium  while it can also contain 
significant fractions of heavier elements as Carbon and Oxygen.   
Dark mirror atoms are considered to interact with the target nuclei 
in the detector via Rutherford-like scattering induced by
kinetic mixing between mirror and ordinary photons,  
both being massless.  
In the present analysis we consider various possible scenarios for the mirror matter chemical composition.
For all the scenarios, the relevant ranges 
for the kinetic mixing parameter have been obtained taking also into account various existing 
uncertainties in nuclear and particle physics quantities.  

\end{abstract}

\vspace{5.0mm}


\vspace{2.0mm}


\section{Introduction}

A peculiar annual-modulation of the counting rate is expected to be induced by  Dark Matter (DM)
particles in the galactic halo in a suitable set-up located deep underground on the Earth. 
In fact, the flux of the DM particles
is modulated during the year as a consequence of the  
Earth revolution around the Sun which is moving in the Galactic frame \cite{Drukier,Freese}. 
The induced signal must satisfy simultaneously several requirements.

The DAMA Collaboration has measured an annual modulation effect over 14 independent annual cycles
by using the highly radiopure NaI(Tl) detectors of the former DAMA/NaI experiment
\cite{prop,allDM1,allDM2,allDM3,allDM4,allDM5,allDM6,allDM7,allDM8,Nim98,Sist,RNC,ijmd,ijma,epj06,ijma07,chan,wimpele,ldm,allRare1,allRare2,allRare3,allRare4,allRare5,allRare6,allRare7,allRare8,allRare9,allRare10,IDM96}
and of the second generation DAMA/LIBRA-phase1 \cite{perflibra,modlibra,modlibra2,modlibra3,bot11,pmts,mu,review,papep,cnc-l,IPP,diu,norole,shadow,mirror2015}. 
By considering the data of the 7 annual cycles 
collected by DAMA/NaI experiment (concluded in 2002) and of the 7 annual cycles 
collected by DAMA/LIBRA-phase1 an exposure of 1.33 ton $\times$ year has been released.
The observed annual modulation effect points out the presence of DM particles 
in the Galactic halo at 9.3$\sigma$ C.L. 
and the modulation amplitude of the single-hit events in the (2-6) keV energy interval 
is: $(0.0112 \pm 0.0012)$ cpd/kg/keV; the measured phase is $(144 \pm 7)$ days and the 
measured period is $(0.998 \pm 0.002)$ yr, values well in agreement with those 
expected for DM particles \cite{modlibra3}. 
No systematic effect or side reaction able to mimic the measured modulation effect,
i.e. able to account for the whole measured modulation amplitude and simultaneously 
satisfy all of many peculiarities of the signature, was found  or suggested by anyone over decades.

An important aspect of the annual-modulation measured by DAMA experiments 
is that this effect  is model-independent. The annual modulation of the event rate 
is an experimental fact and it does not depend on any theoretical interpretations 
of the nature and interaction type(s) of DM particle signal. 
It can be related to a variety of interaction mechanisms of DM particles with the 
detector materials (see for example Ref. \cite{review}).  

The most familiar candidates for DM particles include WIMPs as lightest neutralino and axion, related 
respectively to well-motivated concepts of supersymmetry (+ R-parity) and Peccei-Quinn symmetry 
which are exceptionally promising tools for solving a number of fundamental problems  in particle physics.  
An alternative well-founded idea is that DM particles may a hidden or shadow  gauge sector,  
with particle and interaction content similar to that of known particles.  
In particular, a parallel gauge sector with particle physics exactly identical 
to that of ordinary particles, 
coined as mirror world,  
was introduced long time ago by the reasons related to parity conservation  \cite{Mirror1,Mirror2,Mirror4,Mirror5}. 

Generically, one can consider a theory based on the product $G\times G'$ of two 
identical gauge factors, as two copies of the Standard Model  or two copies of  GUTs   
like $SU(5) \times SU(5)$,  
with ordinary (O)  particles belonging to a sector $G$ and mirror (M)  particles to a parallel sector $G'$. 
In General Relativity the gravity, described by the space-time metric $g_{\mu\nu}$,
is the universal force equally interacting with both sectors.
Therefore, the full dynamics of two sectors is governed by the Einstein-Hilbert action 
\be{S-grav}
S = \int d^4x\,  \sqrt{-g}\,  \left( \frac12 M_P^2 R + {\cal L} + {\cal L}' + 
{\cal L}_{\rm mix} \right) ,
\ee
where $M_P$ is the reduced Planck mass, 
$R$ is the space-time curvature,  
the Lagrangians ${\cal L}= {\cal L}_{\rm gauge} + {\cal L}_{\rm Yuk} + {\cal L}_{\rm Higgs}$ 
and ${\cal L}'= {\cal L}'_{\rm gauge} + {\cal L}'_{\rm Yuk} + {\cal L}'_{\rm Higgs}$ 
describe the interactions in the ordinary and mirror sectors, respectively,  
whereas ${\cal L}_{\rm mix}$ describes the possible interactions between ordinary and mirror 
particles as e.g. photon-mirror photon kinetic mixing  which shall be discussed later. 
 The Lagrangians ${\cal L}$ and  ${\cal L}'$ can be made identical 
by imposing mirror parity, a discrete symmetry under the exchange  
$G \leftrightarrow G'$ when all O particles (fermions, Higgses and gauge fields) 
exchange places with their M twins (`primed' fermions, Higgses and gauge fields).    

Mirror matter, invisible in terms of ordinary photons but gravitationally coupled  to our matter, 
could make a part of cosmological DM.
If mirror parity is an exact symmetry, 
then for all ordinary particles: the electron $e$, proton $p$, neutron $n$, photon $\gamma$, neutrinos $\nu$ etc., 
with interactions described by the Standard Model $SU(3)\times SU(2)\times U(1)$,  
there should exist their mirror twins: $e'$, $p'$, $n'$, $\gamma'$, $\nu'$ etc.\ 
which are sterile to our strong, weak and electromagnetic interactions  
but have instead their own gauge interactions   $SU(3)'\times SU(2)'\times U(1)'$ with exactly the same  coupling constants. 
Thus, we need no new parameters for describing mirror physics: 
ordinary and mirror particles are degenerate in mass, 
and  O and M sectors have identical  microphysics
at all levels from particle to atomic physics. 
In addition, the cosmological fraction of mirror baryons $\Omega'_B$ 
should be related to the dark baryon asymmetry  
as the fraction of ordinary baryons  $\Omega_B$ is related to our baryon asymmetry, 
and baryon asymmetries in two sectors should be related to the same baryogenesis mechanism. 

One could think that O and M worlds, having identical particle physics, 
should also have identical cosmological realisations.
However, if one naively  takes  $\Omega'_B = \Omega_B$, 
then M matter is not sufficient for explaining the whole amount of DM, 
and other type of DM should be introduced to obtain $\Omega_{DM} \simeq 5 \Omega_B$.  
On the other hand, if two sectors have the same temperature, $T'=T$, 
this  would strongly  disagree with  the Big Bang Nucleosynthesis (BBN) limits 
on the effective amount of light degrees of freedom: 
the contribution of M particles in the universe expansion rate at the BBN epoch 
would be equivalent to the amount of extra neutrinos $\Delta N_{\rm eff} = 6.15$, 
while at most  $\Delta N_{\rm eff} \simeq 0.5$ is allowed by the present constraints. 
In addition, due to self-interacting and dissipative nature of mirror baryons, 
$T'=T$ would be in full disagreement with the precision cosmological tests on the 
cosmic microwave background (CMB) anisotropies and the large scale structures (LSS) 
of the Universe, even if mirror baryons constitute a smaller fraction of cosmological DM, 
with $\Omega'_B = \Omega_B$.\footnote{By these reasons, mirror matter was not considered  as a serious candidate 
for Dark Matter for a long time, though some interesting works were done 
\cite{Mirror4,Khlopov:1989fj,Hodges:1993yb}. Also deformed asymmetric versions of mirror 
matter were considered, with electroweak scale larger than the ordinary one, where 
the atoms are heavier and more compact  \cite{Berezhiani:1995am,Berezhiani:1996sz,Mohapatra:1996yy}, 
and where the sterile mirror neutrinos, being much heavier than 
ordinary ones, could also constitute a DM fraction
\cite{Akhmedov:1992hh,Berezhiani:1995yi}. Such models have interesting implications also 
for the axion physics \cite{Berezhiani:2000gh,Berezhiani:1999qh}. 
} 

All these problems problems can be settled at once,
 if we assume that after inflation the two sectors were heated to different temperatures, 
and the temperature of the mirror sector $T'$ remained less than the ordinary one $T$ 
over all stages of the cosmological evolution  \cite{Berezhiani:2000gw}. 
The condition $T' < T$ can be realized by adopting the following {\it paradigm}: 
at the end of inflation the O- and M-sectors  are (re)heated 
in an non-symmetric way, with $T > T'$,  which can naturally occur  
in the context of certain inflationary models; 
the possible particle processes between O and M sectors should be
slow enough and cannot bring two worlds into the 
equilibrium after the (re)heating,  
so that both systems evolve almost adiabatically and the 
temperature asymmetry $T' /T$  remains nearly invariant in 
all subsequent epochs until the present days. 
In this way Mirror matter, with its atoms having the same mass as the ordinary ones,  
could constitute a viable candidate for asymmetric Dark Matter 
despite its collisional and dissipative nature.

Various potential consequences  of mirror world which are worth of theoretical and experimental studies
can be classified in three main parts: 

{\bf A.} {\it Cosmological implications of M baryons.}   
The basic question is, how small the temperature ratio $T'/T$ should be, 
and, on the other hand, how large the ratio  $\Omega'_B/\Omega_B$ between 
M and O baryon fractions can be,  to make the concept of mirror matter cosmologically plausible. 
The BBN limits demand that $T'/T < 0.5$ or so,  which is equivalent to $\Delta N_{\rm eff} = 0.5$ 
 \cite{Berezhiani:2000gw}. 
The stronger limit  $T'/T < 0.3$ or so comes from cosmological considerations,    
by requiring the early enough decoupling of M photons which makes M baryons practically 
indistinguishable from the canonic Cold Dark Matter (CDM) in observational tests 
related to the large scale structure formation and CMB anisotropies  
\cite{Berezhiani:2000gw,Ignatiev,Berezhiani:2003xm,Berezhiani:2003wj,Ciarcelluti:2004ip}. 
The above limits apply independently whether M baryons constitute DM entirely, or only 
about 20 per cent fraction of it, when $\Omega'_B \simeq \Omega_B$ 
\cite{Berezhiani:2003xm,Berezhiani:2003wj}.   In this case the remained 80 per cent of DM 
should come  from other component, presumably some sort of the CDM 
represented by particles belonging to the so-called WIMP class of DM candidates, 
by axion, or by other sort of hidden gauge sectors with heavier shadow baryons 
as in the case of asymmetric mirror matter \cite{Berezhiani:1995am,Berezhiani:1996sz,Mohapatra:1996yy},
considered in our previous paper \cite{mirror2015}. 
On the other hand, if DM is represented entirely by M baryons, i.e. $\Omega'_B \simeq 5 \Omega_B$,  then 
the requirement of the formation of the normal galaxies with masses larger than $10^9 M_\odot$  
gives $T'/T < 0.2$ or so while the power of smaller galaxies will be suppressed by Silk damping 
\cite{Berezhiani:2000gw,Ignatiev}.  
Hence, cosmological evolution of the density perturbations of M matter is compatible 
with the observed pattern of the cosmological large scale power spectrum   
and the CMB anisotropies if M sector is cold enough,  $T'/T < 0.2$ or so,  
while its collisional and dissipative nature can have specific observable implications 
for the  evolution and formation of the structures  at smaller scales, formation of  galaxy halos and stars, etc. 
(for reviews, see e.g. ~\cite{Berezhiani:2003xm,Berezhiani:2005ek,Berezhiani:2008zza}).

Regarding the BBN era in M sector, as far as $T' < T$, its baryonic content should be more neutron rich 
 than in the O world  since the weak interactions freeze out at higher temperatures
and thus the neutron to proton ratio remains large. Hence,  M sector should be helium dominated.    
 In particular, for $T'/T < 0.3$,  
 M world  would have only $25~\%$  mass fraction of mirror hydrogen  and  $75~\%$ of mirror helium-4 
 \cite{Berezhiani:2000gw}, 
 against the observed mass fractions of ordinary hydrogen ($75~\%$) and helium-4 ($25~\%$).   
 In addition, M world can have also somewhat larger primordial metallicity than our sector. 
 All this should have direct implications also for the formation and evolution of 
 mirror stars \cite{Berezhiani:2005vv} which produce also 
 heavier mirror elements as oxygen, carbon etc.   
 Future astrophysical and cosmological observations might accomplish 
a consistent picture of the mirror matter as Dark Matter.

 Interestingly, the condition $T' < T$  have important implications also for the primordial baryogenesis,  
 in the context of  in the context of co-baryogenesis scenarios  
\cite{Bento:2001rc,Bento:2002sj,Berezhiani:2003xm,Berezhiani:2005hv}. 
These scenarios are based  $B$ or $L$ violating interactions which mediate 
the scattering processes that transform O particles into the M ones at the post-inflationary reheat epoch. 
Once these processes violate also CP due to complex coupling constants,  
while their departure from equilibrium  is already implied by the condition $T' < T$,   
all three Sakharov's conditions  can be naturally satisfied. In this way, these scenarios co-generate 
baryon asymmetries in both O and M sectors. Remarkably,  the condition $T' /T < 0.2$ 
leads to a prediction $1 \leq \Omega'_B/\Omega_B \leq 5$ \cite{Berezhiani:2003xm,Berezhiani:2005hv}   
which  sheds a new light to the baryon and dark matter coincidence problem.

 {\bf B.} 
 {\it Particle interactions  between two sectors and oscillation phenomena.} 
A straightforward and experimentally direct  way to establish existence of mirror matter is the 
experimental search for oscillation phenomena between ordinary and mirror particles. 
In fact, any neutral particle, elementary (as e.g. neutrino) or composite (as the neutron or hydrogen atom)  
can mix with its mass degenerate twin from the parallel sector leading to a matter 
disappearance (or appearance) phenomena which can be observable in laboratories. 
E.g., the kinetic mixing between ordinary and mirror photons \cite{Holdom} induces 
the positronium oscillation into mirror positronium which would imitate the invisible channel 
of the positronium decay \cite{Glashow:1985ud,Gninenko:1994dr}. 
The interactions mediated by heavy gauge bosons 
between particles of two sectors, which can have e.g. a common 
flavour gauge symmetry \cite{Berezhiani:1996ii} or common gauge $B-L$ symmetry \cite{Addazi:2016rgo}
can induce mixing of neutral pions and Kaons with their mirror twins. 
The oscillation phenomena between ordinary (active) and mirror (sterile) neutrinos can have 
interesting observational consequences \cite{Foot:1995pa,Berezhiani:1995yi}. 
Interestingly, the present experimental bounds allow 
the neutron oscillation phenomena between two sectors to be rather fast 
\cite{Berezhiani:2005hv}, with interesting astrophysical and experimental implications 
\cite{Mohapatra:2005ng,Berezhiani:2006je,Berezhiani:2011da,Berezhiani:2008bc,Berezhiani:2012rq,Berezhiani:2016ong}. 
In this respect, the relevant interaction terms between O and M particles 
are the ones which violate baryon $B$ and lepton $L$ numbers of both sectors and which 
can be  at the origin of co-baryogenesis scenarios \cite{Bento:2001rc,Bento:2002sj,Berezhiani:2003xm,Berezhiani:2005hv}.

{\bf C.}  {\it Interaction portals and direct detection.}  Mirror matter can interact with ordinary matter via different portals
 in ${\cal L}_{\rm mix}$, 
 e.g. via kinetic mixing of M and O photons, or mass mixing of M and O pions or $\rho$-mesons, 
 or via contact interaction terms $\frac{1}{M} \bar q \gamma_\mu q \bar q' \gamma_\mu q'$ 
between O and M quarks which can be mediated by extra gauge bosons connecting two sectors 
\cite{Berezhiani:1996ii}. 
 In particular, there is not just one Dark Matter 
particle, as in most of well-motivated Dark Matter models, but it could consist of different atoms, 
from the primordial hydrogen and  helium as dominant components, 
to reasonable fractions heavier elements as carbon, oxygen, etc. produced in mirror stars.   
The experimental direct searches of the the particle DM should be concentrated  
on the detection of mirror helium as most abundant mirror atoms. 
In fact, the region  of Dark Matter masses  below 5 GeV  is practically unexplored.  
 In any case,  for any realistic chemical composition of M sector,  
 we know its mass spectrum of possible atomic/nuclear structures 
directly from our physical experience, 
with enormous empirical material available for ordinary atoms.
Therefore, the only unknown in this puzzle is related to the interaction portal.

In this paper we mainly concentrate on this latter issue. 
In particular,  we analyse the annual modulation observed by DAMA in the framework of 
mirror matter, exploiting the interaction portal related to  the 
{\it photon-mirror photon kinetic mixing} term \cite{Holdom} 
\begin{equation}\label{eps}
 \frac{\epsilon}{2} \, F^{\mu\nu} F'_{\mu\nu} 
\end{equation} 
with a small parameter $\epsilon \ll 1$. 
This mixing renders the mirror nuclei mini-charged with respect of ordinary electromagnetic force,  
and thus mediates the scattering of  mirror nuclei with ordinary ones 
with the Rutherford-like cross sections.  
The implications of this detection portal was discussed in Refs. \cite{Foot:2003iv,Foot:2012cs}. 
In our previous paper \cite{mirror2015} we discussed it for the asymmetric mirror dark matter. 
In this paper we perform a detailed analysis of this signal in the NaI(Tl) detectors at DAMA/LIBRA set-up 
for exact mirror matter, for different realistic chemical compositions of mirror sector 
 (while the dominant components should be M hydrogen and mirror helium-4, 
M sector can contain a  mass fraction of heavier mirror atoms as Oxygen, Carbon, etc. up to few per cent),  
for different local temperatures and velocity flows of the mirror gas in the Galaxy.

The paper is organized as follows. In Section 2 we give a brief overview of 
mirror Dark Matter discussing its properties and possible distributions in the Galaxy. 
In Section 3 details of the analysis are given for its 
direct detection  possibilities via photon-mirror photon kinetic mixing 
in the NaI(Tl) detectors of DAMA/LIBRA experiment,  
  while in Section 4 we discuss the obtained results.

\bigskip
\bigskip 
\bigskip

\section{ Mirror matter properties, its distribution and chemical composition in the Galaxy}

{\bf How large fraction of  mirror matter can be produced in baryogenesis?} 
The baryogenesis in the two sectors, ordinary and mirror, emerges by the same mechanism, 
since the particle physics responsible for baryogenesis is the same in the two sectors 
(coupling constants, CP-violating phases, etc.).  
However, the cosmological conditions at the baryogenesis epoch can be 
different (recall that the shadow sector must be colder than the ordinary one). 
One can consider two cases: 

1) {\it Separate baryogenesis}, when the baryon asymmetry in each sector is generated 
independently but by the same mechanism. 
In this case, in the most naive picture when out-of-equilibrium conditions are well satisfied 
in both sectors, one predicts $\eta = n_B/n_\gamma$ and $\eta' = n'_B/n'_\gamma$ 
must be equal, while $n'_\gamma/n_\gamma \simeq x^3 \ll 1$, where $x = T'/T$ is 
the temperature ratio between mirror and ordinary worlds in the early Universe.   
In this case, we have $\Omega'_B/\Omega_B \simeq x^3 \ll 1$.  
Therefore, if e.g. $x = 0.5$,  a limit from BBN,  we have  
$\Omega'_B/\Omega_B \simeq 0.15$ or so.
However, one should remark that  
due to different out-of equilibrium conditions in the two sectors 
situation with $\eta' \gg \eta$ can be also obtained in some specific parameter space, 
where the case $\Omega'_B > \Omega_B$ can be achieved   
\cite{Berezhiani:2000gw}.

2) {\it Co-genesis} of baryon and mirror baryon asymmetries via $B-L$ and 
CP-violating processes between the ordinary and mirror particles, e.g. 
by the terms $ \frac{1}{M}\, l l' H  H' $ in ${\cal L}_{\rm mix}$ which also 
 induce mixing between ordinary (active) and mirror (sterile) neutrinos,  
 and which can be mediated by heavy ``right-handed'' neutrinos  coupled to 
both sectors as e.g. \cite{Bento:2001rc,Bento:2002sj}.  
In perfect out-of equilibrium conditions, 
when $x=T'/T \ll 1$ and so $n'_\gamma/n_\gamma \simeq x^3 \ll 1$,  
this leptogenesis mechanism predicts $n'_B = n_B$ and thus 
$\Omega'_B = \Omega_B$. In this case the cosmological fractions of ordinary 
and mirror baryons are equal, i.e. mirror matter can constitute only about 20 per cent 
of Dark Matter in the Universe, and some other type of Dark Matter
should be invoked for compelling the remaining 80 per cent.   
However, if the out-of-equilibrium is not perfect, 
then generically final $T'/T$ increases and one has $\Omega'_B /\Omega_B> 1$. 
Taking e.g. $T'/T < 0.2$, cosmological  limit  at which mirror matter with 
$\Omega'_B > \Omega_B$ is still allowed by the CMB and large scale tests, 
we get an upper limit $\Omega'_B/\Omega_B < 5 $ or so. 
In this way, mirror matter could represent an entire amount of Dark Matter  
\cite{Berezhiani:2003xm,Berezhiani:2005ek,Berezhiani:2006ac,Berezhiani:2008zza}.

{\bf How large fraction of  mirror matter can be allowed by cosmological constraints?} 
Interestingly, for $T'/T < 0.2$, the cosmological tests (LSS and CMB) are compatible 
with the situation when DM is entirely represented by mirror baryons, and 
mirror Silk-damping allows formation of the normal size galaxies 
\cite{Berezhiani:2000gw,Ignatiev,Berezhiani:2003xm,Berezhiani:2003wj}. 

More difficult question is the distribution of the mirror matter in the galaxy and halo 
problem. At first glance M baryons, having the same physics as O matter, 
cannot form extended galactic halos but instead should form the disk, as usual matter does. 
If so, the situation with $\Omega'_B \simeq 5 \Omega_B$ is excluded by observations, 
however $\Omega'_B \simeq \Omega_B$ remains acceptable. 
There should exist two disks in the Milky Way (MW), one visible and another invisible 
and perhaps of different radius and thickness,  with comparable amount of O and M components.   
It is known that the total surface density of matter in the MW disk at the region of the sun 
is about $(68 \pm 4)~ M_\odot$/pc$^2$ \cite{bovy2013}, while the ordinary matter can account for 
a fraction $ (38 \pm 4) M_\odot$/pc$^2$ or so \cite{bovy2013}.
Therefore, the surface density of mirror matter can be $ (30 \pm 6) M_\odot$/pc$^2$, perfectly compatible 
for the presence of dark disk similar to ours in MW. In fact, this 
would not contradict to the shape of the rotational velocities if the dark mirror  
disk is somewhat more thick than ordinary disk, and the mirror bulge is more extended 
than ours. 

In this case, the remaining fraction of DM which should form galactic halos 
could come from particles belonging to the so-called WIMP class of DM candidates, from axions or from some other 
parallel gauge sector, like asymmetric mirror matter considered in our previous paper 
\cite{mirror2015}. 
Interestingly, if there may be particles belonging to the so-called WIMP class of DM candidates
of ordinary sector, then mirror ``WIMPs'' should give 
less contribution since M sector is colder, as well as contribution of mirror neutrinos 
should be smaller than that of ordinary ones \cite{Berezhiani:2000gw}.  
Ordinary and mirror axions could give comparable contributions in DM.
In any case, in what follows, we do not require that mirror baryons 
provide entire amount of DM, but  we assume that it provides some fraction $f$ 
of DM which we shall keep as an arbitrary parameter, taking $f=0.2$ as a benchmark 
value.

The case whether mirror matter could be entirely Dark Matter, 
is difficult and it requires additional discussion. 
The main problem is related to galactic halos. At first glance mirror matter, 
having the same microphysics as ordinary matter, cannot form extended 
galactic halos. However,  this can be possible if mirror stars are formed earlier 
than ordinary stars, and before the mirror matter collapsed into the disk.\footnote{
One can consider also the possibility of the modified gravity in the context of 
bigravity theories \cite{Berezhiani:2007zf,Berezhiani:2009kx} 
when O and M sectors have their own 
gravities described by two different metrics $g_{\mu\nu}$ and $g'_{\mu\nu}$, 
and instead of universal Hilbert-Einstein action (\ref{S-grav}), the theory is described 
by the action of the form $S = \int d^4x  \left[  \sqrt{-g} \big(\frac12 M_P^2 R + {\cal L}\big) +
\sqrt{-g'} \big(\frac12 M_P^2 R' + {\cal L'}\big) +  \sqrt[4]{g g'}\big({\cal V}_{\rm mix} + {\cal L}_{\rm mix} \big)\right]$ 
where ${\cal V}_{\rm mix}(g^{\mu\rho}g'_{\rho\nu})$ is a mixed function 
of two metrics.  
In this situation one could have anti-gravitation phenomena between ordinary and Dark Matter 
at short distances and the galactic rotational curves can be well described without the need of halos, 
when mirror matter is entirely distributed in the disk \cite{Berezhiani:2009kv}.  
}
  
However, one has to take into account the possibility that in the galaxy evolution
dissipative M matter,  during its cooling and contraction fragments  into 
molecular clouds in which cool rapidly and form the stars. Star formation, 
and moreover of the first stars, is a difficult question, however, by formal Jeans criteria, 
in M matter which is cooler and also helium dominated, the Jeans mass is smaller 
and star formation could be more efficient. In this way, mirror matter forming the 
stars could form, during the collapse, dark elliptical galaxies, perfectly imitating halos, 
 while some part of survived gas could form also a dark mirror  disk.     
 In other words, we speculate on the possibility that due to faster star formation 
 M baryons mainly form elliptical galaxies. For comparison, in MW less than one per mille of 
 mass is contained in globular clusters  and halo stars which were formed 
 before disk formation. In MW there are up to 200 globular clusters  
 orbiting in the Galaxy halo at distances of 50 kpc while 
 some giant elliptical galaxies, particularly those at the centers of galaxy clusters 
 can have as many as $10^4$ globular clusters containing the overall mass 
 $\sim 10^9-10^{10}~M_\odot$. In mirror sector, if fragmentation in molecular clouds and stars 
 is more efficient, stars are smaller and evolving faster, the elliptical galaxy can be formed 
 by mirror stars in which ordinary matter goes mainly into disk 
 (and also faster stellar evolution is important.) 
 It is also possible that the mass function and chemical composition of these stars 
 is balanced so that many of them could form black holes with masses $10-30~M_\odot$, 
 and among those binary black holes.  
 This can be interesting also in view of the recent publication about gravitational 
 wave signals from such a heavy black holes in the galaxies
  \cite{Abbott:2016blz}. 
 Also this can have implications for central black hole 
 formation \cite{Berezhiani:2000gw}. 
 
For Dark Matter direct detection experiments, it is important that mirror matter, being self-interacting and 
dissipative, cannot have the same density and velocity distributions in the Galaxy as canonical Cold Dark Matter. 
As far as a big fraction of mirror matter can exist in the form of mirror stars, one can rather expect that only 
the gas contained in the disk component is relevant for direct detection. In principle, the mirror disk 
can be co-rotating or counter-rotating with respect to ordinary disk, while the mirror gas at the present locality 
of the sun in the Galaxy can exist in the same forms that we know for the ordinary interstellar gas. 
Namely, it can be present in the form varying from cold molecular cloud, with temperatures $T\sim 10$~K,  
to warm neutral medium with $T \sim 10^4$~K  and hot ionized medium with $T \sim 10^7$~K.  
This medium can have a local peculiar flow velocity in the galactic frame which 
can be dependent on the galactic coordinates and can have 
a value of few hundreds of km/s and certain orientation with respect to sun's velocity.  
In addition, in the rest frame of this medium the mirror particles will have thermal velocities
which will be dependent on the particle mass.
In this case the angle $\alpha$ between the Sun velocity and the local peculiar flow velocity 
can be tested by the phase of the experimental signal in a way independent on the thermal distribution velocity.
In the following we consider  situations with different benchmark values of the local peculiar flow velocity and 
of the thermal velocities. In view that mirror Dark Matter is supposed to be multi-component, 
consisting of not only hydrogen and helium but containing also some significant amount of heavier 
mirror atoms, the dependence of thermal velocity on the particle mass makes the predictions 
different from the CDM case when dark particles would have  the same 
pseudo-maxwellian velocity distribution independent on their masses.

{\bf Chemical composition of mirror matter.} 
As far as at the mirror BBN epoch the universe expansion rate is dominated 
by O matter density, the weak interaction's freezing in M sector occurs earlier 
and frozen ratio of neutrons to protons is larger than in O nucleosynthesis. 
As a result, primordial chemical content of M sector is helium dominated, 
with $^4$He$'$ constituting up to 80 per cent of mass fraction of M baryons 
in the limit $T'/T \to 0$ \cite{Berezhiani:2000gw}. 
In the following we take mirror helium-4 benchmark mass fraction as 75 \%,
and mirror hydrogen as 25 \%. The primordial chemical content in mirror sector 
should also have larger metallicity that in ordinary one, but the primordial mass fraction 
of the heavier elements is anyway negligible. 

\begin{table}
\begin{center}
\caption{Abundance of elements in the Solar System.}
\begin{tabular}{cccc}
\hline
Isotope & $(Z,A)$  & Mass fraction  & Atom fraction \\
        &   & (in per cents) & (in per cents)  \\
\hline \hline
H &  (1,1)  & 70.57   & 91.0 \\
He & (2,4)  & 27.52 & 8.87\\
C & (6,12)  & 0.30  & 0.032\\
N & (7,14) & 0.11  & 0.010\\
O & (8,16) & 0.59   & 0.048\\
Ne & (10,20) & 0.15  & 0.010 \\
Si & (14,28)  & 0.065 & 0.0030 \\
Fe & (26,56)  & 0.117  & 0.0027\\
\end{tabular}
\label{ssystemiso}
\end{center}
\end{table}

However, heavier elements should be produced in stars and thrown in the galaxy via 
supernova explosions. In  O sector, the chemical elements with $A\sim 16$ as Oxygen, Carbon,  
Nitrogen and Neon account for about a per cent of mass fraction, 
while heavier elements are less abundant,  accounting in whole for about 4 per mille 
of mass fraction. 
In mirror sector, these proportions can be quite different. 
One can imagine one extreme possibility that mirror stars are typically light and do not end up 
as supernovae, or the gravitational collapse of heavier mirror stars typically leads to black hole formation 
rather than to supernova at the final stage. In this case the chemical content of mirror gas will 
be essentially the same as the primordial content. i.e. dominated by helium and hydrogen. 
On the other extreme, one can imagine that the star formation in M sector  can be more efficient, 
including the heavier stars with mass $> 10~M_\odot$. 
As it was studied in Ref. \cite{Berezhiani:2005vv}, the evolution of the latter is at least an order of 
magnitude faster than for ordinary heavy stars,  they can produce many supernovae 
and so the heavier elements in M sector could be more abundant than in ordinary sector. 

\begin{table}[!b]
\caption{Typical chemical compositions of mirror matter; the mass fraction 
of different mirror atoms for some benchmark scenarios is reported.}
\begin{center}
\begin{tabular}{cccccc}
\hline
Mirror matter composition & H (\%) & He (\%)& C (\%)& O (\%)& Fe (\%)  \\
     &       &    &   &   &      \\
\hline\hline
     &       &    &   &   &      \\
 H$'$, He$'$     &  25  &  75  & -  & -  &  -    \\
     &       &    &   &   &      \\
 H$'$, He$'$, C$'$, O$'$  &  12.5  &  75.  & 7.  & 5.5  &  -     \\
     &       &    &   &   &      \\
 H$'$, He$'$, C$'$, O$'$, Fe$'$    & 20  &  74   & 0.9  & 5.  &  0.1     \\
     &       &    &   &   &      \\
\end{tabular}
\label{tb:specie}
\end{center}
\end{table}

We assign to the mirror atoms a cosmological abundances directly rescaled from the abundances in 
ordinary sector (for reference, Table \ref{ssystemiso} shows the benchmark values 
for mass and atom fractions of different elements in the solar system). 
Table \ref{tb:specie} reports typical chemical composition of mirror matter under different
assumptions: a) only primordial nuclei (H$'$, He$'$); b) CNO elements also present; 
c) also Fe$'$ generated by mirror supernovae explosion present.

\section{Analysis}

In the framework of the considered mirror model, the Dark Matter particles are expected
to form, in the Galaxy, clouds and bubbles with diameter which could be 
even as the size of the solar system. In this modeling a dark halo, at the present epoch, is 
crossing a region close to the Sun with a velocity
in the Galactic frame that could be, in principle, arbitrary.
Hereafter we will refer to such local bubbles simply as halo. 
The halo can be composed by different species of mirror DM particles (different
mirror atoms) that have been thermalized 
and in a frame at rest with the halo. They have a velocity distribution
that can be considered maxwellian with the characteristic velocity 
related to the temperature of the halo and to the mass
of the mirror atoms. We assume that the halo has its own local equilibrium temperature, 
$T$, and the velocity parameter of the $A'$ mirror atoms is given by
$ \sqrt{2 k_{\scriptscriptstyle B} T / M_{\scriptscriptstyle A'} }$.
In this scenario lighter mirror atoms have bigger velocities than 
the heavier ones, on the contrary of the CDM model where the velocity distribution is
mass independent.  
If we extrapolate this assumption for electrons, in the case of hot ionized plasma  
with $T \sim 1$ keV, electron recoils 
due to  elastic scattering of mirror electrons 
and ordinary electrons could also be relevant. 
In this case even some reasonable fraction of hot ionized mirror medium
could give a contribution to the signal in the detector.
However this contribution is model dependent since generically in the astrophysical plasma
the temperature of the electrons can be different from that of the ions. 
Therefore, in this paper we do not concentrate on  this contribution. 

The expected phase of the annual modulation signal induced by
the mirror particles depends on the halo velocity (module and direction) with respect to 
the laboratory in the Galactic frame.
A detailed study on the behaviour of the annual modulation phase as a 
function of the halo velocity will be presented
in the next section where we will show -- without loosing generality -- that we can consider 
the case of a dark halo moving either parallel or anti-parallel to the Earth
in the Galactic frame.

\subsection{The study of the annual modulation phase}
\label{annph}

We will use the Galactic coordinate frame, 
that is $x$ axis towards the Galactic 
center, $y$ axis following the rotation of the Galaxy
and the $z$ axis towards the Galactic North pole. 
In the following the velocity of any object can be presented as a 
vector $\vec{v} = (v_{x},v_{y},v_{z})$.

The velocity of the DM particles in the laboratory frame (reference system 
related to the Earth) can be written as: 
\begin{equation}
\vdm = \vdmp - \ve  
\label{eq:vdm}
\end{equation}
where $\vdmp$ and $\ve$ are
the velocities of the DM particles 
and of the Earth in the Galactic frame, respectively.
The DM particles, as described before, are enclosed inside a halo 
which is moving in the Galaxy with a constant velocity, $\vhalo$. 
In a frame at rest with the halo, 
the DM particles have a velocity, $\vdms$, 
that follow a maxwellian distribution, $F$, depending on the assumed temperature of the system:
\begin{equation}
F_{A',halo} (\vdms) = \mathcal{A} \; e^{-\frac{\vec{v_{\scriptscriptstyle DM}^{''2}}}{\vzh^2}}
\end{equation}
where $ \mathcal{A}$ is a normalization constant and $\vzh$ is the velocity parameter 
of the distribution related to the temperature, $T$, of the halo. 
If one considers a halo composed by mirror atoms of specie $A'$ with $M_{\scriptscriptstyle A'}$ mass
then $ \vzh = \sqrt{2 k_{\scriptscriptstyle B} T / M_{\scriptscriptstyle A'} }$,
where $k_{\scriptscriptstyle B}$ is the Boltzmann constant.

Since $\vdmp = \vdms + \vhalo$, by Eq. \ref{eq:vdm} one gets:
\begin{equation}
\vdms = \vdm + \ve - \vhalo  
\end{equation}
The Earth velocity $\ve$ in the Galactic frame can be expressed as the sum of the 
Sun velocity, $\vs$, and of the revolution velocity of the Earth 
around the Sun, $\vr$. Here we neglect the contribution
of the rotation of the Earth around its axis which gives a very small effect 
on the annual modulation phase (it gives also rise to a diurnal modulation
effect which is not of interest in this paper; see Ref. \cite{diu} for more details). 
Note that $\vr$ depends on the sidereal time, $t$.

The velocity $\vs = \vec{v}_{LSR} + \vec{v}_{\odot} $ can be written as the sum of the 
velocity of the Local Standard of Rest (LSR) 
due to the rotation of the Galaxy  (local rotation velocity of matter in the Milky Way) 
$\vec{v}_{LSR}=(0,v_0,0)$, and of the Sun peculiar velocity with respect to LSR
$\vec{v}_{\odot} = (11.10, 12.24, 7.25)$ km/s \cite{scho09,delh65}. 
The parameter $v_0$ is the Galactic local velocity;
the estimate of $v_0$ ranges from $(200\pm20)$ km/s and $(279\pm33)$ km/s
depending on the model of rotational curve used in its evaluation \cite{mcmi10}.
Although the interval of possible values of $v_0$ is rather large, in the present analysis we adopt
for illustration $v_0 = (220\pm50)$ km/s \cite{allDM4,astr_v0} (uncertainty at 
90\% C.L.). In such a case, 
one has $\left |\vs \right | = (232 \pm 50)$ km/s. 

Hence, the velocity distribution of the DM particles ($A'$ mirror atoms) in the laboratory frame becomes:
\begin{equation}
F_{A'} (\vdm) = \mathcal{A} \; e^{-\frac{ \left ( \vdm+\ve-\vhalo \right)^2} {\vzh^2}}
\label{eq:maxw}
\end{equation}

The annual modulation of the counting rate and its phase depend on the relative velocity 
distribution of the DM particles
with respect to the laboratory frame (Eq. \ref{eq:maxw}). Thus, once averaging over the angles,
they depend on the module of $\vrel = \ve - \vhalo $. 
Since $\left| \vrel \right|$ depends on the 
time revolution of the Earth around the Sun,   
the counting rate shows the typical modulation behaviour: $\mathcal{S}(t) = \mathcal{S}_0 + \mathcal{S}_m cos\omega(t-t_0)$
where t$_{0}$ is the phase of the annual modulation and $T_p=2\pi/\omega= 1$ sidereal year is the period.

In the following we calculate the expected phase t$_{0}$ as a function of the halo velocity.

The motion of the Earth around the Sun can be worked out
by using the ecliptic coordinate system $(\hat{e}^{ecl}_1, \hat{e}^{ecl}_2, \hat{e}^{ecl}_3)$,
where the $\hat{e}^{ecl}_1$ axis is directed towards the vernal equinox and $\hat{e}^{ecl}_1$
and $\hat{e}^{ecl}_2$ lie on the ecliptic plane. The right-handed convention is used.
In the Galactic coordinates, we can write (see Ref. \cite{diu} for details):
\begin{eqnarray}
\hat{e}^{ecl}_1 & = & (-0.05487,  0.49411, -0.86767), \nonumber \\
\hat{e}^{ecl}_2 & = & (-0.99382, -0.11100, -0.00035), \nonumber \\
\hat{e}^{ecl}_3 & = & (-0.09648,  0.86228,  0.49715). 
\end{eqnarray}
The ecliptic plane is tilted with respect to the galactic plane by $\approx 60^o$,
as $\hat{e}^{ecl}_3 \cdot (0,0,1) = 0.49715$. So the evolution of the Earth in the ecliptic plane can be described as:
\begin{equation}
\vr = v_{ov} \; (\hat{e}^{ecl}_1\sin\lambda(t) - \hat{e}^{ecl}_2\cos\lambda(t))
\label{eq:verv}
\end{equation}
where $ v_{ov} $ is the orbital velocity of the Earth which has a weak dependence on time 
due to the ellipticity of the Earth orbital motion around the Sun. Its value ranges
between 29.3 km/s and 30.3 km/s; for most purposes it can be assumed constant 
and equal to its mean value $\simeq$ 29.8 km/s.
On the other hand, when more accurate calculations 
are necessary, the routines in Ref. \cite{starlink}  can be used:
they also take into account the ellipticity of the Earth orbit and the gravitational influence of
other celestial bodies (Moon, Jupiter, and etc.)
Moreover, the phase in Eq. \ref{eq:verv} can be written as $\lambda(t) = \omega (t-t_{equinox})$,
where $t$ is the sidereal time and $t_{equinox}$ 
is the spring equinox time ($\approx$ March 21).

The time-independent part of $\vrel$ is given by 
$ \vec{v}_{ti} = \vec{v}_{LSR} + \vec{v}_{\odot} - \vhalo$,
while the time-dependent one is $\vr$. Thus:
\begin{equation}
\left| \vrel \right| = \sqrt{v_{ti}^2 + v_{ov}^2 + 2\, \vec{v}_{ti} \cdot \vr}  
\label{eq:mvrel}
\end{equation}
The scalar product in the previous equation can be written as:
\begin{equation}
\vec{v}_{ti} \cdot \vr = v_{ov}(\vec{v}_{ti} \cdot \hat{e}^{ecl}_1\sin\lambda(t) - \vec{v}_{ti} \cdot \hat{e}^{ecl}_2\cos\lambda(t)).
\label{eq:scp1}
\end{equation}
Defining $\hat{v}_{ti} \cdot \hat{e}^{ecl}_1 = A_m \sin \beta_m $ and
$-\hat{v}_{ti} \cdot \hat{e}^{ecl}_2 = A_m \cos \beta_m $
which depend on the assumed DM halo velocity in the Galaxy $\vhalo$,
Eq. \ref{eq:scp1} becomes:
\begin{equation}
\vec{v}_{ti} \cdot \vr = v_{ov}\, v_{ti} \, A_m \cos(\lambda(t) - \beta_m) = 
v_{ov}\, v_{ti} \, A_m \cos \omega(t - t_0).
\label{eq:scp2}
\end{equation}
$A_m$ and $\beta_m$ can be calculated once the halo velocity and the $v_0$ value are fixed.
Then, substituting the Eq. \ref{eq:scp2} in Eq. \ref{eq:mvrel}, one gets:
\begin{equation}
\left| \vrel \right| = v_{med}\,\sqrt{1+2\delta \, \cos{\omega(t-t_0)}}
\label{eq:mvrel2}
\end{equation}
where 
\begin{equation}
\delta = \frac{v_{ov} \, v_{ti} \,A_m }{v_{ti}^2 + v_{ov}^2}.
\end{equation}
and $v_{med}=\sqrt{v_{ti}^2 + v_{ov}^2}$.

For those values of $\vhalo$ so that 
$v_{ti} \gg v_{ov} \simeq 30$ km/s, one gets $\delta \ll 1$, and:
\begin{equation}
\left| \vrel \right| \simeq v_{med}\,(1+\delta \, \cos{\omega(t-t_0)})
\label{eq:limvrel}
\end{equation}
that is the usual case of a DM halo at rest in the Galactic frame.  

In the general case the phase of the DM annual modulation is determined at the 
time when the argument of the cosine in Eq. \ref{eq:mvrel2} is null:
\begin{equation}
 t_0 = t_{equinox} + \beta_m/\omega,
\end{equation}
and $ \left| \vrel \right|$ assumes its maximal value.

In conclusion, the annual modulation phase depends on the module of the halo velocity (i) and 
on the relative direction of the halo 
with respect to the Earth velocity (ii).
The case of a mirror DM halo with 
a null velocity corresponds to the description generally considered for 
the DM halo in which it is at rest in the Galactic frame; in particular, in this case the 
expected phase of the annual modulation is around June 2$^{nd}$. 

In the present analysis 
we are interested only in scenarios compatible with the annual modulation phase measured experimentally
by DAMA. We recall that, considering the annual cycles collected with DAMA/NaI and
the annual cycles of DAMA/LIBRA-phase1, the best fit value of the phase obtained by the measured 
residual rate in 2-6 keV energy range is $144 \pm 7$ days \cite{modlibra3}. 

\begin{figure}[!th]
	\begin{center}
		\includegraphics [width=0.45\textwidth]{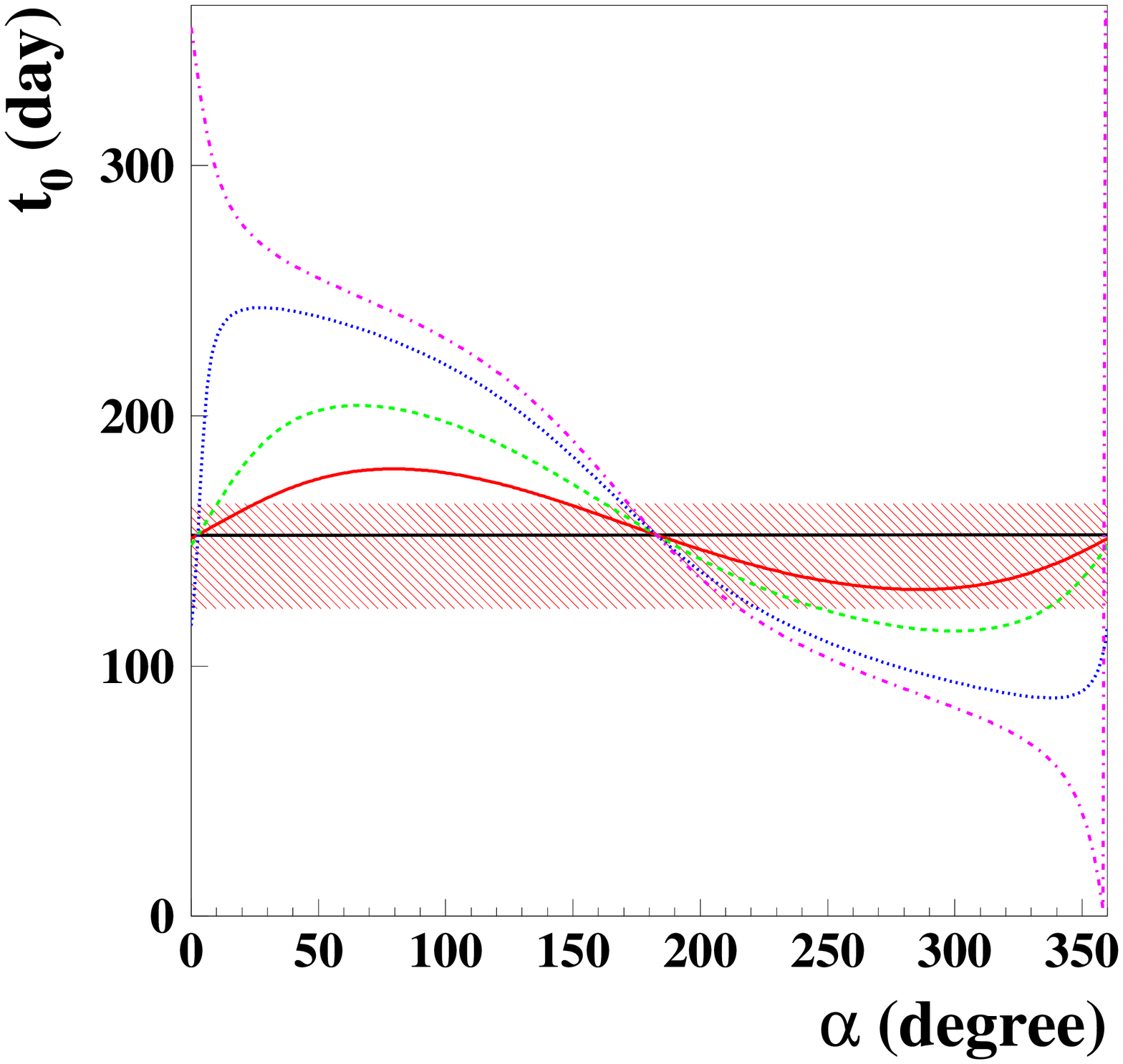}
		\includegraphics [width=0.45\textwidth]{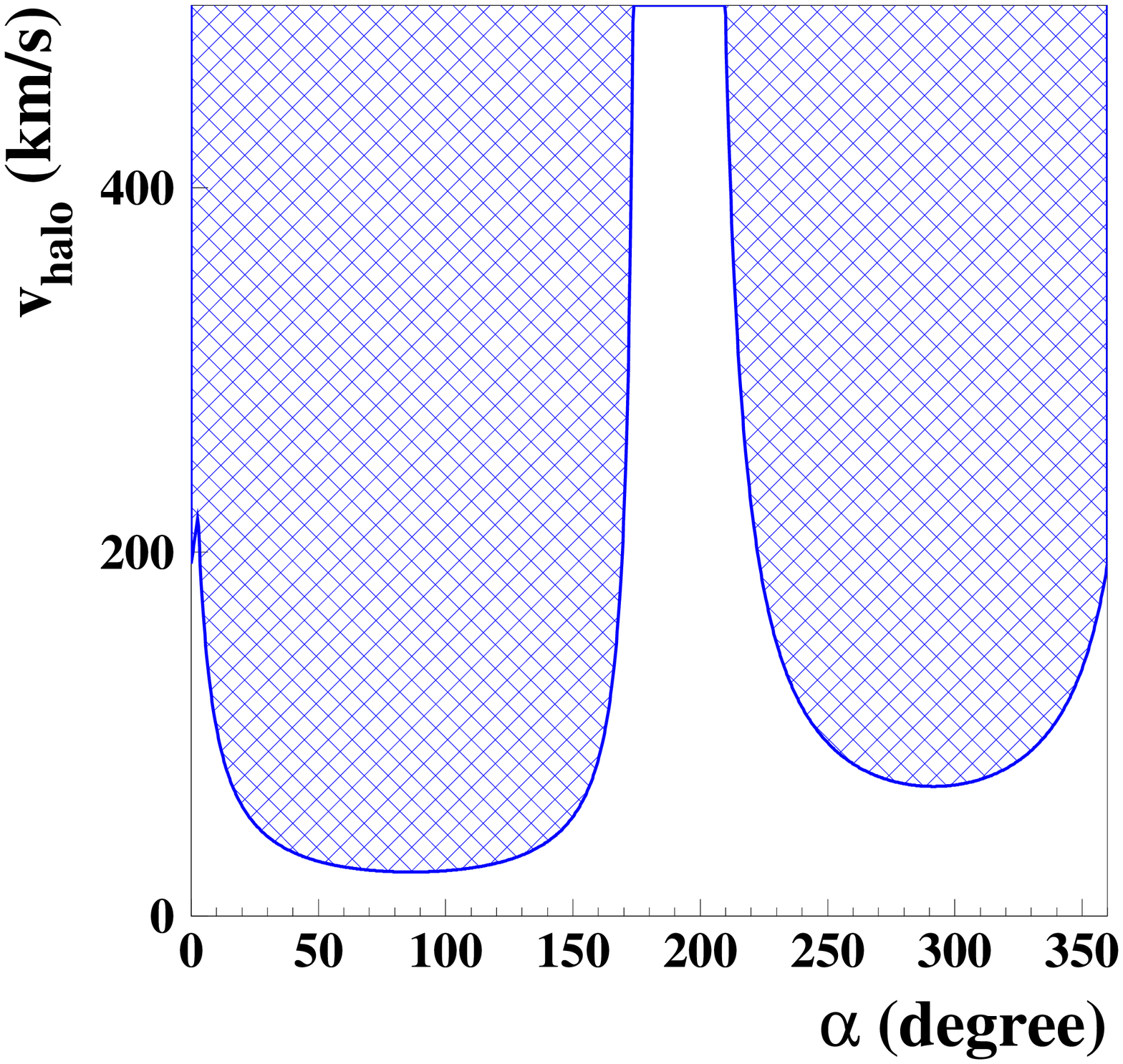}
	\end{center}
	\vspace{-0.8cm}
	\caption{{\it Left:} examples of expected phase of the annual modulation signal as a function of
		the angle, $\alpha$, between the Sun velocity and the velocity of the DM mirror halo moving in the Galactic plane.
		The different curves refer to different values of the module of the halo velocity: 300 km/s (dashed-dotted),
		200 km/s (dotted), 100 km/s (dashed), 50 km/s (solid). The shaded area (red on-line) is defined by
		the values of the phase of the annual modulation signal allowed at 3$\sigma$ by DAMA. For each
		halo velocity, only the values of $\alpha$ included inside the shaded area
		are allowed. The solid horizontal black line corresponds
		to a halo at rest in the Galactic frame ($\vhalo=0$) giving a phase equal to 152.5 day
		(June 2$^{nd}$).
		{\it Right:} the shaded regions in the plane $v_{halo}$ $vs$ $\alpha$ correspond to
		halo velocities (module and direction) giving a phase that differs more than 3$\sigma$ 
		from the phase of the annual modulation effect measured by DAMA. 
		These velocities in the shaded regions are thus excluded by the DAMA results at 3$\sigma$ C.L..
	}
	\label{fg:phase}   
\end{figure}

The curves in Fig. \ref{fg:phase}-{\it left} show, as examples for halos moving in the galactic plane, 
the expected phase of the annual modulation signal as a function of
the angle, $\alpha$, between the Sun velocity and the halo velocity: 
$\cos{\alpha} = \hat{v}_{\scriptscriptstyle S} \cdot \hat{v}_{halo}$;
they have been obtained for four different values of the halo velocity module.

As it can be easily inferred, when the halo velocity is anti-parallel to the Sun velocity
($\alpha \simeq \pi$) the phase of the annual modulation is $\simeq$ June 2$^{nd}$ for any module of
$\vhalo$. For parallel halo velocity ($\alpha \simeq 0$) depending whether or not $v_{halo}$ is larger 
than $v_S$ the phase of annual modulation can be even reversed.
The 3$\sigma$ region compatible with the DAMA annual modulation phase is also 
reported as shaded area (red on-line); the points included inside the shaded area
are allowed by the DAMA result. The solid horizontal black line corresponds
to a halo at rest in the Galactic frame ($\vhalo=0$) giving a phase equal to 152.5 day (June 2$^{nd}$).

The module of the halo velocity that corresponds to a phase compatible at 3$\sigma$ C.L. 
with the annual modulation phase measured by DAMA can be worked out for each $\alpha$ value.
The result is reported in Fig. \ref{fg:phase}-{\it right} where
the configurations giving a phase that exceed by 3$\sigma$ from the one measured by DAMA 
are shaded in the plot. 

\begin{figure}[!ht]
	\begin{center}
		\includegraphics [width=0.98\textwidth]{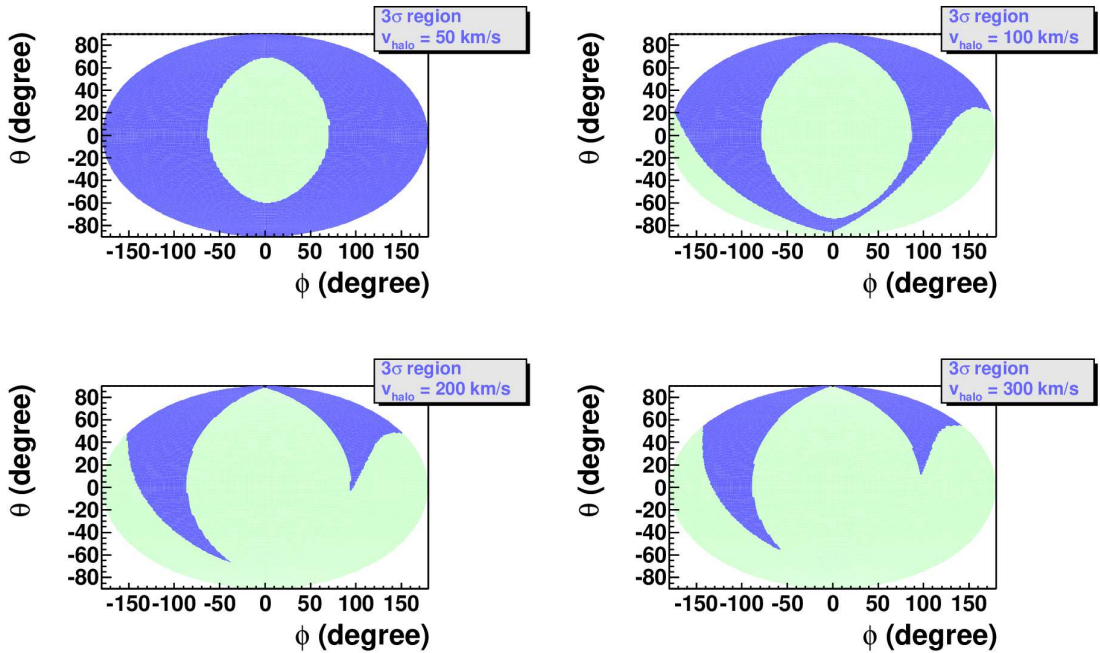}
	\end{center}
	\vspace{-1.2cm}
	\caption{The dark (blue on-line) regions correspond to directions of the halo velocities in Galactic 
		Coordinate ($\theta$,$\phi$) giving a phase compatible at 3$\sigma$ C.L.
                with the annual modulation phase measured by DAMA. The four panels refer to different
		values of the velocity module: 50 km/s, 100 km/s, 200 km/s, 300 km/s.}
	\label{fg:velgf1}   
\end{figure}

Finally, in Fig. \ref{fg:velgf1} the directions of the halo velocity in Galactic Coordinate
compatible with the DAMA annual modulation phase are reported for four different 
values of the velocity module. 

The results shows that many scenarios exist that are compatible with the annual modulation
observed by DAMA. Without losing generality, in the rest of the paper we will
consider only halo velocities parallel or anti-parallel
to the Sun ($\alpha \simeq 0$ and $\simeq \pi$, respectively). 
For these configurations (for $\alpha \simeq 0$ when $v_{halo} < v_S$)
the expected phase is $\simeq$ June 2$^{nd}$,
as in the case of a halo at rest with respect to the Galactic Center.
The only parameter whose value will be varied in the analysis is the module of the velocity.
For convention positive velocity will correspond to halo moving in the same direction 
of the Sun while negative velocity will correspond to opposite direction.

\subsection{Interaction Rates}

The low-energy differential cross-section of the scattering between the ordinary and mirror atoms
is essentially the same as the cross section $\mathcal{N}'+\mathcal{N} \rightarrow \mathcal{N}'+\mathcal{N}$
between the respective nuclei, mirror $\mathcal{N}'$ with a mass $M'_A$ and mirror electric charge $Z'$
and an ordinary $\mathcal{N}$  with a mass $M_A$ and an electric charge $Z$,
mediated by the photon -- mirror photon kinetic mixing; it has the Rutherford-like form \cite{Foot:2003iv}:
\be{dsigmaR}
\frac{d\sigma_{A,A'}}{dE_{R}}=
\frac{\mathcal{C}_{A,A'}}{E_{R}^{2}v_{\scriptscriptstyle DM}^{2}},
\ee
where $E_{R}$ is the recoil energy of the ordinary nucleus, $v_{\scriptscriptstyle DM}$ is
the relative velocity between the nuclei $\mathcal{N}'$ and $\mathcal{N}$,  and
\be{C}
\mathcal{C}_{A,A'}=\frac{2\pi \epsilon^{2}\alpha^{2}Z^{2}Z'^{2}}{M_A}\mathcal{F}^2_{A}\mathcal{F}^2_{A'} \, ,
\ee
where $\alpha$ is the fine structure constant,
and  $\mathcal{F}_{\scriptscriptstyle X}(qr_{\scriptscriptstyle X})$ $(X=A,A')$
are the Form-factors of ordinary and
mirror nuclei, which depend on the momentum transfer, $q$,
and on the radius of $X$ nucleus.
The effect of the $e'$ screening will be negligible since the mirror atoms are not compact,
i.e. the inverse radius of the mirror atom $1/a' \simeq \alpha m_e$
is smaller than the transfer momentum $q= \sqrt{2M_A E_R}$.
In particular, for Na target in DAMA,
considering that the relevant recoil energy range  is 2--6 keV electron equivalent
which corresponds to $E_R \simeq 6-20$ keV
when one takes into account a quenching factor value around 0.3
we have $q>20$ MeV, so that the condition $1/q < a'$ is fully satisfied.

The differential interaction rate of mirror nuclei of different species on a target 
composed by more than one kind of nucleus is: 
\be{RateR}
\frac{d\mathcal{R}}{dE_{R}}=\sum_{A,A'}N_{A}\chi_{A'}\int 
\frac{d\sigma_{A,A'}}{dE_{R}} F_{A'}(\vdm) v_{\scriptscriptstyle DM} d^{3}v_{\scriptscriptstyle DM} = \ee
$$= \sum_{A,A'}N_{A}\chi_{\scriptscriptstyle A'}\frac{\mathcal{C}_{A,A'}}{E_{R}^{2}}\int_{v_{\scriptscriptstyle DM}>v_{min}(E_{R})}
\frac{F_{A'}(\vdm)}{v_{\scriptscriptstyle DM}}d^{3}v_{\scriptscriptstyle DM},$$
where: i) $N_{A}$ is the number of the target atoms of specie $A$ per kg of detector; ii) 
$\chi_{\scriptscriptstyle A'}=\rho_{\scriptscriptstyle DM}\Upsilon_{A'}/M_{A'}$ with $\rho_{\scriptscriptstyle DM}$ 
halo mirror matter density, $\Upsilon_{A'}$ fraction of the
specie $A'$ in the dark halo, and $M_{A'}$ mass of the mirror nucleus $A'$; iii) the sum 
is performed over the mirror nuclei involved in the interactions
($A'$) and over the target nuclei in the detector ($A$).  We can normalize $\rho_{\scriptscriptstyle DM}$ 
to a reference value $\rho_0 = 0.3$ GeV/cm$^3$ as
$\rho_{\scriptscriptstyle DM} = f\,\rho_0$; thus all numerical results
presented below will be written in terms of  $\sqrt{f}\epsilon$.

The lower velocity limit 
$v_{min}(E_R)$ is
\be{vmin}
v_{min}(E_R)=\sqrt{\frac{(M_{A}+M_{A'})^{2}E_{R}}{2M_{A}M_{A'}^2}}.
\ee

The theoretical differential counting rate can be written as:
\be{RateG}  
\frac{d\mathcal{R}}{dE}=\sum_A \int \mathcal{K_A}(E|E_{R})\frac{d\mathcal{R_A}}{dE_{R}}dE_{R},
\ee
where $\frac{d\mathcal{R_A}}{dE_{R}}$ is the differential interaction rate on the $A$ nucleus in the detector.
The $\mathcal{K_A}(E|E_{R})$ kernel can be written as \cite{mirror2015}:
\be{kernelK}
\mathcal{K_A}(E|E_{R})=\int \mathcal{G}(E|E') \mathcal{Q_A}(E'|E_{R}) dE',
\ee
where $\mathcal{G}(E|E')$ takes into account the energy resolution of the detector, while
$\mathcal{Q_A}(E'|E_{R})$ takes into account the energy transformation 
of the nuclear recoil energy in keV electron equivalent (hereafter indicated simply as keV)
through the quenching factor (see later).  
For example, the latter kernel can be written in the simplest case of a constant quenching factor $q_A$ as:
$\mathcal{Q_A}(E'|E_{R})= \delta(E'-q_AE_R)$.

Defining $\eta(t) = v_{rel}(t) / v_0 $, when the Eq. \ref{eq:limvrel} holds, one gets:
$\eta(t) =  \eta_0 +  \Delta\,\eta cos\omega(t-t_0)$,
where $\eta_0$ is the yearly average of $\eta$ and $\Delta\eta$ is its maximal variation
along the year. Since, in this case, $\Delta\eta\ll\eta_0$, the expected counting
rate can be expressed by the first order Taylor expansion:
\begin{equation}
\frac{d\mathcal{R}}{dE}[\eta(t)] = \frac{d\mathcal{R}}{dE}[\eta_0] +
\frac{\partial}{\partial \eta} \left( \frac{d\mathcal{R}}{dE} \right)_{\eta =
	\eta_0} \Delta \eta \cos\omega(t - t_0) .
\end{equation}
Averaging this expression in a given energy interval one obtains:
\begin{equation}
\mathcal{S}\lbrack\eta(t)\rbrack = \mathcal{S}\lbrack\eta_0\rbrack
+ \left[\frac{\partial  \mathcal{S}}{\partial \eta}\right]_{\eta_0}
\Delta\eta cos\omega(t-t_0) =\mathcal{S}_0 + \mathcal{S}_m cos\omega(t-t_0),
\label{eq:sm}
\end{equation}
with the contribution from the higher order terms less than 0.1$\%$;
$\mathcal{S}_{m}$ and $\mathcal{S}_{0}$ are the modulated and the unmodulated
part of the expected differential counting rate, respectively.

The cross-section (Eq. \ref{dsigmaR}) strongly depends on on the kinetic mixing parameter $\epsilon$.
On the other hand, there are direct experimental limits on it
 from the ortopositronium oscillation into mirror ortopositronium
\cite{Glashow:1985ud,Gninenko:1994dr}. The latest limit on the experimental
search reads  $\epsilon < 4 \times 10^{-7}$  \cite{Badertscher:2006fm}.
The cosmological limits are more stringent, from the condition that $e^+ e^- \to e^{\prime +} e^{\prime -}$
process mediated by this kinetic mixing will not heat  too much the mirror bath \cite{Carlson:1987si}.
Namely, the condition $T'/T < 0.3$ implies  $\epsilon < 3\times 10^{-9}$ or so \cite{Berezhiani:2008gi}.
As we see below, our results for Dark Matter detection are compatible with the existing
limits on the Dark Matter particle mini-charges, or in some situation in some tension with the
cosmological limit.\footnote{In principle, the parameter $\epsilon$ can be a dynamical
degree of freedom which varied during the evolution of the universe \cite{BKK}.
Thus, one cannot exclude the situation that $\epsilon \sim 10^{-7}$ today but
it was $< 10^{-9}$ at the BBN epoch, and thus one take as a limit direct experimental bound
$\epsilon < 4 \times 10^{-7}$  \cite{Badertscher:2006fm}.
 It is interesting that for larger values of $\epsilon$,  the electron drag due to
relative rotation of ordinary and mirror plasma components during the galaxy formation
can can give rise to circular electric currents which can originate galactic magnetic fields
\cite{Berezhiani:2013dea}.}

\begin{figure}[!ht]
	\begin{center}
		\includegraphics [width=0.49\textwidth]{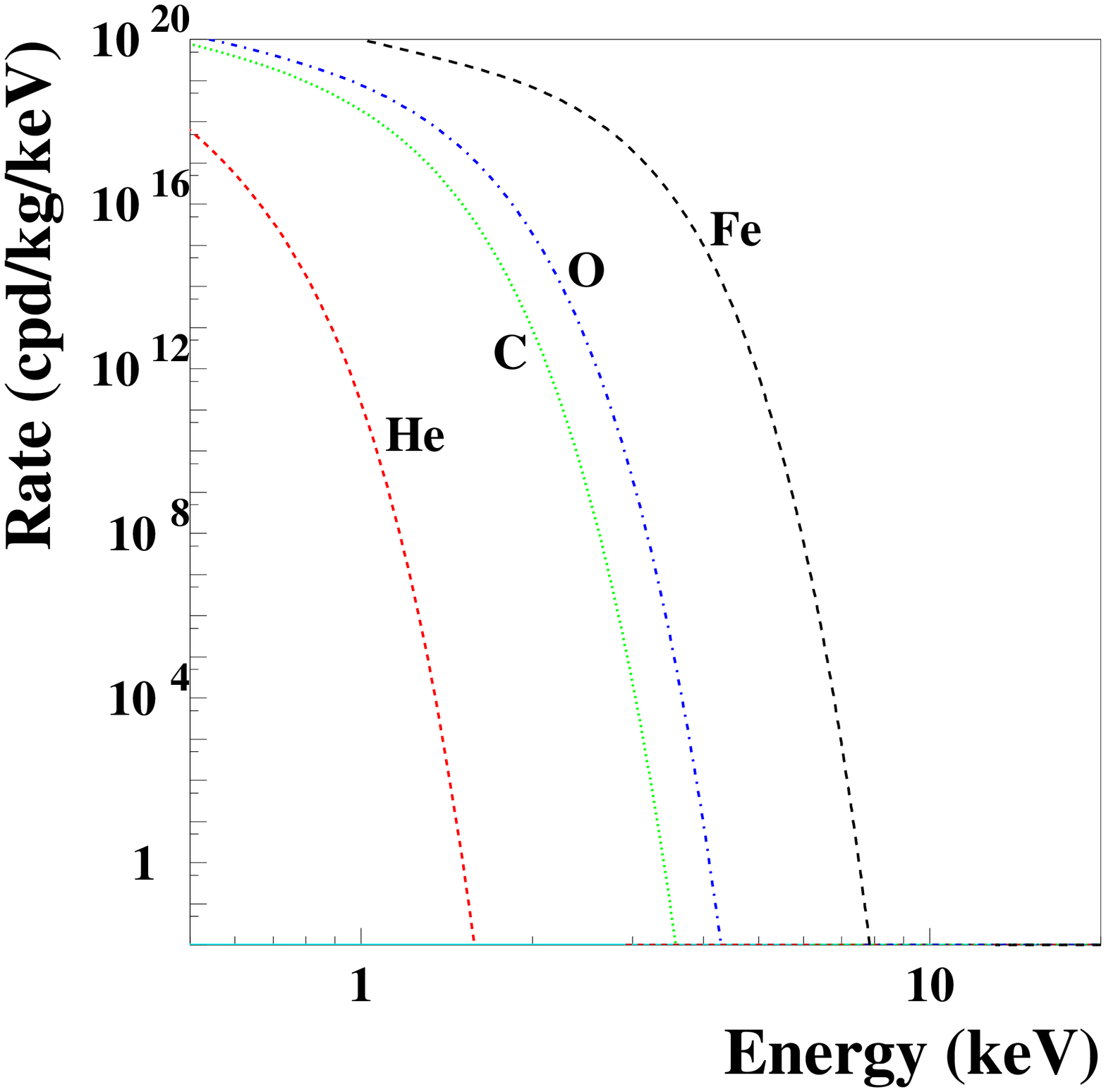}
		\includegraphics [width=0.49\textwidth]{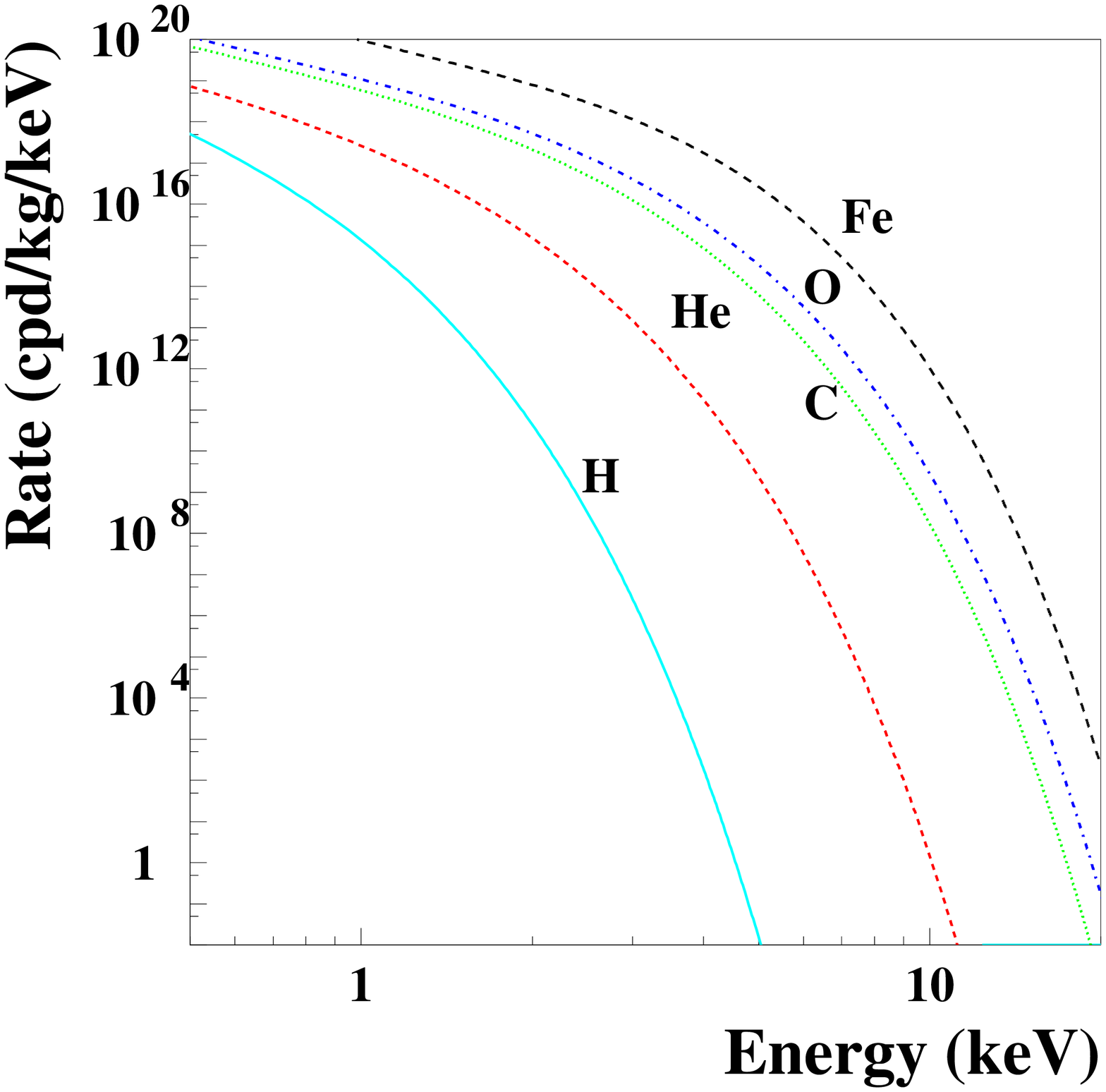}
	\end{center}
	\vspace{-0.8cm}
	\caption{Constant part of the annual modulation signal expected 
		for only one mirror atom specie in a NaI(Tl)
		detector, by considering $\sqrt{f}\epsilon = 1$. Few different mirror atoms are reported.
		The two panels refer to the case of cold ($T=10^{4}$ K) (left panel) 
		and hot ($T=10^{7}$ K) (right panel) halo with $v_{halo} = 100$ km/s.
		The considered scenario is the case $a$ of Table \ref{tb:scenarios} in
		set A (see Sect. \ref{setabc}).
	}
	\label{fg:s0mirror}   
\end{figure}

In Fig. \ref{fg:s0mirror} the behaviour of the unmodulated 
part of 
the signal expected
for only one mirror atom specie in a NaI(Tl) detector in a template case is reported. 
In this Fig. $\sqrt{f}\epsilon = 1$, few mirror atoms and two different halo temperatures have been
considered.

\section{Details of the analysis}

The data analysis in the symmetric mirror DM model considered here allows the determination of the
$\sqrt{f}\epsilon$ parameter. As mentioned this corollary analysis is model dependent.
The main aspects which enter in the $\sqrt{f}\epsilon$ determination and the related uncertainties
are pointed out in Ref. \cite{mirror2015}. Here we just remind few items.

\subsection{Phase-space distribution functions of DM mirror particles in the dark halo}
\label{halo}
Mirror dark halo is composed by dark atoms of different species having maxwellian velocity 
distribution in a frame where the halo is at rest. The halo has its own equilibrium temperature 
$T$ and the velocity parameter of the $A'$ mirror atoms is given by
$ \sqrt{2 k_{\scriptscriptstyle B} T / M_{\scriptscriptstyle A'} }$.
In the analysis we have considered different temperature regimes:
cold halo ($T \simeq 10^{4}-10^{5} K$) and hot halo ($T \gsim 10^{6}-10^{8} K$).
For simplicity the escape velocity of mirror atoms in the halo has been considered infinite.

\subsection{Nuclei and Dark Matter Form factors}

As regard the nuclei and DM form factors, entering in the determination of the expected 
signal counting rate, a {\it Helm form factor} \cite{Helm1,Helm2} has been 
considered\footnote{It should be noted that 
	the Helm form factor is the less favorable one e.g. for iodine and requires larger SI cross-sections for 
	a given signal rate; in case 
	other form factor profiles, considered in the literature, would be used \cite{RNC}, the allowed 
	parameters space would extend.} for each X ordinary and mirror nucleus. 
Details on the used form factors can also be found in Ref. \cite{mirror2015}.
In the analysis some uncertainties on the nuclear radius and on the 
nuclear surface thickness parameters in the Helm SI form factors 
have been included (see e.g. \cite{RNC,bot11}).

\subsection{Quenching factors and Channeling effect}
\label{qf}

Following the procedure reported in Ref. \cite{bot11,ldm,mirror2015}, 
in the present analysis three possibilities for the Na and I quenching factors 
have been considered: Q$_I$) the quenching factors of Na and I ``constants''
with respect to the recoil energy $E_{R}$: $q_{Na}\simeq 0.3$ and $q_{I}\simeq 0.09$ as measured by 
DAMA with neutron source integrated over the $6.5-97\, \rm keV$ and the $22-330\, \rm keV$ 
recoil energy range, respectively \cite{allDM1}; Q$_{II}$) the quenching factors evaluated as in Ref. \cite{Tretyak}
varying as a function of $E_R$; Q$_{III}$) the quenching factors with the same behaviour of Ref. \cite{Tretyak},
but normalized in order to have their mean values  
consistent with Q$_{I}$ in the energy range considered there.

A detailed discussion about the uncertainties in the quenching factors 
has been given in section II of Ref. \cite{bot11}
and in Ref. \cite{mirror2015}.
In fact, the related uncertainties affect all the results both in terms of exclusion 
plots and in terms of allowed regions/volumes; thus, comparisons with a fixed 
set of assumptions and parameters values are intrinsically strongly uncertain. 

Another important effect is the {\it channeling} of low energy ions along axes and planes of the NaI(Tl) DAMA crystals.
This effect can lead to an important deviation, in addition to the other uncertainties discussed in section II of Ref. \cite{bot11}
and in Ref. \cite{mirror2015}. In fact, the 
{\it channeling} effect in crystals implies that a fraction of nuclear recoils are channeled and experience
much larger quenching factors than those derived from neutron calibration (see \cite{chan,bot11} for a discussion
of these aspects). 
The channeling effect in solid crystal detectors is not a well fixed issue. There are a lot of uncertainties in the modeling. 
Moreover, the experimental approaches (as that in Ref. \cite{collar_qnai}) are rather difficult since
the channelled nuclear recoils are -- even in the most optimistic model -- a very tiny fraction of the not-channeled recoils. 
In particular,
the modeling of the {\it channeling} effect described by DAMA in Ref. \cite{chan} is able to reproduce the recoil spectrum
measured at neutron beam by some other groups (see Ref. \cite{chan} for details). 
For completeness, we mention the alternative {\it channeling} model of Ref. \cite{Mat08}, where larger probabilities of the
planar channeling are expected, and the analytic calculation where the {\it channeling} effect holds
for recoils coming from outside a crystal and not from recoils from lattice sites, due to the blocking effect \cite{gelmini}.
Nevertheless, although some amount of blocking effect could be present, the precise description of the 
crystal lattice with dopant and trace contaminants is quite difficult and analytical calculations require 
some simplifications which can affect the result.
Because of the difficulties of experimental measurements and of theoretical estimate of the {\it channeling} effect,
in the following it will be either included using the procedure given in Ref. \cite{chan} 
or not in order to give idea on the related uncertainty.

\subsection{Further uncertainties}
\label{setabc}

In case of low mass DM particles giving rise to nuclear recoils it is also necessary to account 
for the Migdal effect. A detailed discussion of its impact in the corollary analyses in terms of some 
DM candidates is given in Ref. \cite{ijma07,mirror2015}.

Moreover, to take into account the uncertainty on the local velocity, $v_0$, 
following the discussion in Sect. \ref{annph} we have considered the discrete values: 170, 220 and 270 km/s.

Finally, some discrete cases 
are considered to account for the uncertainties on the measured quenching factors and on the 
parameters used in the nuclear form factors, as already done in previous analyses for other DM candidates
and scenarios. 
The first case (set A) considers the mean values of the parameters of the used nuclear form factors \cite{RNC}
and of the quenching factors. The set B adopts the same procedure as in Refs. \cite{allDM6,allDM7},
by varying (i) the mean values of the $^{23}$Na and $^{127}$I quenching factors as 
measured in Ref. \cite{allDM1} up to +2 times the errors; (ii) the nuclear 
radius, $r_A$, and the nuclear surface thickness parameter, $s$,
in the form factor from their central values down to -20\%.
In the last case (set C) the Iodine nucleus parameters are fixed at the values of case B,
while for the Sodium nucleus one considers: (i) $^{23}$Na quenching factor at the lowest value measured
in literature; (ii) the nuclear radius, $r_A$, and the nuclear surface thickness parameter, $s$,
in the SI form factor from their central values up to +20\%.

\subsection{Analysis procedures}

The analysis procedure has been described in Ref. \cite{mirror2015}.
Here we just remind that the obtained $\chi^2$ for the considered mirror DM model
is function of only one parameter: $\sqrt{f}\epsilon$;
thus, we can define:
$$\Delta \chi^{2}\{\sqrt{f}\epsilon\}=\chi^{2}\{\sqrt{f}\epsilon\}-\chi^{2}\{\sqrt{f}\epsilon=0\}.$$
The $\Delta \chi^{2}$ is a $\chi^{2}$ with one degree of freedom and is used to determine the allowed interval of the $\sqrt{f}\epsilon$ parameter 
at $5\sigma$ from the {\it null signal hypothesis}.

\section{Results}

In the data analysis we have taken into account all the uncertainties discussed in the previous sections.
The scenarios summarized in Table \ref{tb:scenarios} have been considered depending on:
i) the adopted quenching factors; ii) either inclusion or not of the channeling effect;
iii) either inclusion or not of the Migdal effect. For each scenario the different halo compositions
reported in Table \ref{tb:specie} have been considered,
\begin{table}[!hbt]
	\caption{Summary of the scenarios considered in the present analysis of the DAMA data 
                 in terms of  mirror DM framework as discussed in the text.
	}
	\begin{center}
		\resizebox{0.6\textwidth}{!}{ 
			\begin{tabular}{|c|c|c|c|c|r|}\hline
				Scenario & Quenching        & Channeling & Migdal   \\
				& Factor           &            &          \\
				\hline
				&                  &            &        \\
				$a$  & $Q_{I}$ \cite{allDM1}    & no         & no   \\
				&                  &            &            \\
				$b$  & $Q_{I}$ \cite{allDM1}    & yes        & no   \\
				&                  &            &            \\
				$c$  & $Q_{I}$ \cite{allDM1}    & no         & yes  \\
				&                  &            &            \\
				$d$  & $Q_{II}$ \cite{Tretyak}   & no         & no  \\
				&                  &            &            \\
				$e$  & $Q_{III}$ \cite{Tretyak}-normalized& no & no \\
				&                  &            &            \\
				\hline
			\end{tabular}
		}
		\label{tb:scenarios}
	\end{center}
\end{table}
with halo temperature in the range \mbox{$10^{4}-10^{8}$ K} and with halo velocity from -400 to +300 km/s.
The uncertainties, described in the three
sets given in Sect. \ref{setabc}, have been considered.
\begin{figure}[!ht]
	\begin{center}
		\includegraphics [width=0.49\textwidth]{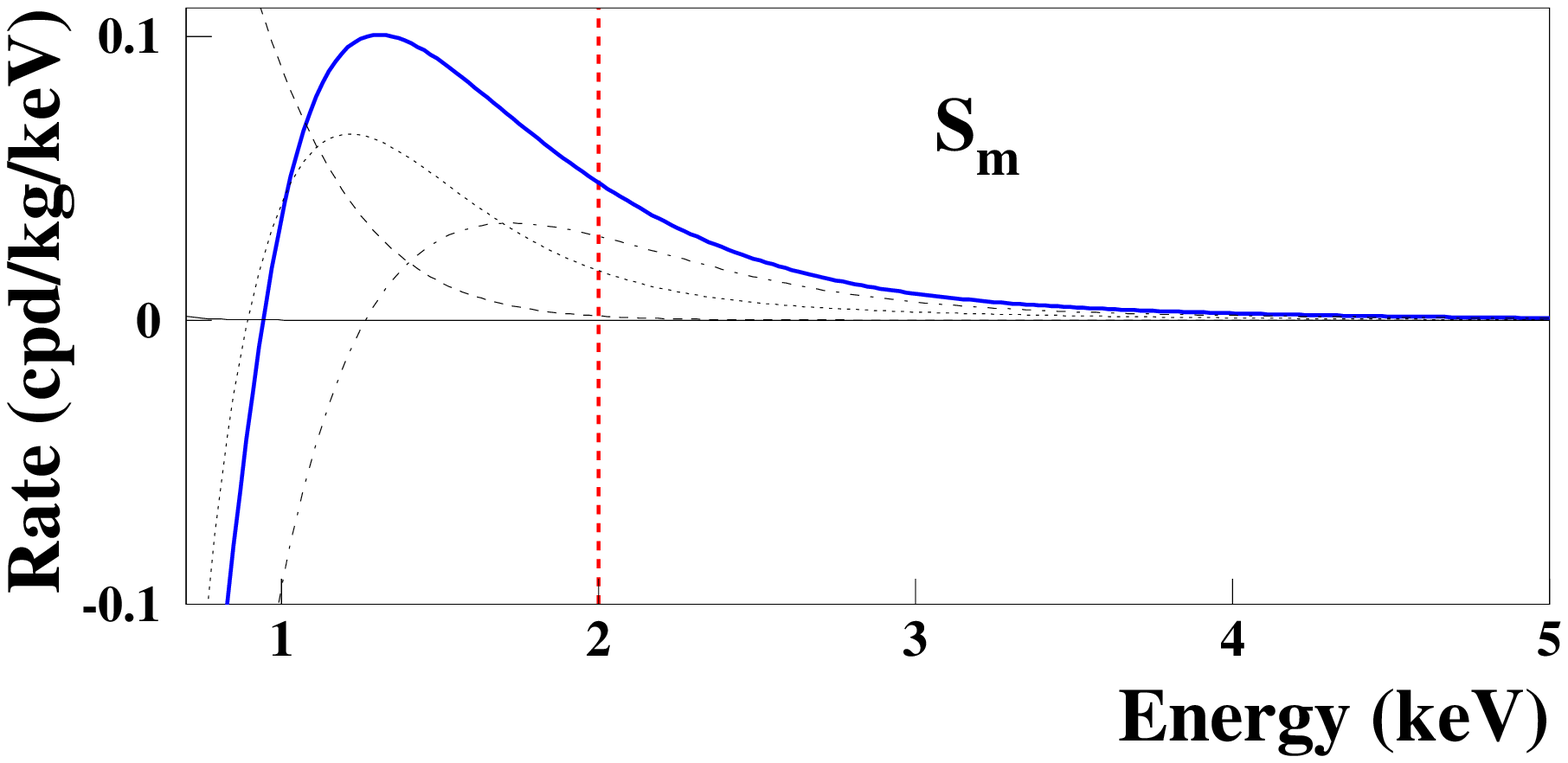}
		\includegraphics [width=0.49\textwidth]{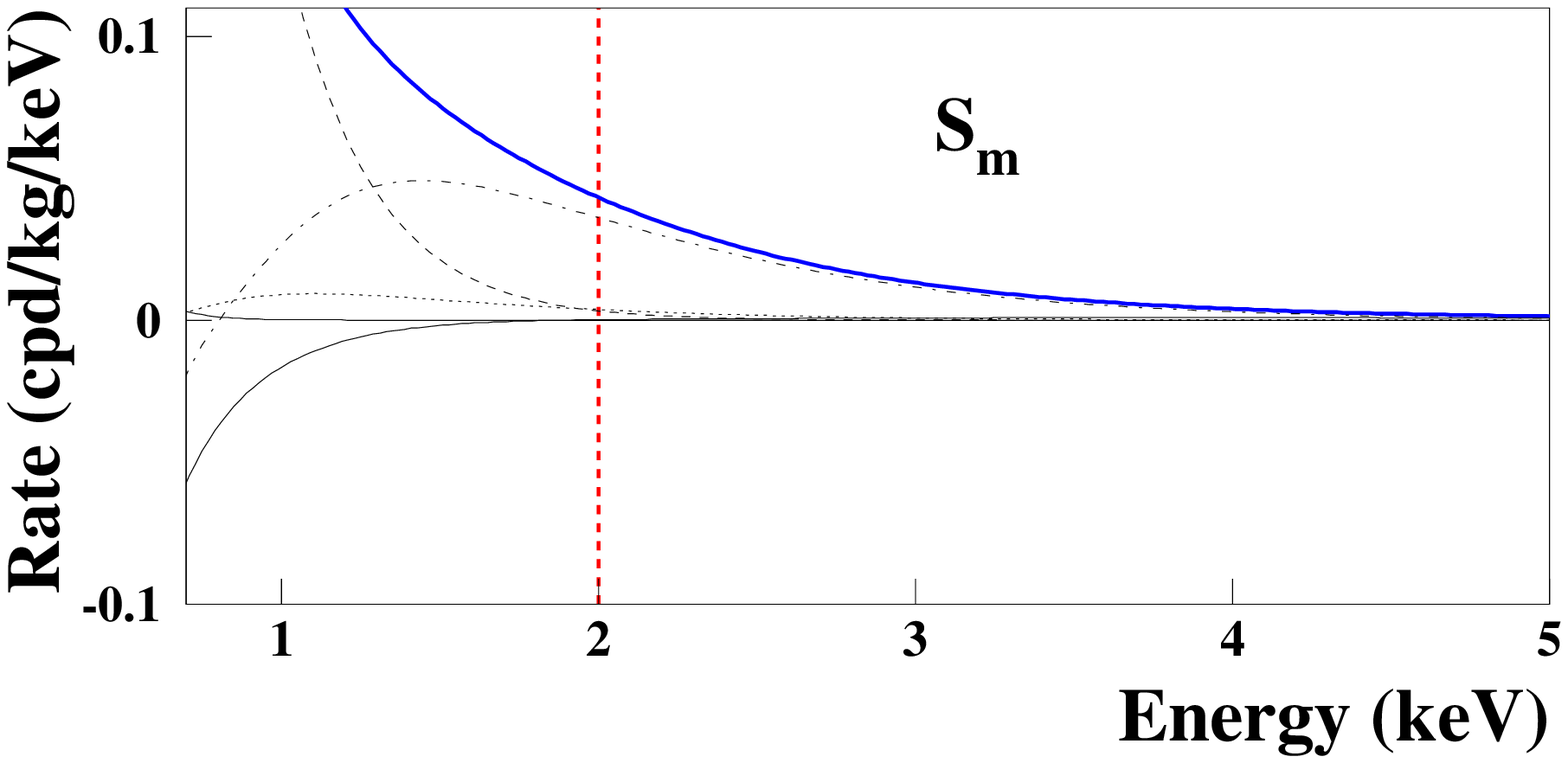}
	\end{center}
	\vspace{-4.cm}
	\caption{Examples of expected modulation amplitude, $S_{m}$, of the DM signal
		for the mirror DM candidates in the scenario $(b)$ (left)  and $(a)$ (right) of Table \ref{tb:scenarios}, 
		for  two different halo compositions.
		{\it Left}: composite dark halo:  H$'$(12.5\%), He$'$(75\%), C$'$(7\%), O$'$(5.5\%),
		with halo velocity $v_{halo} = 30$ km/s, temperature $T=10^{6} {\textrm K} $, 
		$v_{0} = 220$ km/s and parameters in the set A. The contributions to the signal 
		(solid line, blue on-line) of the different dark atoms are depicted: H$'$ (not visible), He$'$ (dashed), C$'$ (dotted),
		O' (dashed-dotted). 
		{\it Right}: composite dark halo: H$'$(20\%), He$'$(74\%), C$'$(0.9\%), O$'$(5\%), Fe$'$(0.1\%),
		with halo velocity $v_{halo} = 0$ km/s, temperature $T=10^{7} {\textrm K}$, 
		$v_{0} = 220$ km/s and parameters in the set A. The contributions to the signal 
		(solid line, blue on-line) due to the different dark atoms are depicted: 
                H$'$ (solid line, visible well below 1 keV), He$'$ (dashed), C$'$ (dotted),
		O$'$ (dashed-dotted), Fe$'$ (solid and negative below 2 keV). 
	}
	\label{fg:sm}   
\end{figure}

Firstly we show in Fig. \ref{fg:sm} the behaviour of the 
modulated part, $S_m$
of the Dark Matter signal obtained by fitting the considered DM mirror model with the DAMA
annual modulation data. Two composite halo models (left: H$'$(12.5\%), He$'$(75\%), C$'$(7\%), O$'$(5.5\%),
right: H$'$(20\%), He$'$(74\%), C$'$(0.9\%), O$'$(5\%), Fe$'$(0.1\%)) having different temperatures in different frameworks
have been considered as examples.
The contribution to the signal coming from each mirror atom species are reported. 
In both case the most relevant contribution comes from the $O'$ dark atoms while the contribution of the H$'$ is
negligible. It is interesting to note that the profile of the modulated signal below 2 keV is 
different for the two halo models; this can be studied by DAMA/LIBRA, now running in its phase2 
with a software energy  threshold down to 1 keV.

In the following, we present the $\sqrt{f}\epsilon$ values allowed by DAMA 
in different halo modeling and various scenarios. In particular,   
we present two different plots for each halo composition. We report: 
i) allowed regions for the $\sqrt{f}\epsilon$ parameter
as a function of the halo temperature for different values of the halo velocity in the Galactic frame;
ii) allowed regions for the $\sqrt{f}\epsilon$ parameter
as a function of the halo velocity in the Galactic frame for different 
halo temperature. All the allowed intervals reported identify the $\sqrt{f}\epsilon$ values corresponding to C.L.
larger than $5\sigma$ from the {\it null hypothesis}, that is $\sqrt{f}\epsilon=0$.

\begin{figure}[!ht]
	\begin{center}
		\includegraphics [width=0.49\textwidth]{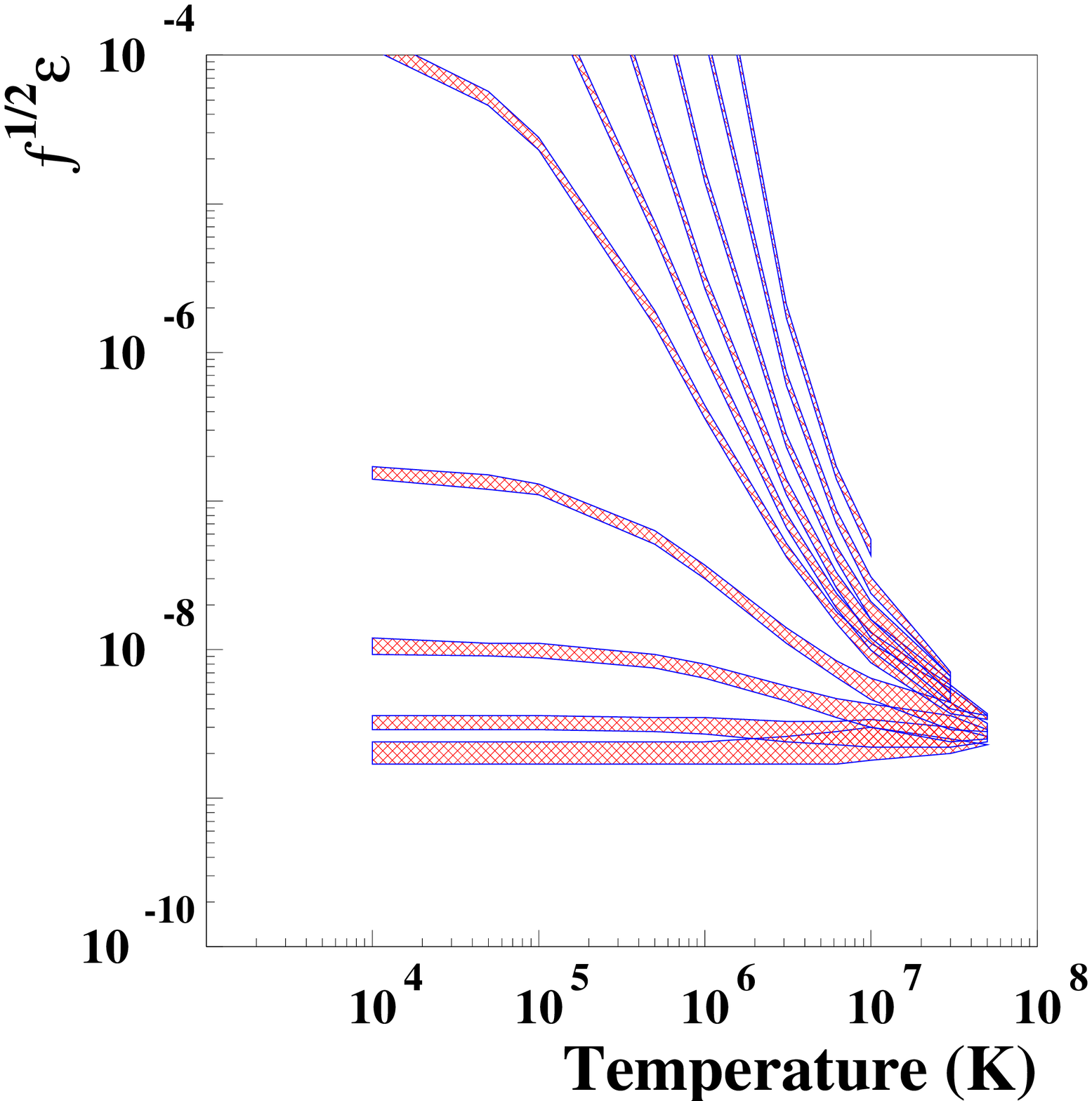}
		\includegraphics [width=0.49\textwidth]{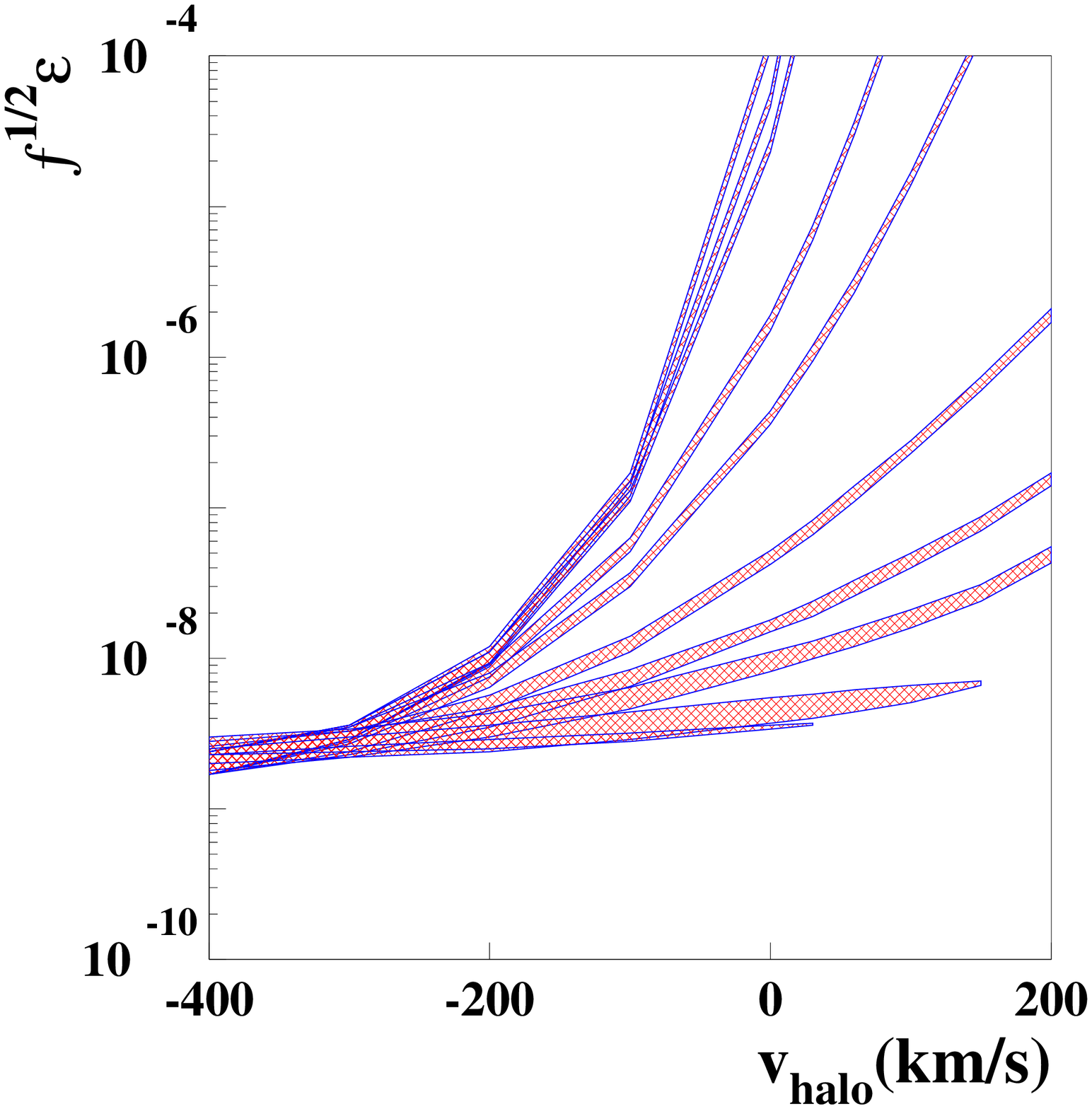}
	\end{center}
	\vspace{-0.8cm}
	\caption{Case of halo composed by pure He' dark atoms in the scenario $(a)$ of Table \ref{tb:scenarios}
		with $v_{0} = 220$ km/s and parameters in the set A (see text). 
		{\it Left}: allowed regions for the $\sqrt{f}\epsilon$ parameter
		as a function of the halo temperature for different values of the halo velocity in the Galactic frame:
                -400, -300, -200, -100, 0, 30, 60, 100, 150, 200 km/s. Increasing the halo velocity 
                the allowed regions e.g. at temperature of $10^{4}$ K 
		move to higher values of $\sqrt{f}\epsilon$ parameter. 
		{\it Right}: allowed regions for the $\sqrt{f}\epsilon$ parameter
		as function of the halo velocity in the Galactic frame for different 
		halo temperature: 
                $10^4, 5 \times 10^4, 10^5, 5 \times 10^5, 10^6, 3.1 \times 10^6, 6.2 \times 10^6, 10^7, 3 \times 10^7, 5 \times 10^7$ K
		Increasing the temperature the allowed region at large positive $v_{halo}$ 
		move to small values of $\sqrt{f}\epsilon$ parameter.
		These allowed intervals identify the $\sqrt{f}\epsilon$ values corresponding to C.L. larger than 5$\sigma$
		from the null annual modulation hypothesis, that is $\sqrt{f}\epsilon = 0$. 
	}
	\label{fg:pureHe}   
\end{figure}
\begin{figure}[!ht]
	\begin{center}
		\includegraphics [width=0.49\textwidth]{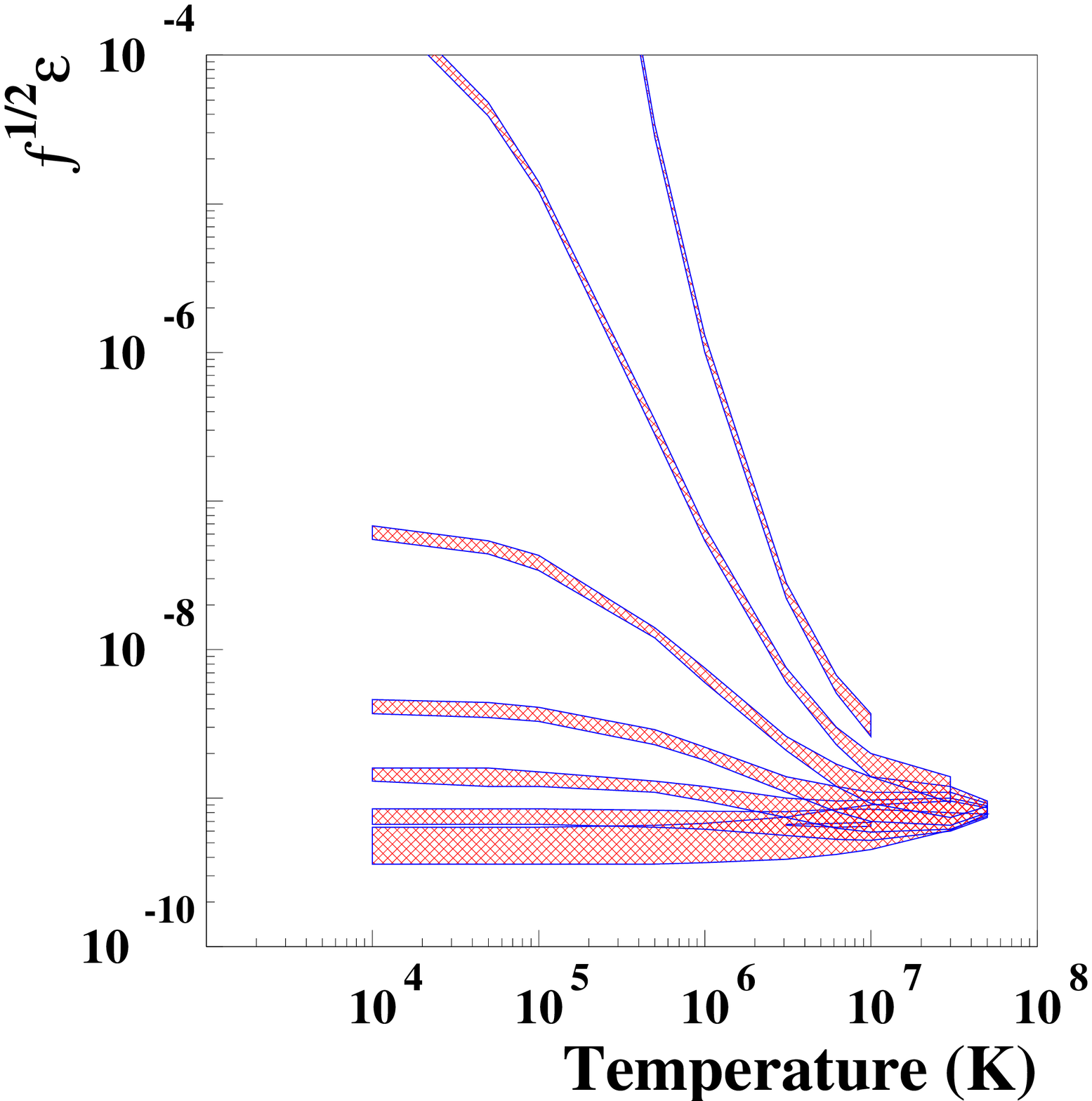}
		\includegraphics [width=0.49\textwidth]{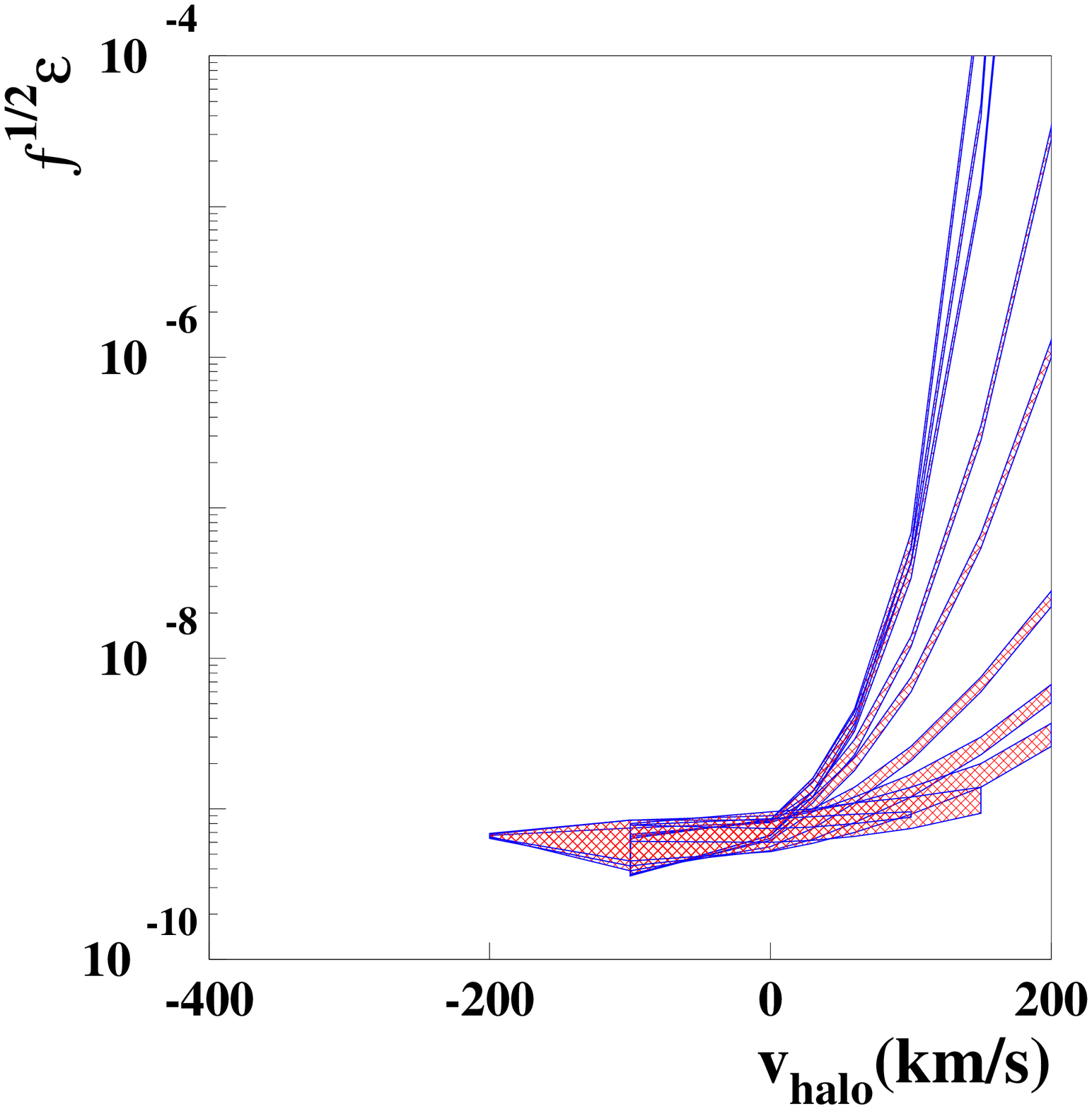}
	\end{center}
	\vspace{-0.8cm}
	\caption{Case of halo composed just by pure C' dark atoms in the scenario $(a)$ of Table \ref{tb:scenarios}
		with $v_{0} = 220$ km/s and parameters in the set A (see text and Fig. \ref{fg:pureHe}). 
                The different values of the halo velocity in the left plot are:
                -200, -100, 0, 30, 60, 100, 150, 200 km/s. 
                The different values of the halo temperature in the right plot are as those of Fig. \ref{fg:pureHe}.
	}
	\label{fg:pureC}   
\end{figure}
\begin{figure}[!ht]
	\begin{center}
		\includegraphics [width=0.49\textwidth]{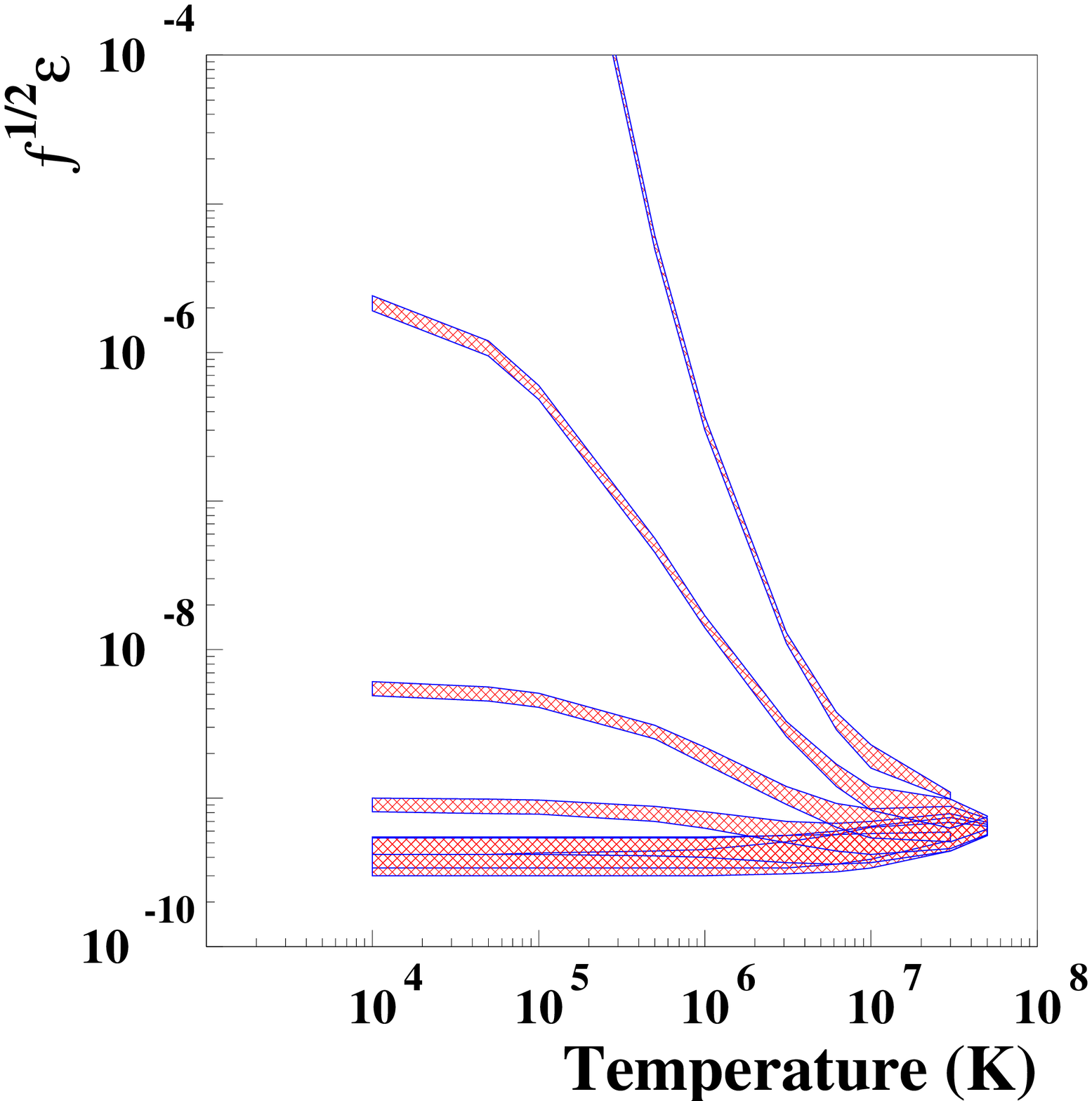}
		\includegraphics [width=0.49\textwidth]{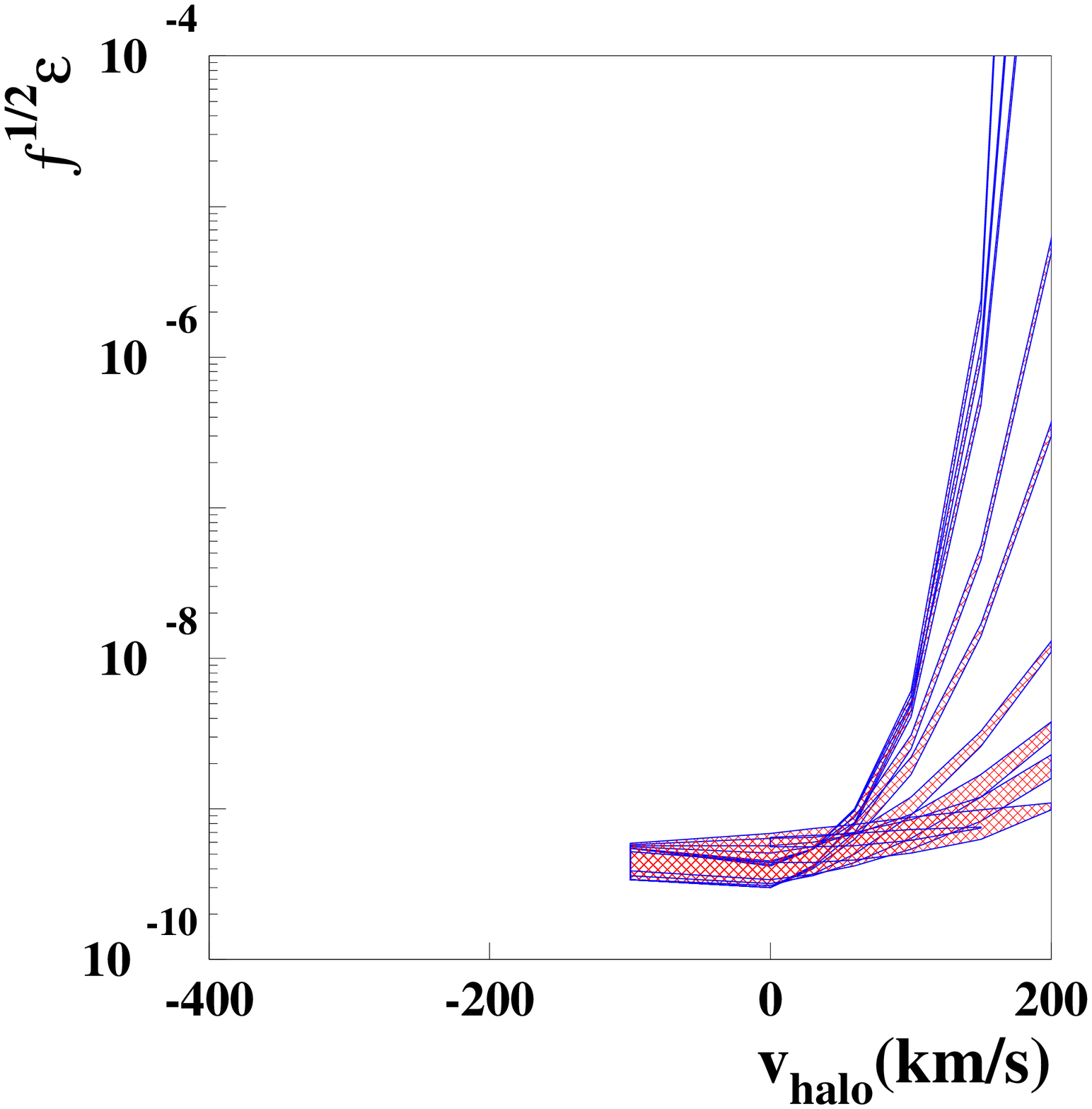}
	\end{center}
	\vspace{-0.8cm}
	\caption{Case of halo composed just by pure O' dark atoms in the scenario $(a)$ of Table \ref{tb:scenarios}
		with $v_{0} = 220$ km/s and parameters in the set A (see text and Fig. \ref{fg:pureHe}). 
                The different values of the halo velocity in the left plot are:
                -100, 0, 30, 60, 100, 150, 200 km/s. 
                The different values of the halo temperature in the right plot are as those of Fig. \ref{fg:pureHe}.
	}
	\label{fg:pureO}   
\end{figure}
\begin{figure}[!ht]
	\begin{center}
		\includegraphics [width=0.49\textwidth]{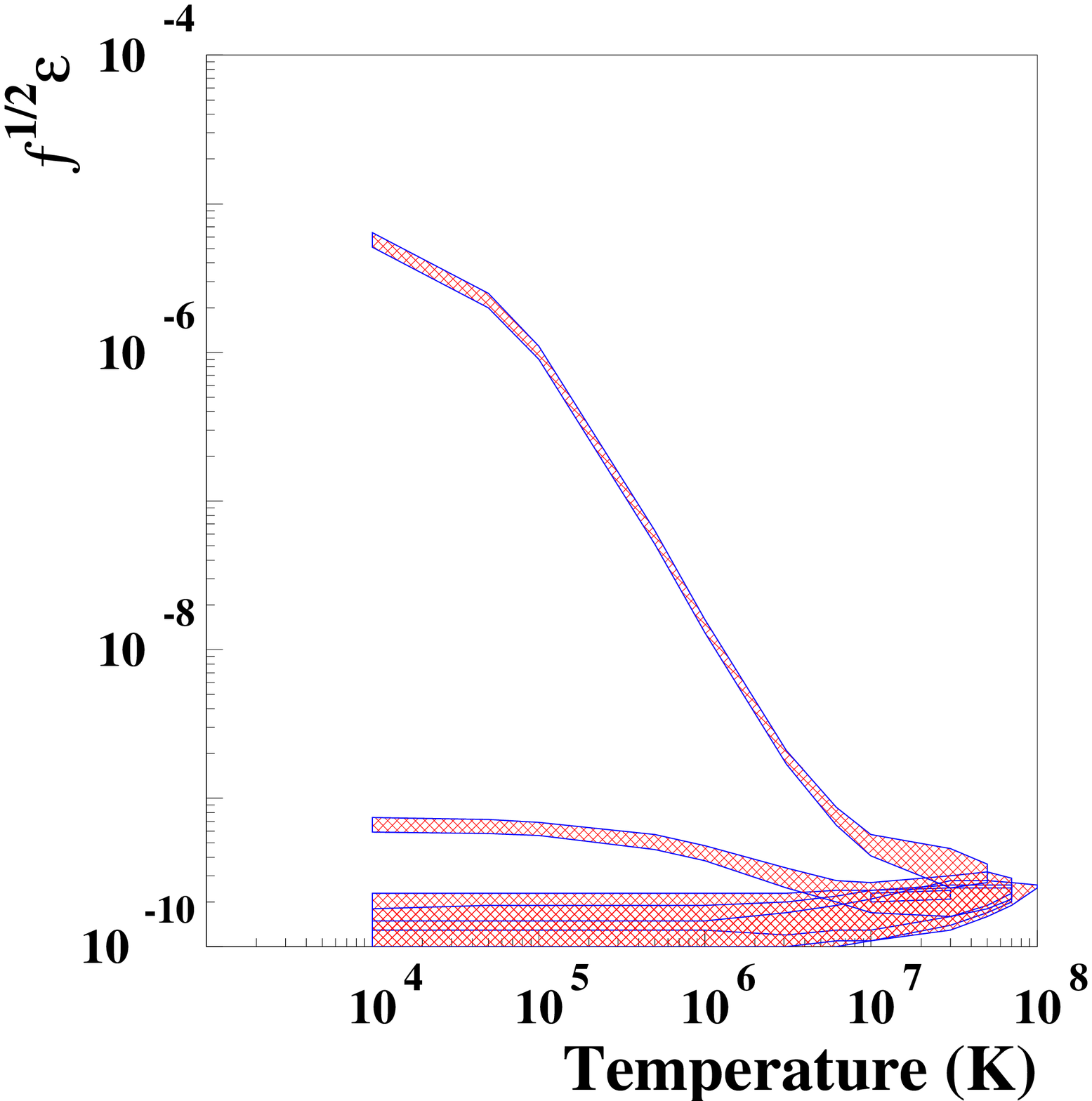}
		\includegraphics [width=0.49\textwidth]{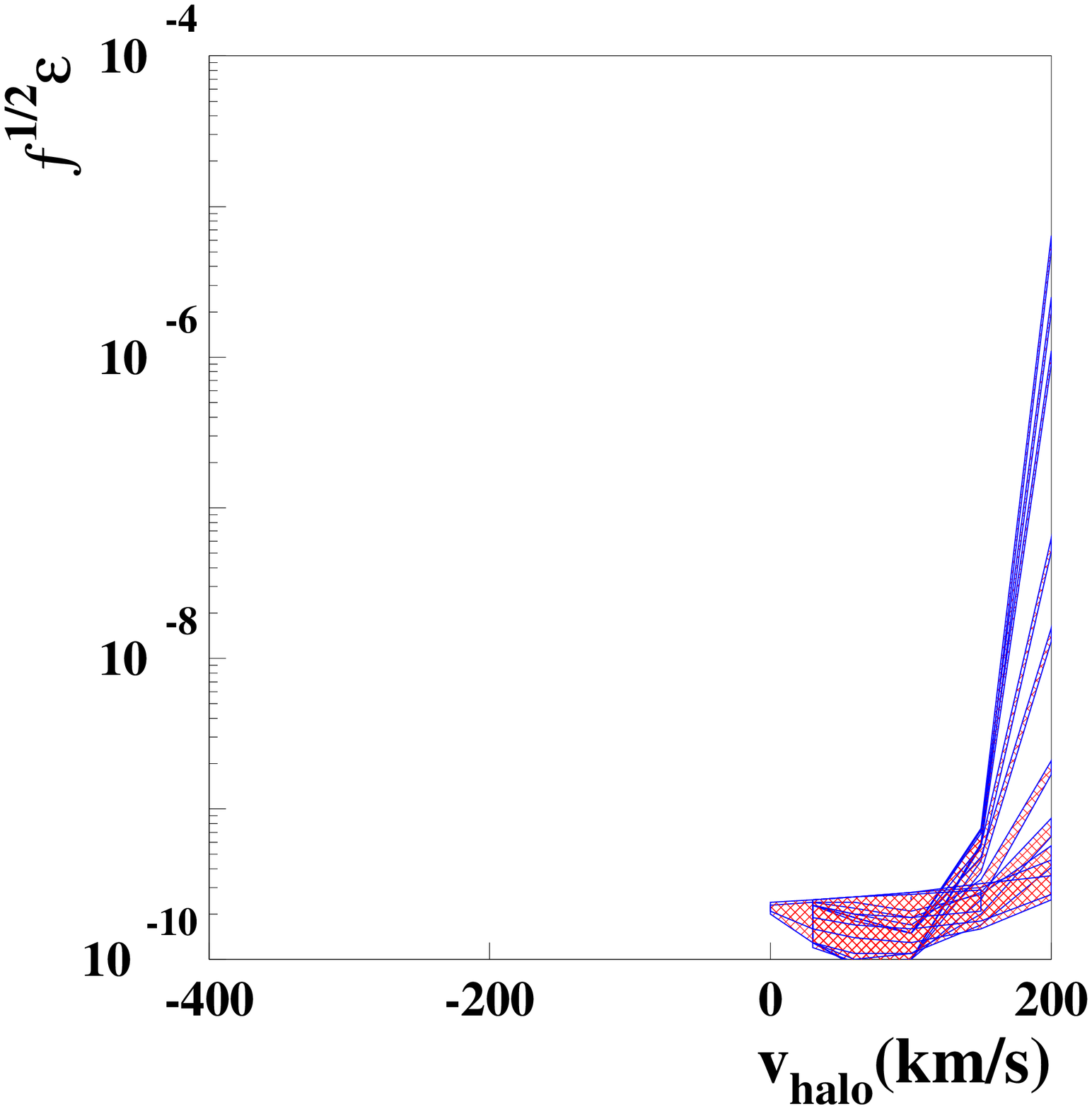}
	\end{center}
	\vspace{-0.8cm}
	\caption{Case of halo composed just by pure Fe' dark atoms in the scenario $(a)$ of Table \ref{tb:scenarios}
		with $v_{0} = 220$ km/s and parameters in the set A (see text and Fig. \ref{fg:pureHe}). 
                The different values of the halo velocity in the left plot are:
                0, 30, 60, 100, 150, 200 km/s. 
                The different values of the halo temperature in the right plot are as those of Fig. \ref{fg:pureHe} plus
                $7 \times 10^7, 10^8$ K.
	}
	\label{fg:pureFe}   
\end{figure}
In Fig. \ref{fg:pureHe} for template purpose only the case set A and $v_0 = 220$ km/s  is depicted
considering a halo composed only by He' dark atoms. The cases of halos made either 
only of O', or only of C' or only of Fe' are reported in Fig. \ref{fg:pureC},
Fig. \ref{fg:pureO} and Fig. \ref{fg:pureFe}, respectively. 

\begin{figure}[!ht]
	\begin{center}
		\includegraphics [width=0.49\textwidth]{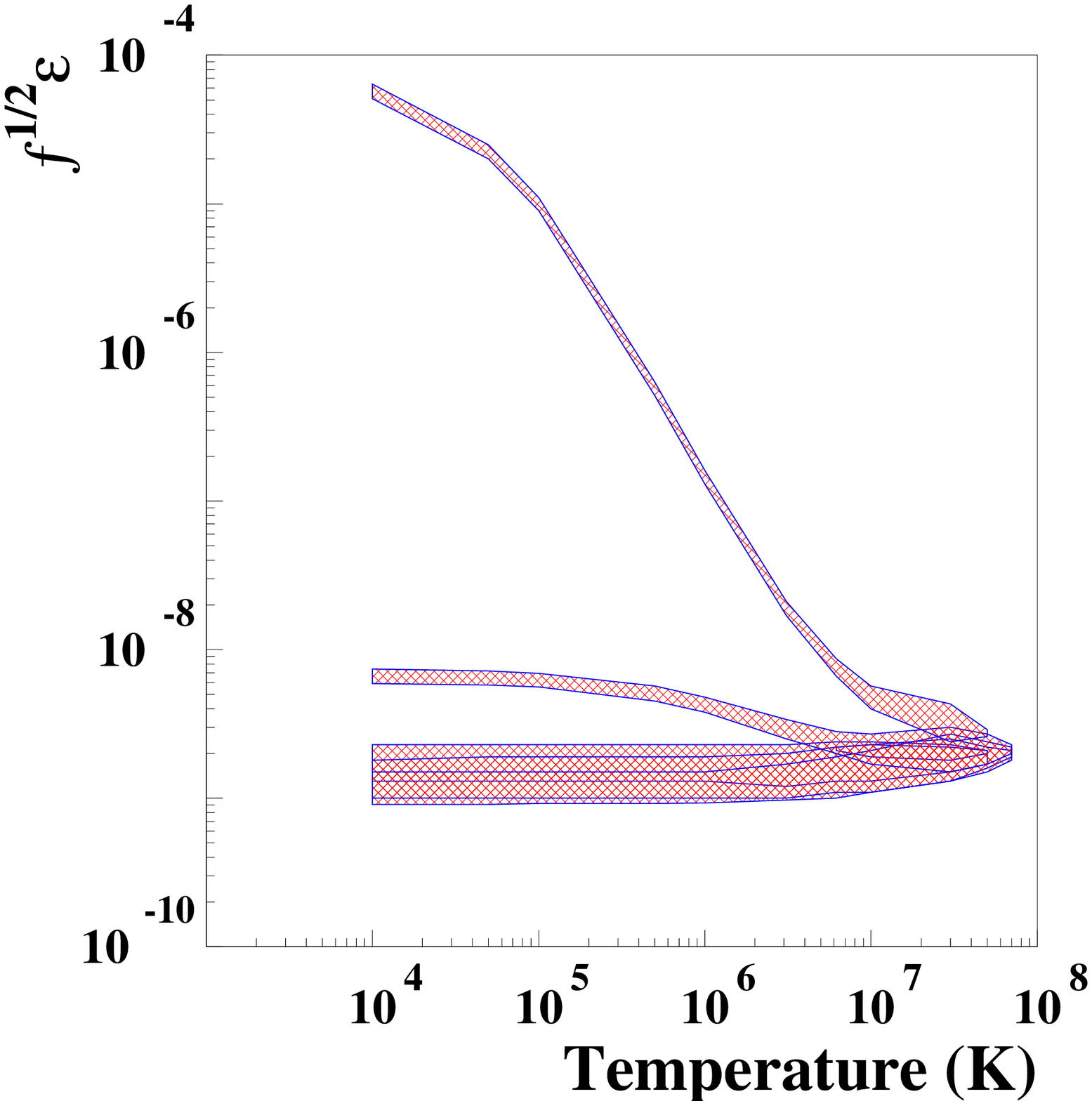}
		\includegraphics [width=0.49\textwidth]{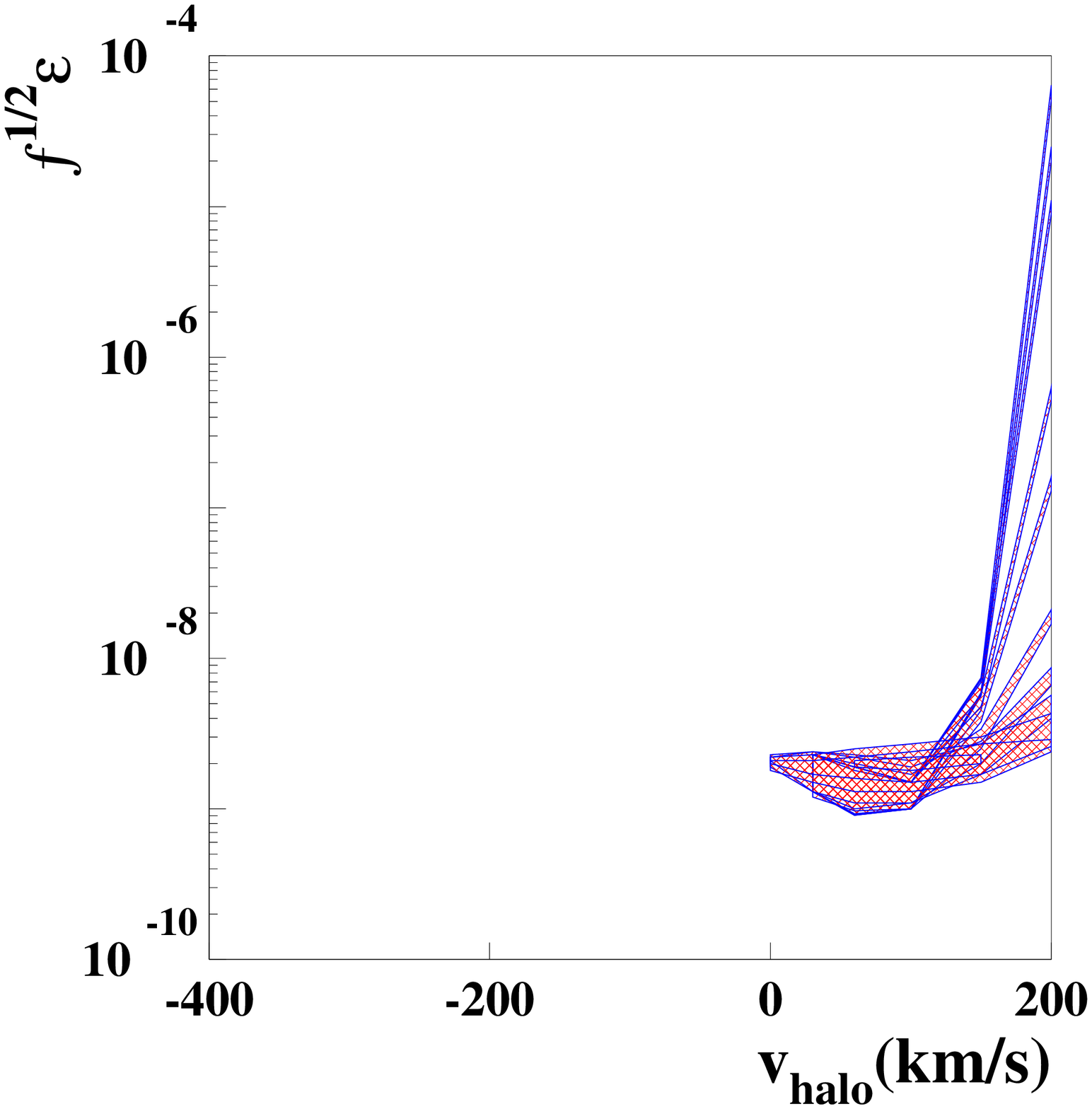}
	\end{center}
	\vspace{-0.8cm}
	\caption{Case of composite dark halo: H$'$(24\%), He$'$(75\%), Fe$'$(1\%), in the 
		scenario $(a)$ of Table \ref{tb:scenarios}
		with $v_{0} = 220$ km/s and parameters in the set A (see text and Fig. \ref{fg:pureHe}). 
                The different values of the halo velocity in the left plot are:
                0, 30, 60, 100, 150, 200 km/s. 
                The different values of the halo temperature in the right plot are as those of Fig. \ref{fg:pureHe} plus
                $7 \times 10^7$ K.
	}
	\label{fg:case_H_He_Fe}   
\end{figure}

\begin{figure}[!ht]
	\begin{center}
		\includegraphics [width=0.49\textwidth]{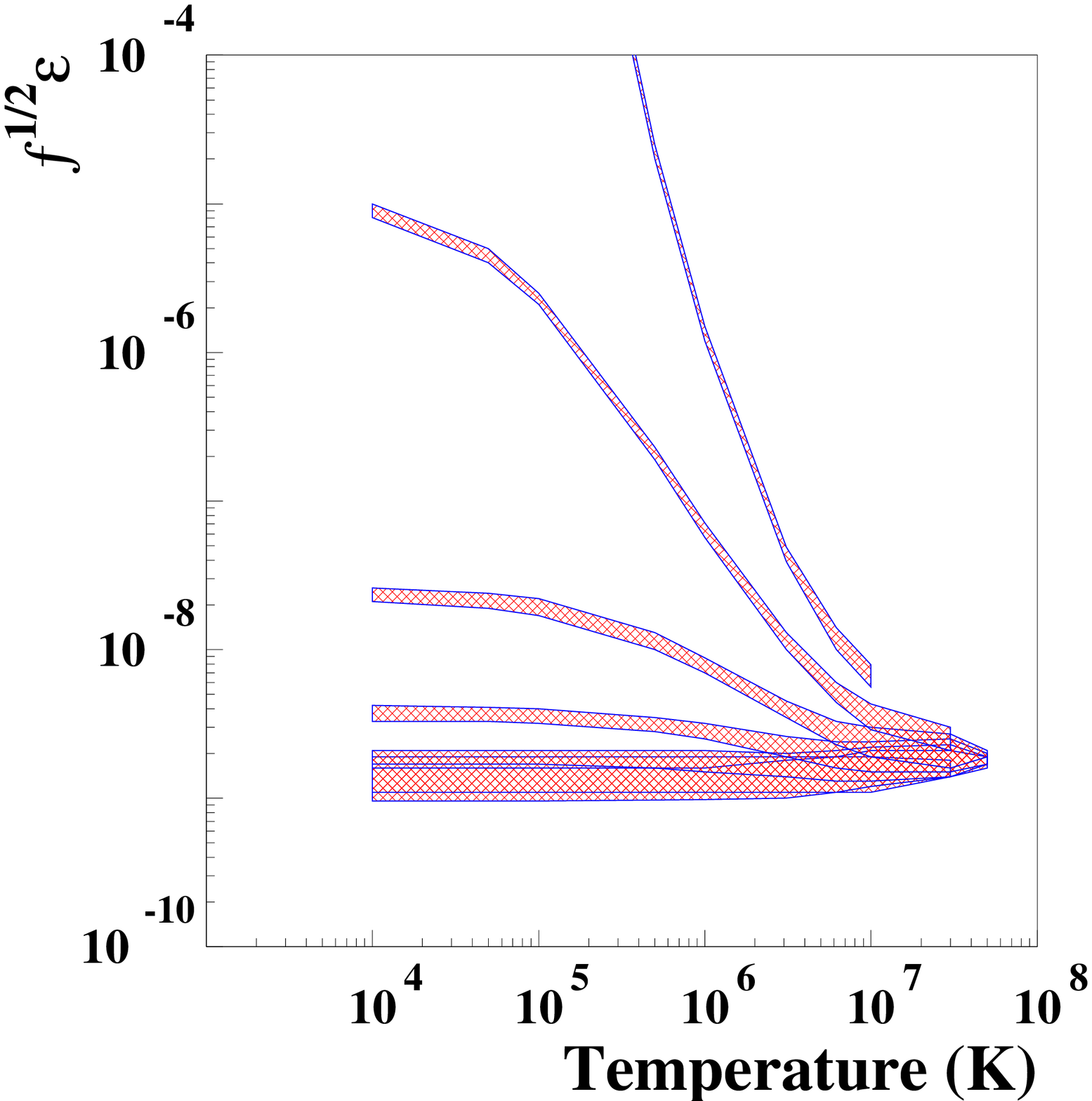}
		\includegraphics [width=0.49\textwidth]{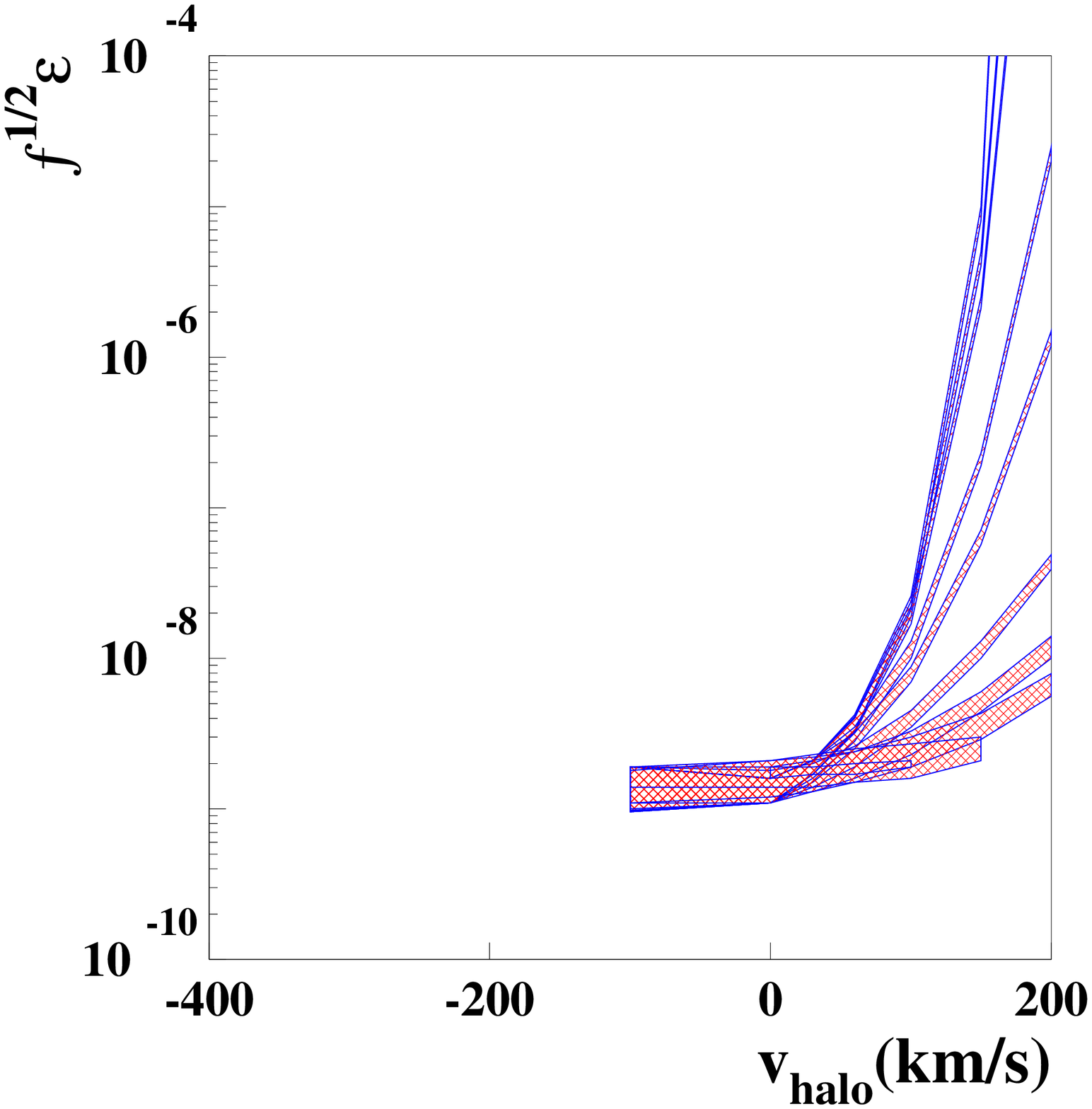}
	\end{center}
	\vspace{-0.8cm}
	\caption{Case of composite dark halo: H$'$(12.5\%), He$'$(75\%), C$'$(7\%), O$'$(5.5\%), in the 
		scenario $(a)$ of Table \ref{tb:scenarios}
		with $v_{0} = 220$ km/s and parameters in the set A (see text and Fig. \ref{fg:pureHe}). 
                The different values of the halo velocity in the left plot are:
                -100, 0, 30, 60, 100, 150, 200 km/s. 
                The different values of the halo temperature in the right plot are as those of Fig. \ref{fg:pureHe}.
	}
	\label{fg:case_H_He_C_O}   
\end{figure}

\begin{figure}[!ht]
	\begin{center}
		\includegraphics [width=0.49\textwidth]{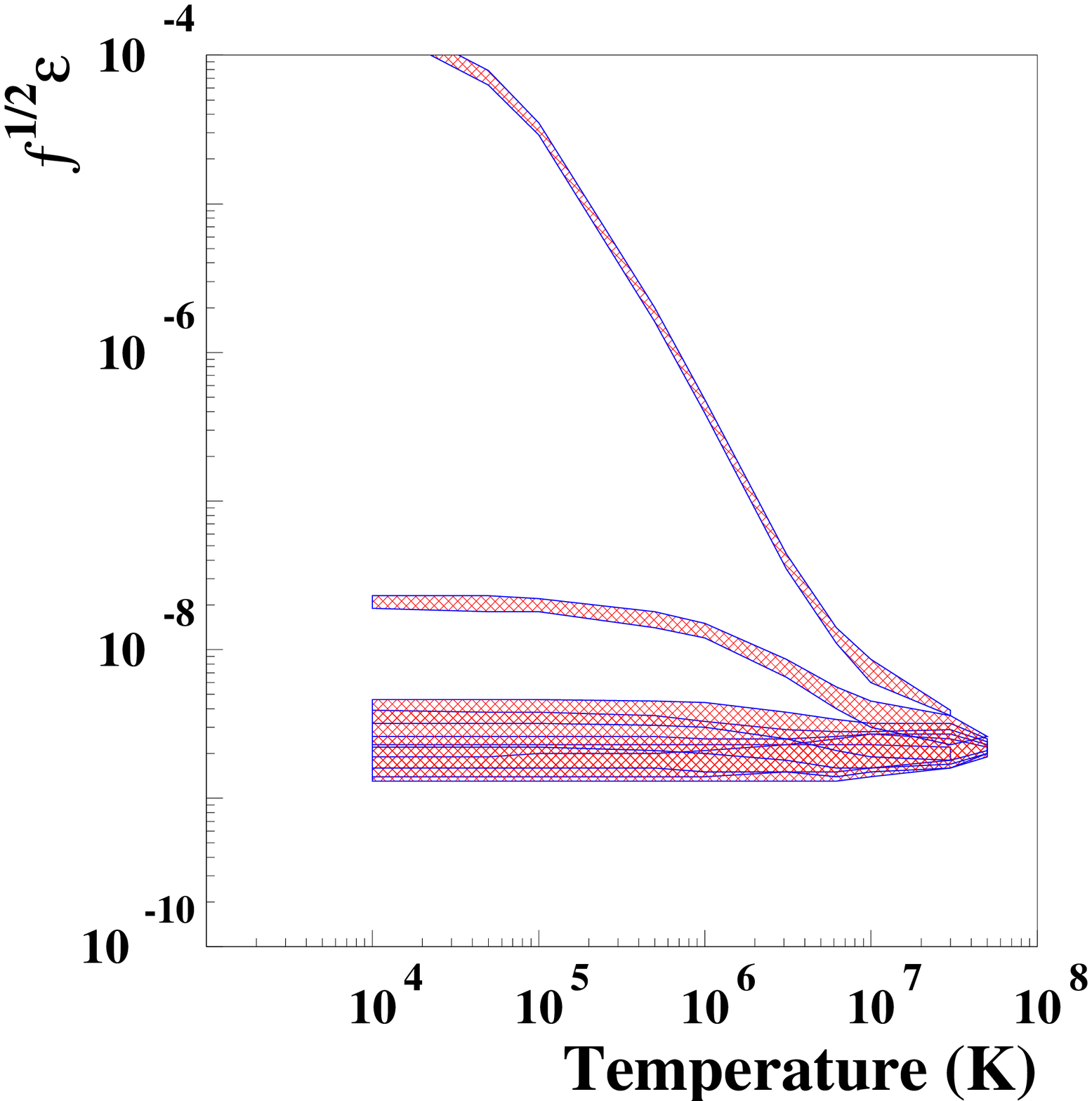}
		\includegraphics [width=0.49\textwidth]{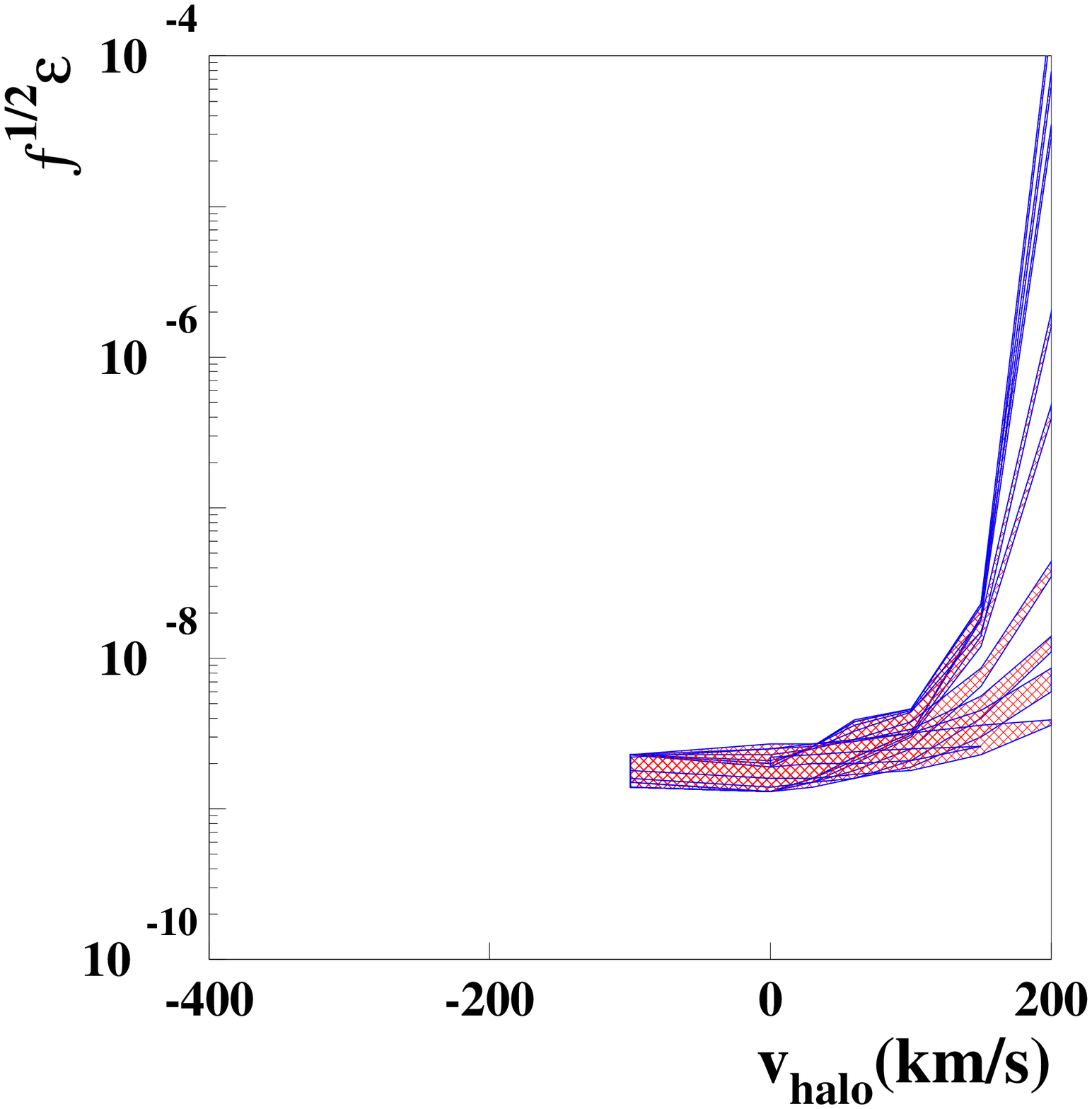}
	\end{center}
	\vspace{-0.8cm}
	\caption{Case of composite dark halo: H$'$(20\%), He$'$(74\%), C$'$(0.9\%), O$'$(5\%), Fe$'$(0.1\%),
		in the  scenario $(a)$ of Table \ref{tb:scenarios}
		with $v_{0} = 220$ km/s and parameters in the set A (see text and Fig. \ref{fg:pureHe}). 
                The different values of the halo velocity in the left plot are:
                -100, 0, 30, 60, 100, 150, 200 km/s. 
                The different values of the halo temperature in the right plot are as those of Fig. \ref{fg:pureHe}.
	}
	\label{fg:case_H_He_C_O_Fe}   
\end{figure}

The result corresponding to composite halos are reported in Fig. \ref{fg:case_H_He_Fe},
in Fig. \ref{fg:case_H_He_C_O}  and in Fig. \ref{fg:case_H_He_C_O_Fe} 
where the cases: 
i) H$'$(24\%), He$'$(75\%), Fe$'$(1\%); 
ii) H$'$(20\%), He$'$(74\%), C$'$(0.9\%), O$'$(5\%), Fe$'$(0.1\%),
iii) H$'$(12.5\%), He$'$(75\%), C$'$(7\%), O$'$(5.5\%), have been considered respectively.
In particular, in the case i) we introduce 1\% of Fe$'$ for demonstrating how much heavier 
nuclei can influence the signal. 

As it can be expected, considering for example the behaviour of unmodulated
part of the dark atom signal depicted in Fig. \ref{fg:s0mirror}, the allowed regions 
- in all the considered scenarios - 
move toward lower value of $\sqrt{f}\epsilon$ parameter when the dark atoms of the halo
are heavier with higher charge numbers; in this case the interaction cross section
increases and, in order to keep the same strength of the DM signal, lower value of coupling 
are preferred. The lowest allowed regions is obtained for a pure Fe' halo. 
For each scenario there are two regimes: for cold halo the allowed $\sqrt{f}\epsilon$ parameter 
increases with the halo velocity while the parameter converges to a lower value for
hot halo regardless its velocity in the Galactic frame. 
In cold scenario the dark atoms kinetic energy in the halo is small
and the relative velocity of the halo with respect to the Earth
is the dominant contribution to the average velocity of the particles 
in the laboratory frame.
Thus, for large positive halo velocity the kinetic energy of the dark atoms
in the laboratory frame is small and, to have recoils with sufficient energy to fit the DAMA signal,
large value of $\sqrt{f}\epsilon$ parameter are favoured.
On the contrary, when the velocity of the halo is opposite to the Earth motion,
the kinetic energy of the dark atoms in the laboratory increases and  
lower $\sqrt{f}\epsilon$ values are favoured.  
In hot scenario, the velocity of the dark atoms in the halo is high and it 
becomes the dominant contribution to the velocity of the particles in the laboratory frame.
In this regime the allowed $\sqrt{f}\epsilon$ parameters converge to lower values
for any halo velocity. 
\begin{figure}[!p]
	\begin{center}
		\includegraphics [width=0.32\textwidth]{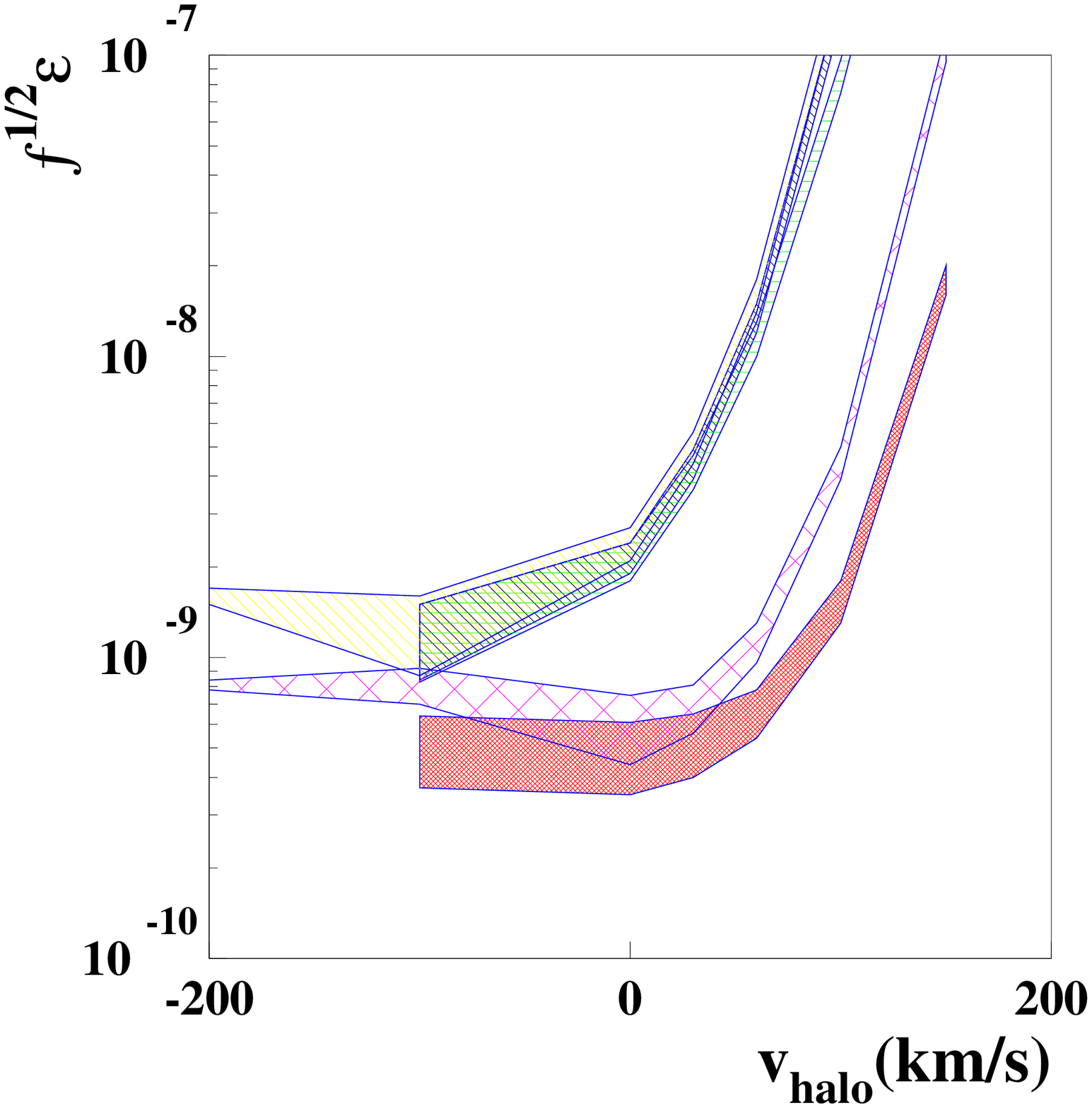}
		\includegraphics [width=0.32\textwidth]{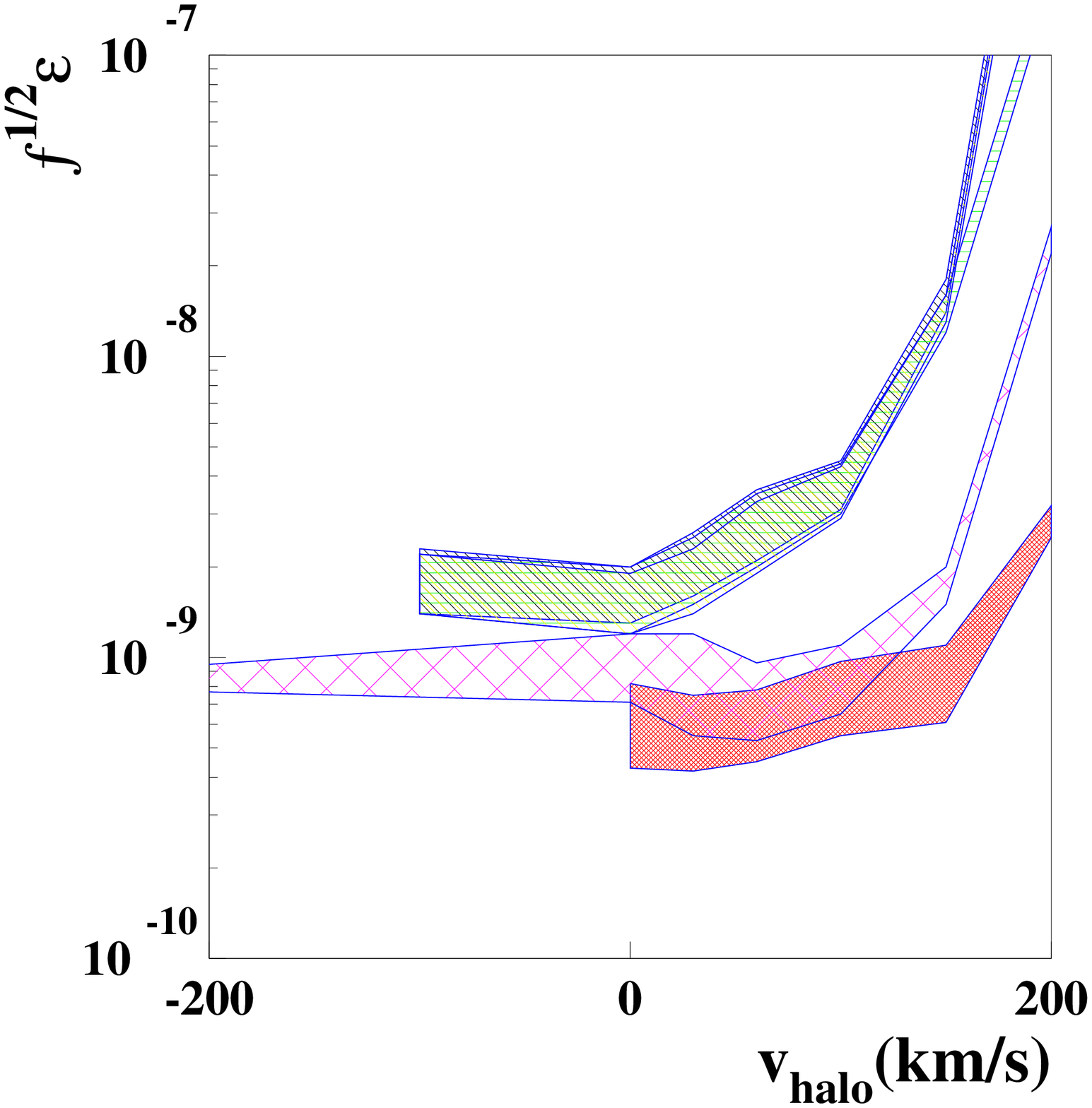}
		\includegraphics [width=0.32\textwidth]{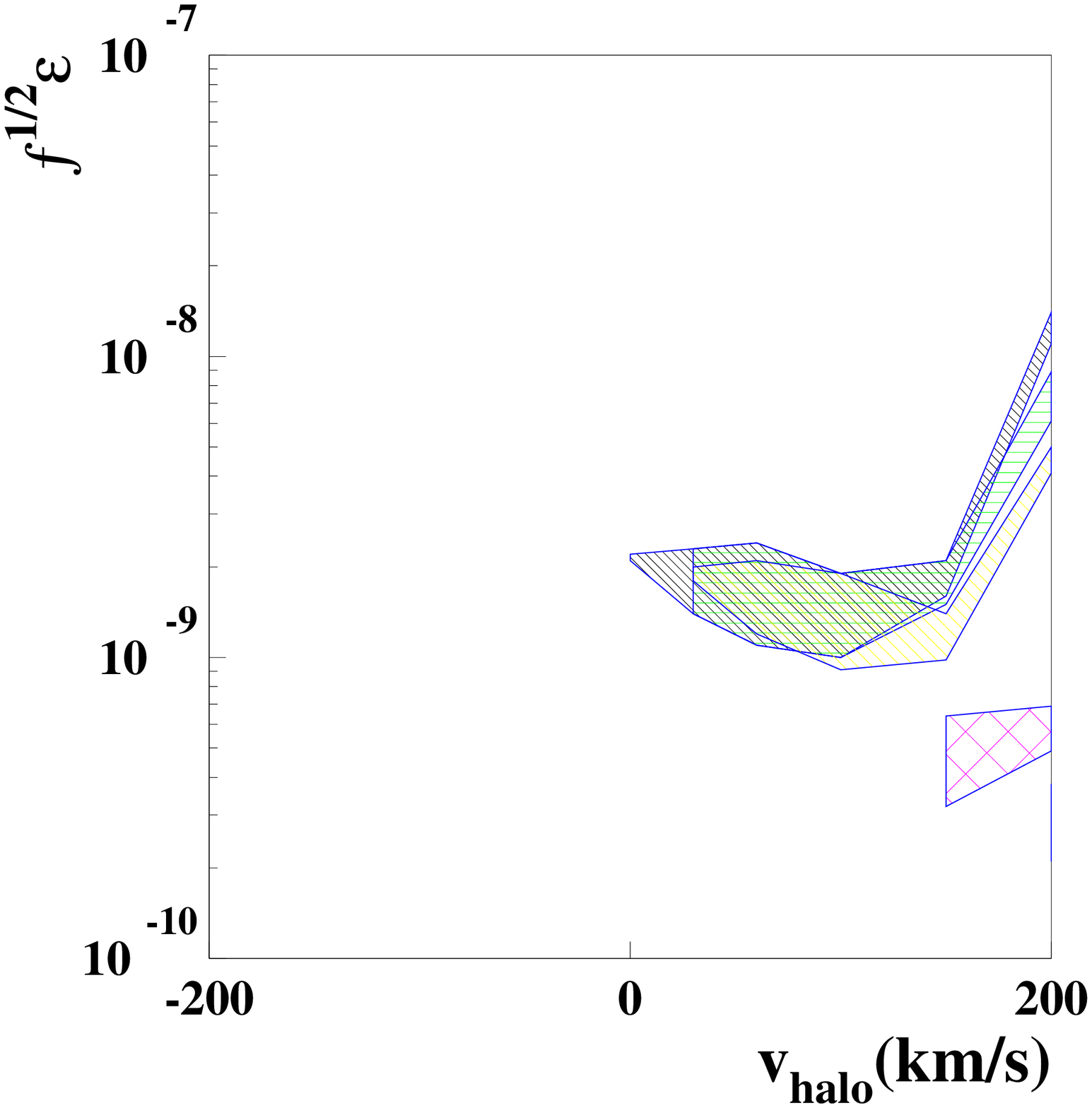}
		\includegraphics [width=0.32\textwidth]{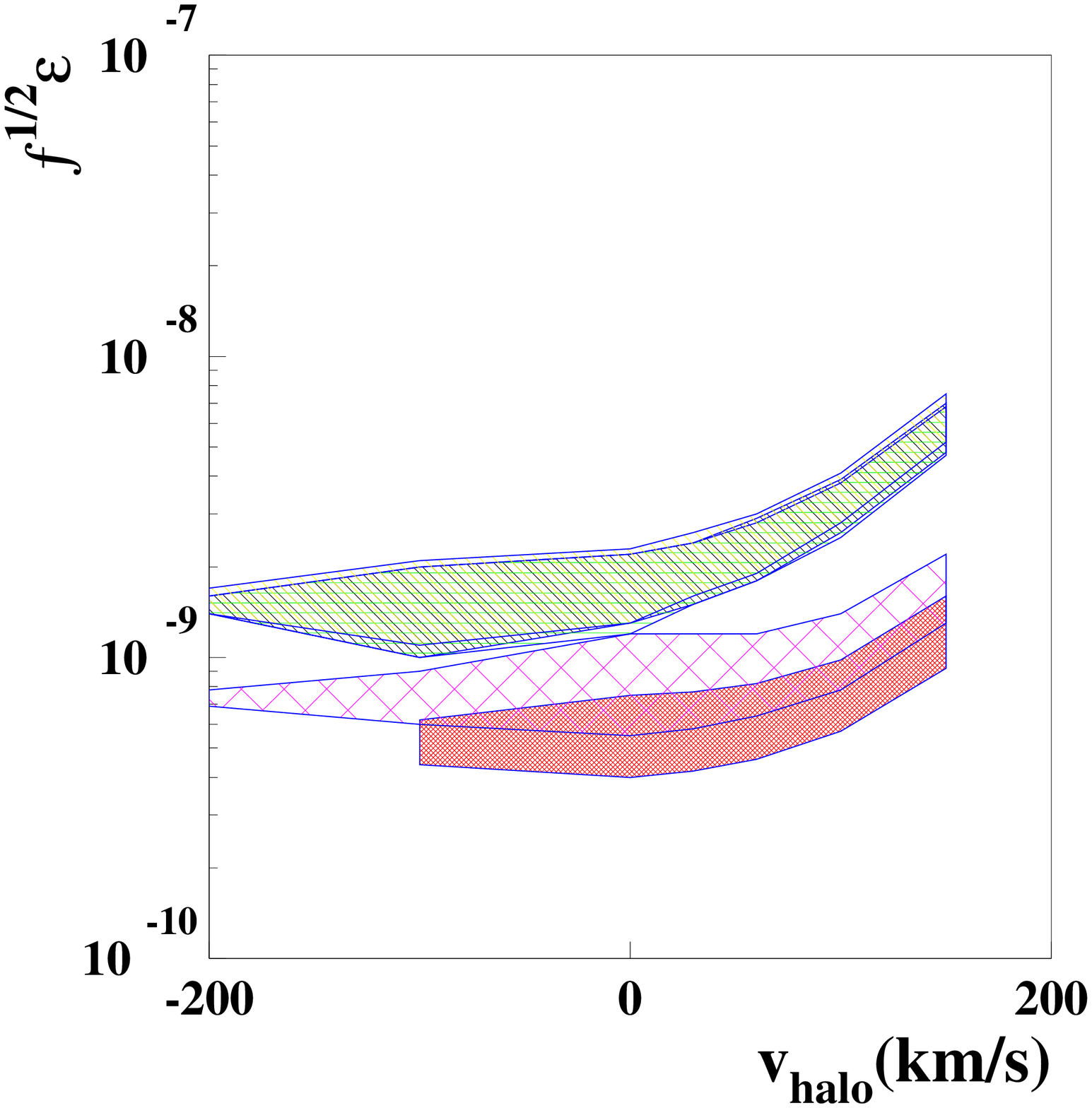}
		\includegraphics [width=0.32\textwidth]{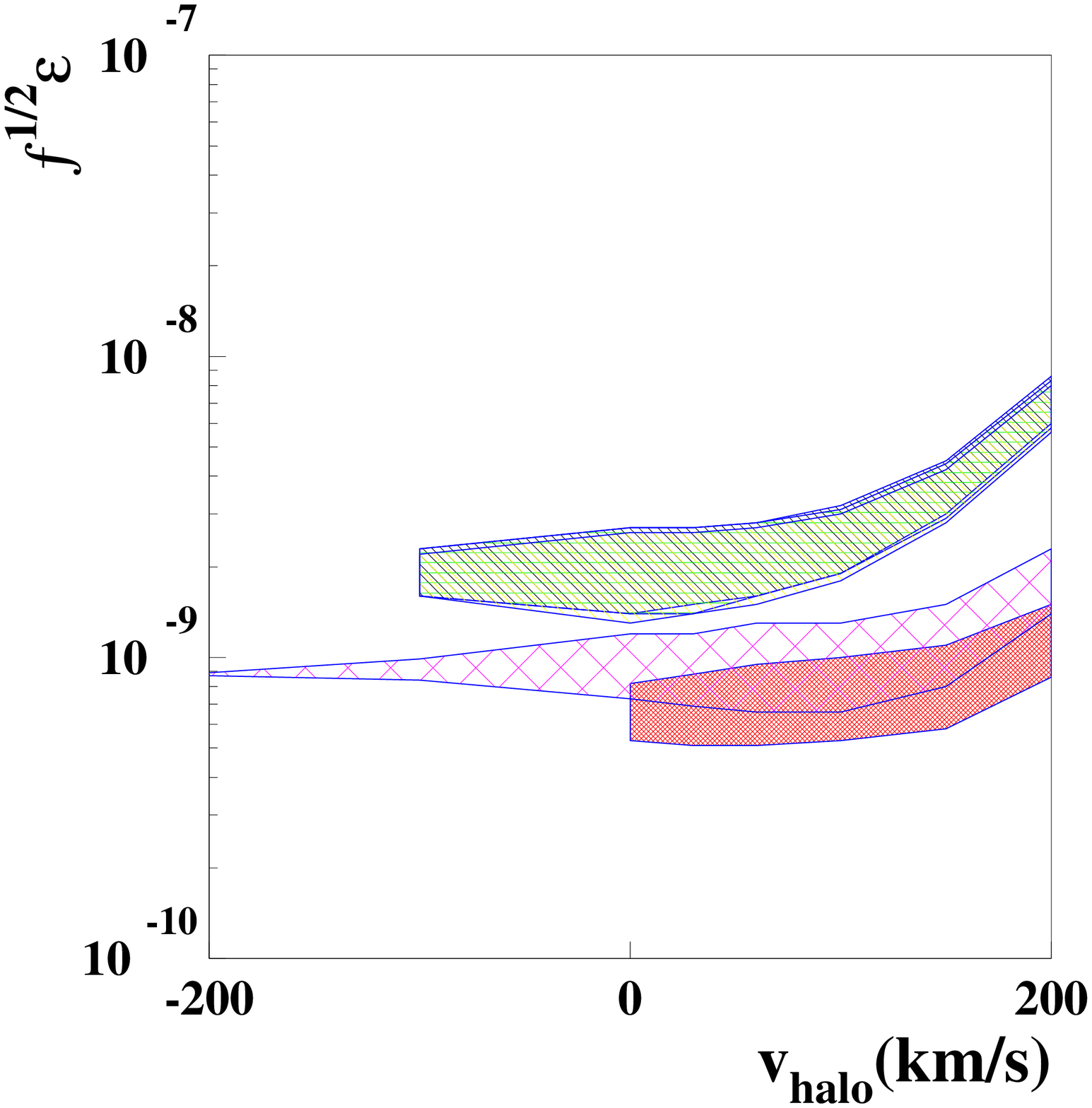}
		\includegraphics [width=0.32\textwidth]{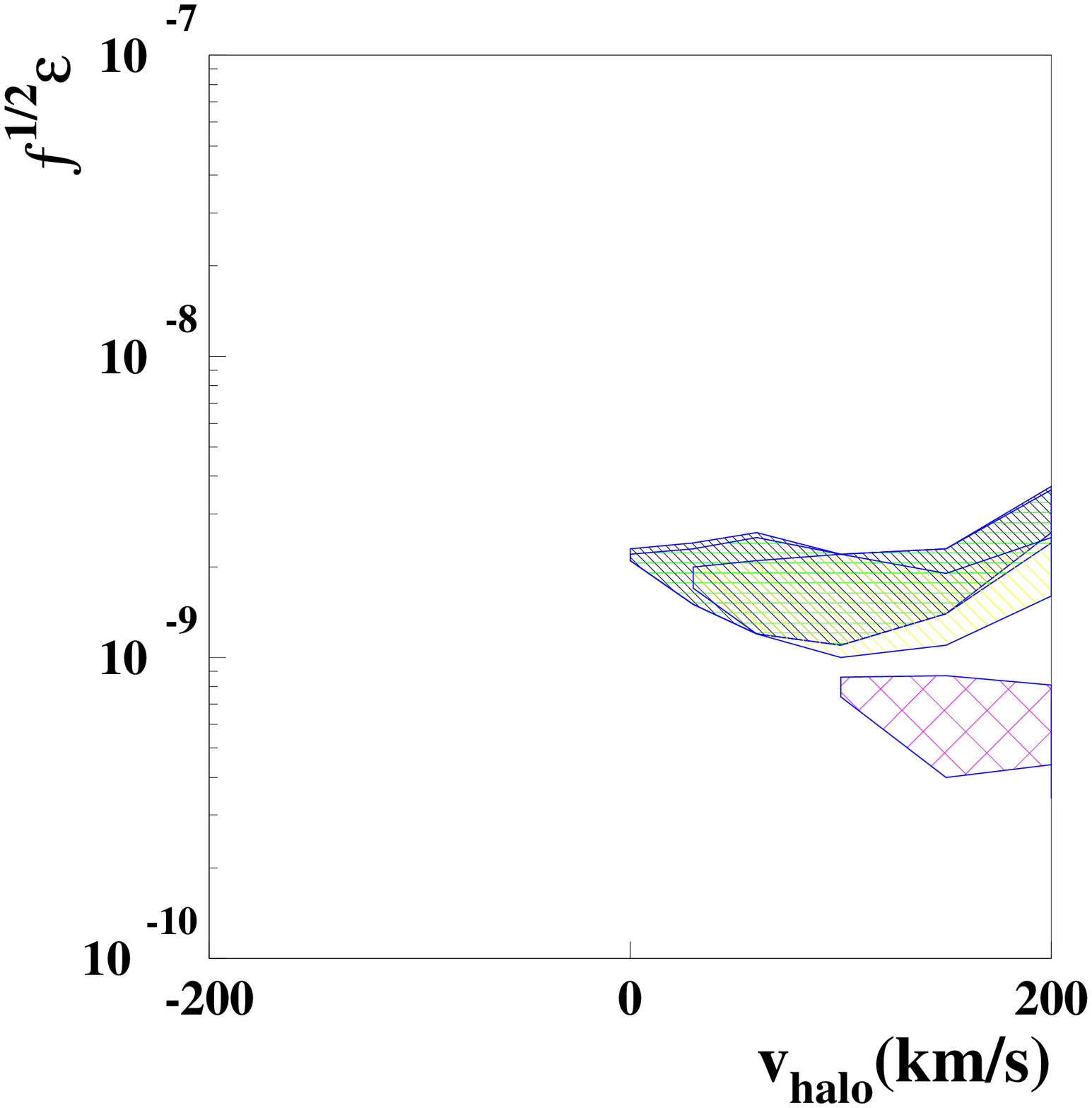}
	\end{center}
	\vspace{-0.8cm}
	\caption{Allowed regions for the $\sqrt{f}\epsilon$ parameter
		as a function of $v_{halo}$. 
                The three graphs in the top (bottom) have been obtained by considering
                a halo temperature $T=5 \times 10^{5} \;{\textrm K}$ ($T=10^{7} \;{\textrm K}$).
		The results of three different dark halo compositions and frameworks have been reported.
		{\it Left}: composite dark halo H$'$(12.5\%), He$'$(75\%), C$'$(7\%), O$'$(5.5\%),
		with $v_{0} = 170$ km/s  and parameters in the set B. 
		{\it Center}:  composite dark halo  H$'$(20\%), He$'$(74\%), C$'$(0.9\%), O$'$(5\%), Fe$'$(0.1\%),
		with $v_{0} = 220$ km/s  and parameters in the set A. 
		{\it Right}:  composite dark halo H$'$(24\%), He$'$(75\%), Fe$'$(1\%),
		with $v_{0} = 270$ km/s  and parameters in the set C. 
		Each graph has five contours corresponding to the scenarios of Table \ref{tb:scenarios}:
                from bottom to top the regions (see for example at $v_{halo} = 150$ km/s) refer to
		the cases (b), (d), (c), (a), (e), respectively.
	}
	\label{fg:cfr_que1}   
\end{figure}
When the velocity of the halo is high and opposite to the Earth, its contribution
to the kinetic energy of the dark atoms in the laboratory frame is dominant 
with respect to the velocity distribution of the particles in the halo. In this case
the allowed $\sqrt{f}\epsilon$ parameters is independent on the temperature of the halo.
In the case of a pure Fe' halo there are no allowed region for negative halo velocity.
In fact, the coupling of the Fe' mirror atoms with ordinary matter is high and the 
expected signal in case of a particle with high kinetic energy is too large to fit 
the DAMA observed annual modulation effect. 

\begin{figure}[!p]
	\begin{center}
		\includegraphics [width=0.32\textwidth]{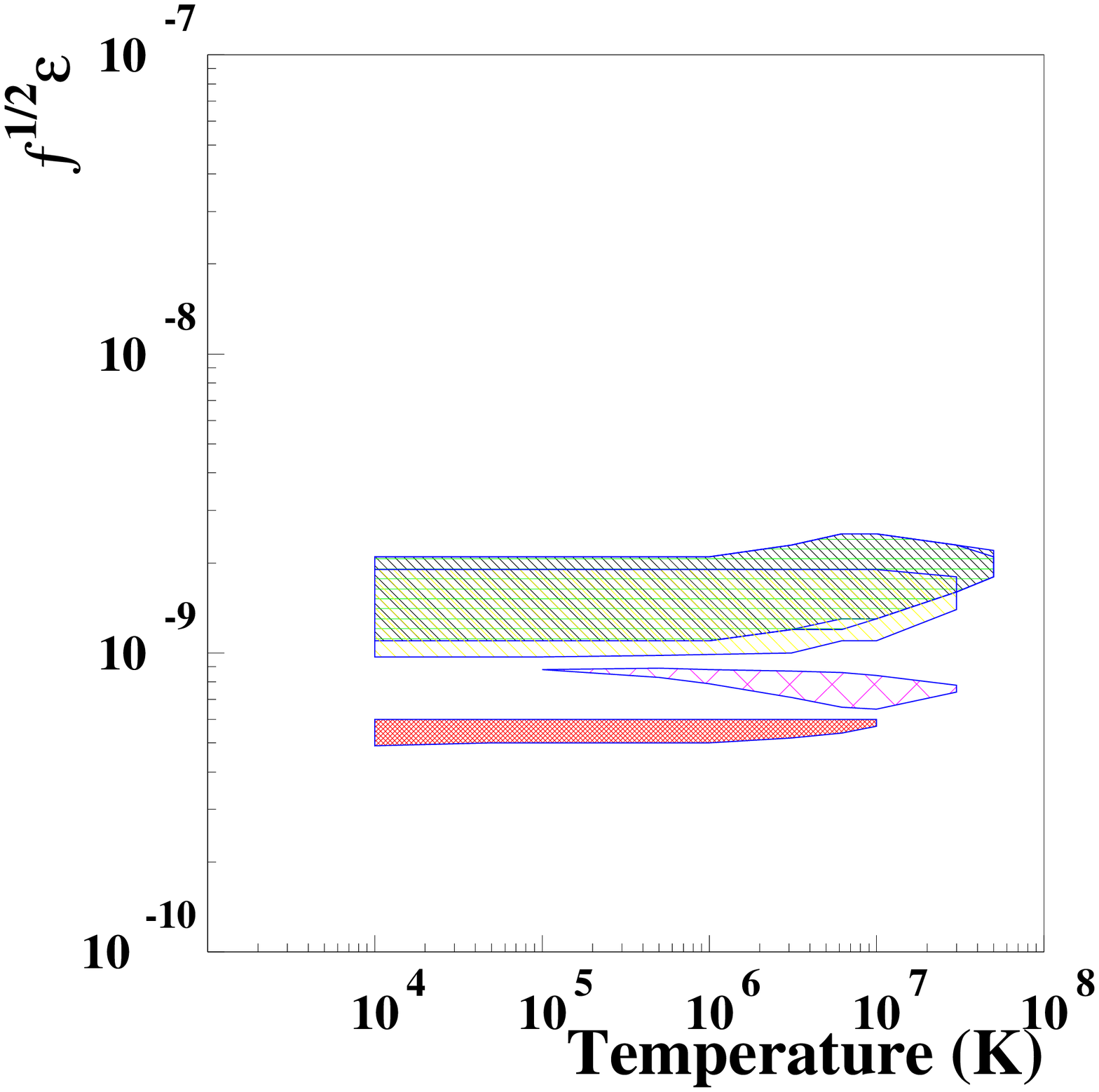}
		\includegraphics [width=0.32\textwidth]{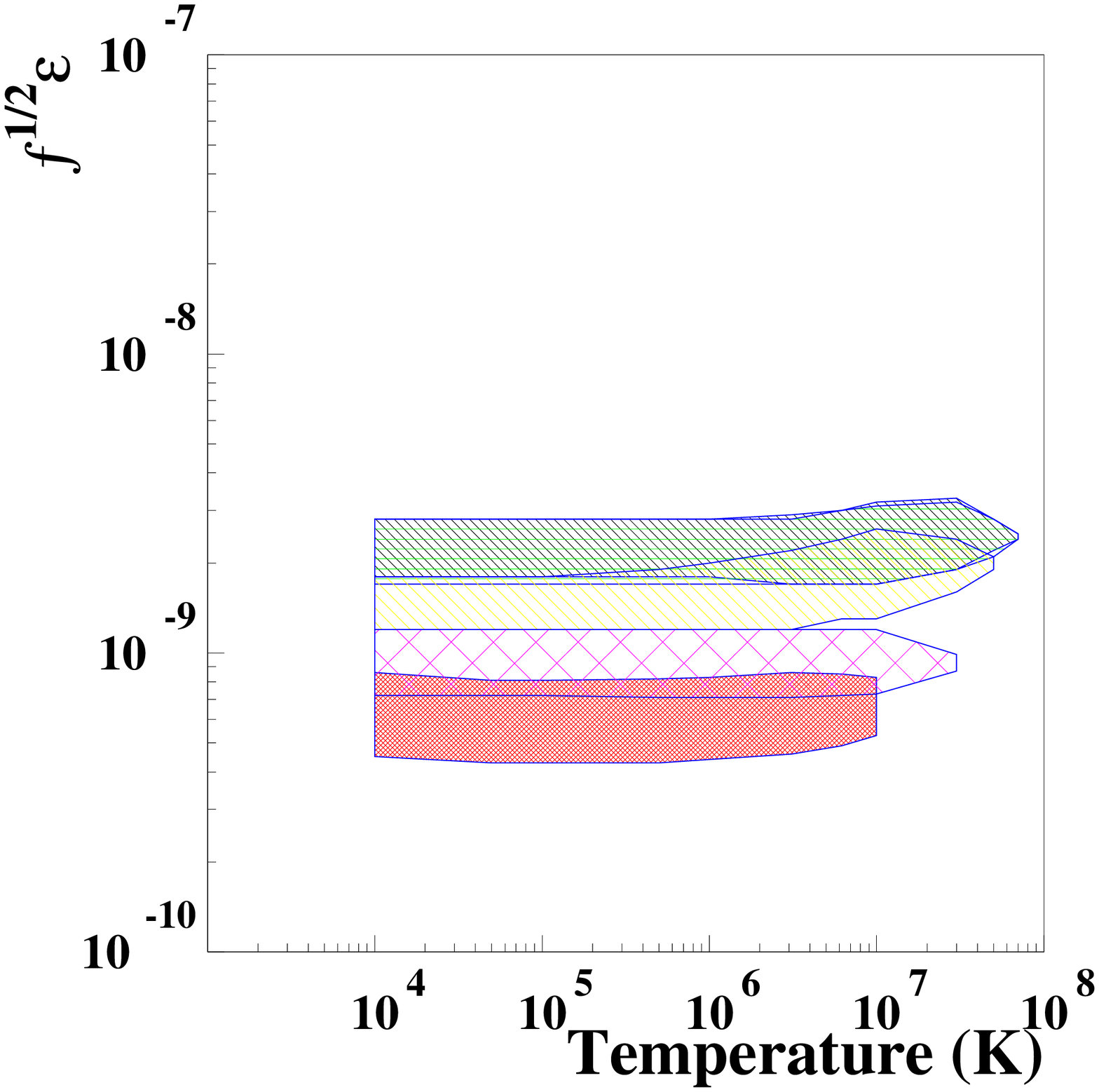}
		\includegraphics [width=0.32\textwidth]{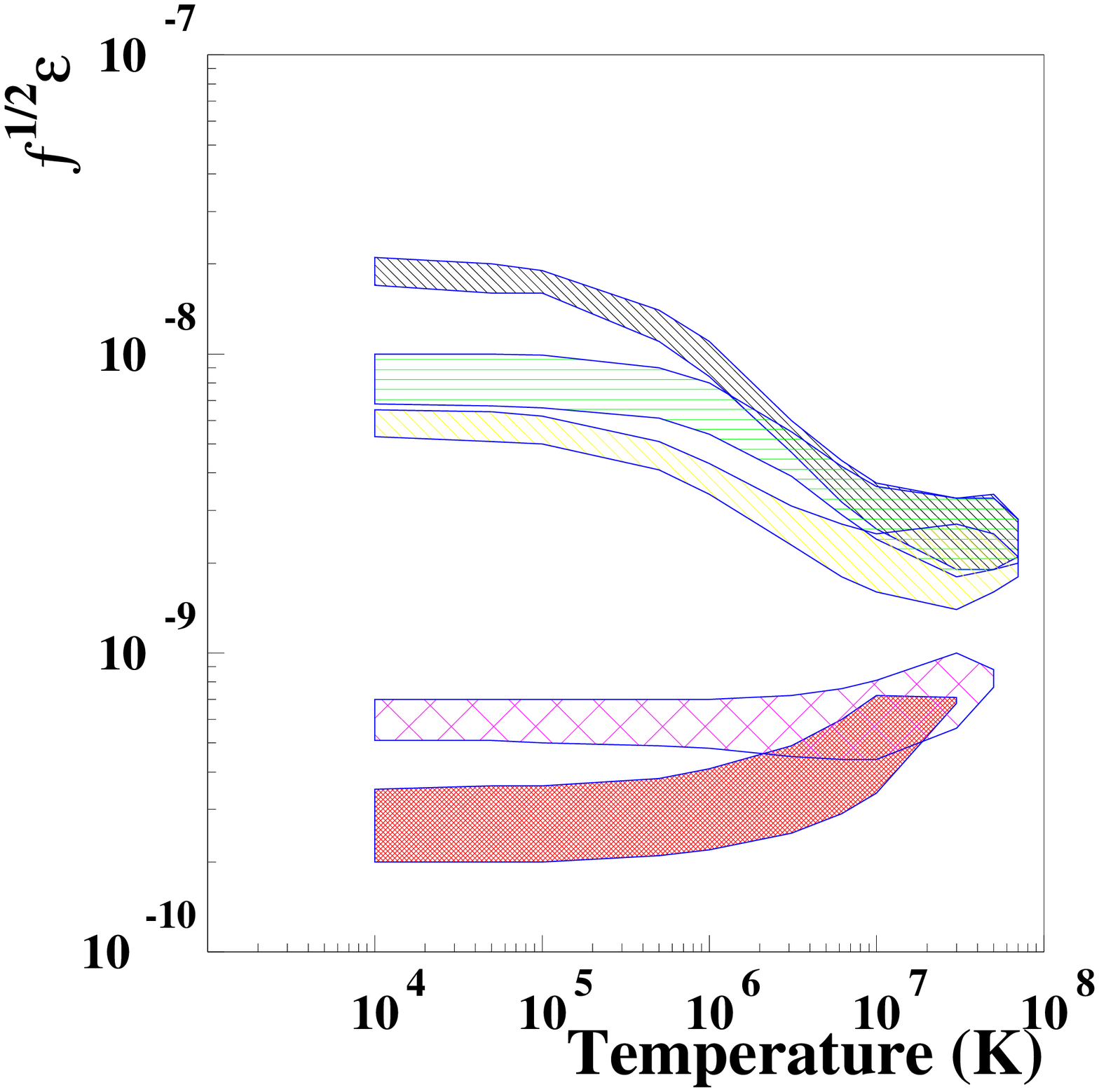}
	\end{center}
	\vspace{-0.8cm}
	\caption{Allowed regions for the $\sqrt{f}\epsilon$ parameter
		as function of the halo temperature. 
		The three graphs refer to different dark halo composition and allow 
		to compare the results obtained by considering the different scenarios of Table \ref{tb:scenarios}.
               	The five contours in each plot correspond, from the bottom to the top, to  
                the cases (b), (d), (c), (a), (e), respectively.
		{\it Left}:  composite dark halo  H$'$(12.5\%), He$'$(75\%), C$'$(7\%), O$'$(5.5\%),
		with $v_{0} = 220$ km/s, $v_{halo} = -100$ km/s and parameters in the set C. 
		{\it Center}: composite dark halo H$'$(20\%), He$'$(74\%), C$'$(0.9\%), O$'$(5\%), Fe$'$(0.1\%),
		with $v_{0} = 220$ km/s, $v_{halo} = 0$ km/s and parameters in the set C. 
		{\it Right}:  composite dark halo H$'$(24\%), He$'$(75\%), Fe$'$(1\%),
		with $v_{0} = 220$ km/s, $v_{halo} = 150$ km/s and parameters in the set C. 
	}
	\label{fg:cfr_que3}   
\end{figure}

As discussed in the previous section we have considered many uncertainties regarding the models and the parameters
needed in the calculation of the expected dark atoms signal. To show the impact of these uncertainties 
we have reported in the following the different allowed regions obtained for the same dark halo when different
parameters and scenarios are considered. All the figures will have three plots corresponding
to the following composite dark halo:  
i) H$'$(12.5\%), He$'$(75\%), C$'$(7\%), O$'$(5.5\%) (left plot); 
ii) H$'$(20\%),  He$'$(74\%), C$'$(0.9\%), O$'$(5\%), Fe$'$(0.1\%) (central plot);
iii) H$'$(24\%), He$'$(75\%), Fe$'$(1\%) (right plot). 
These plots show the impact of the considered model framework and parameters 
to the $\sqrt{f}\epsilon$ allowed values.

In Fig. \ref{fg:cfr_que1} 
the impact of the different adopted quenching factor
is reported. The figures in the top (bottom) have been obtained by considering a halo temperature of
$5 \times 10^{5} \;{\textrm K}$ ($10^{7}$ K);
in each plots, the five scenarios of Table \ref{tb:scenarios}
have been considered for the three different halo models and different model frameworks. 
As it can be noted the allowed $\sqrt{f}\epsilon$ region can span over orders of magnitudes
depending on the considered scenario. 

In Fig. \ref{fg:cfr_que3} allowed regions for the $\sqrt{f}\epsilon$ parameter
as function of the halo temperature are reported to show the impact of 
the different scenarios of Table \ref{tb:scenarios}.
The three panels refer to three different halo models and model framework.

\begin{figure}[!ht]
	\begin{center}
		\includegraphics [width=0.32\textwidth]{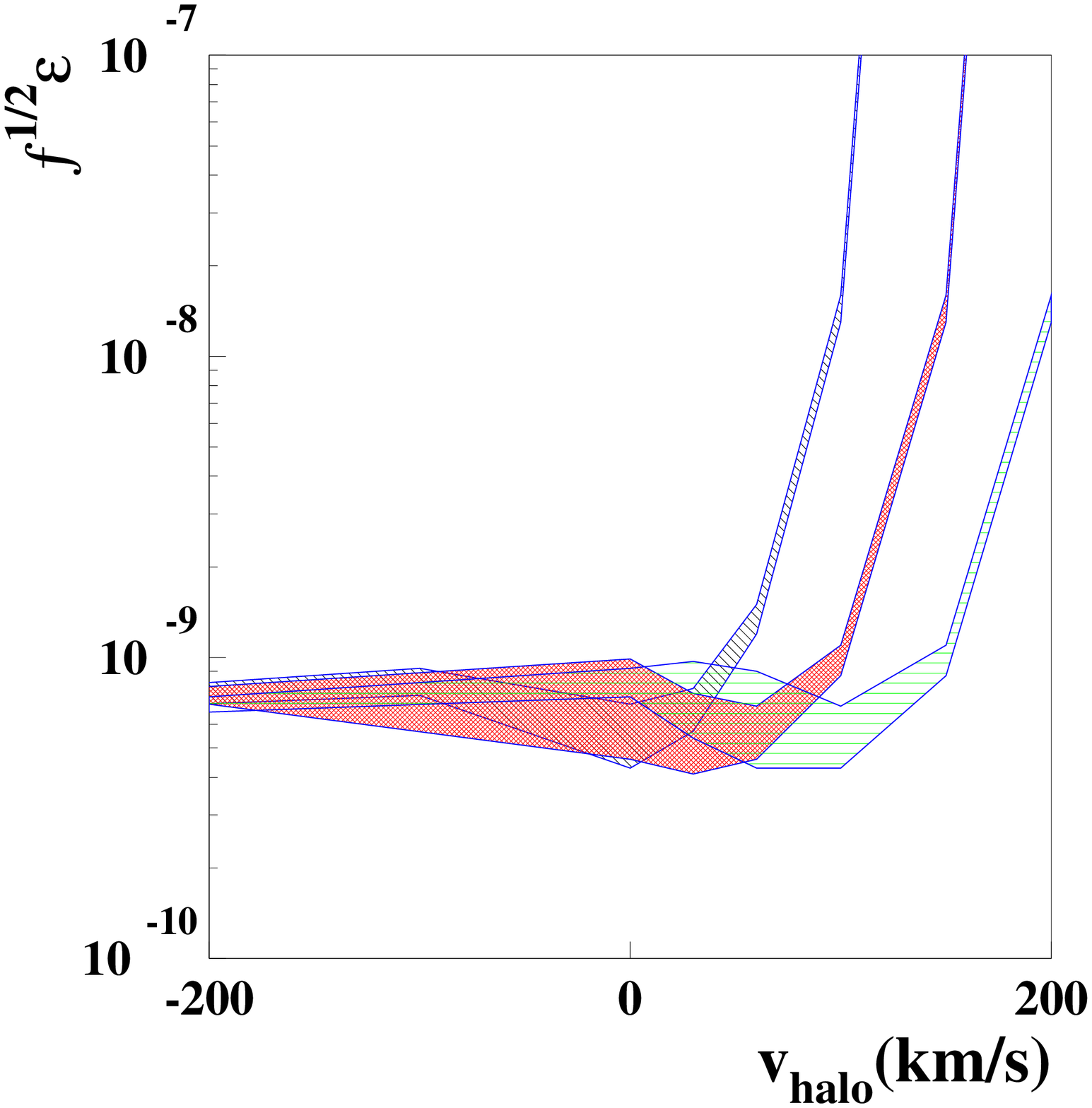}
		\includegraphics [width=0.32\textwidth]{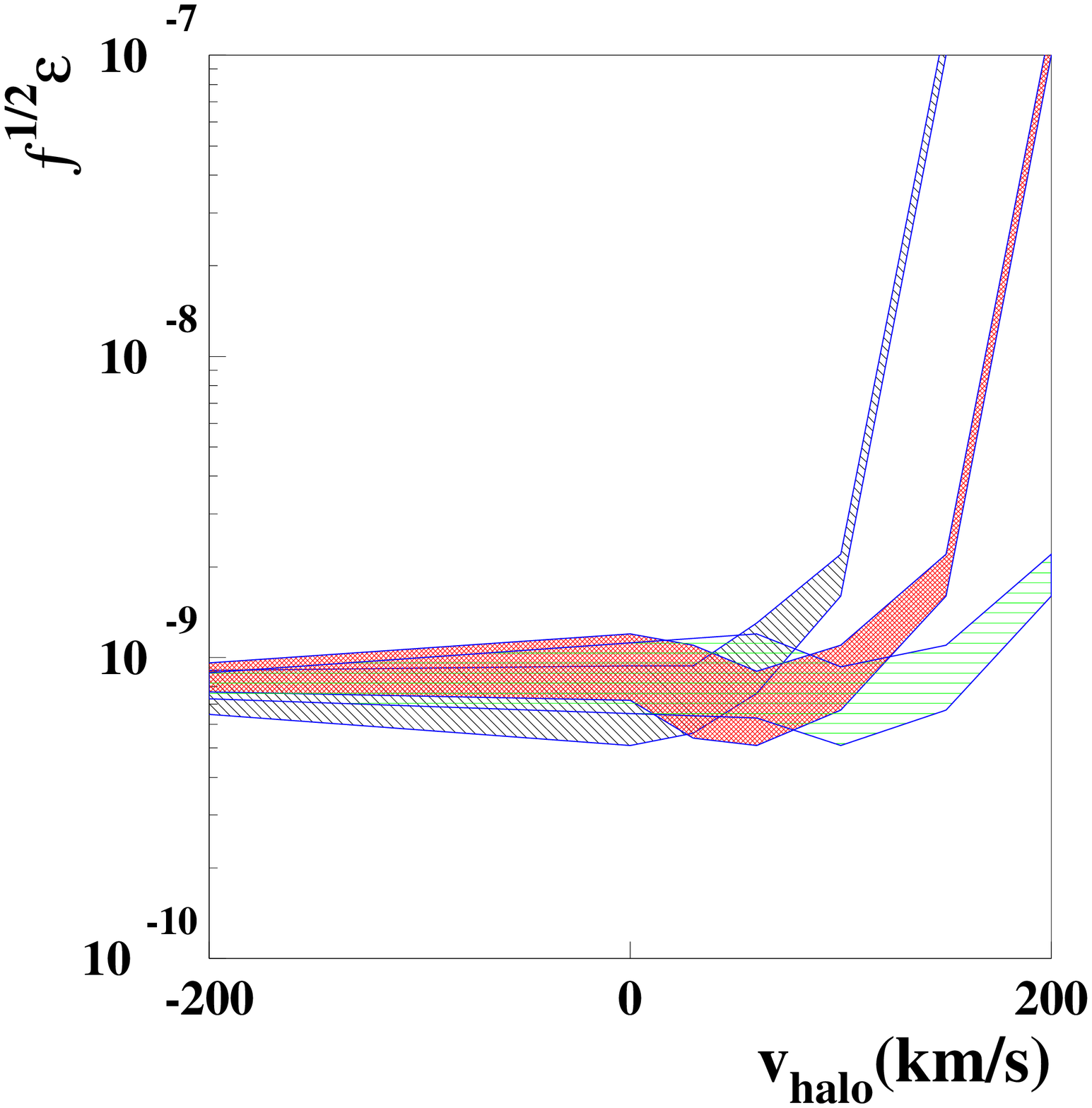}
		\includegraphics [width=0.32\textwidth]{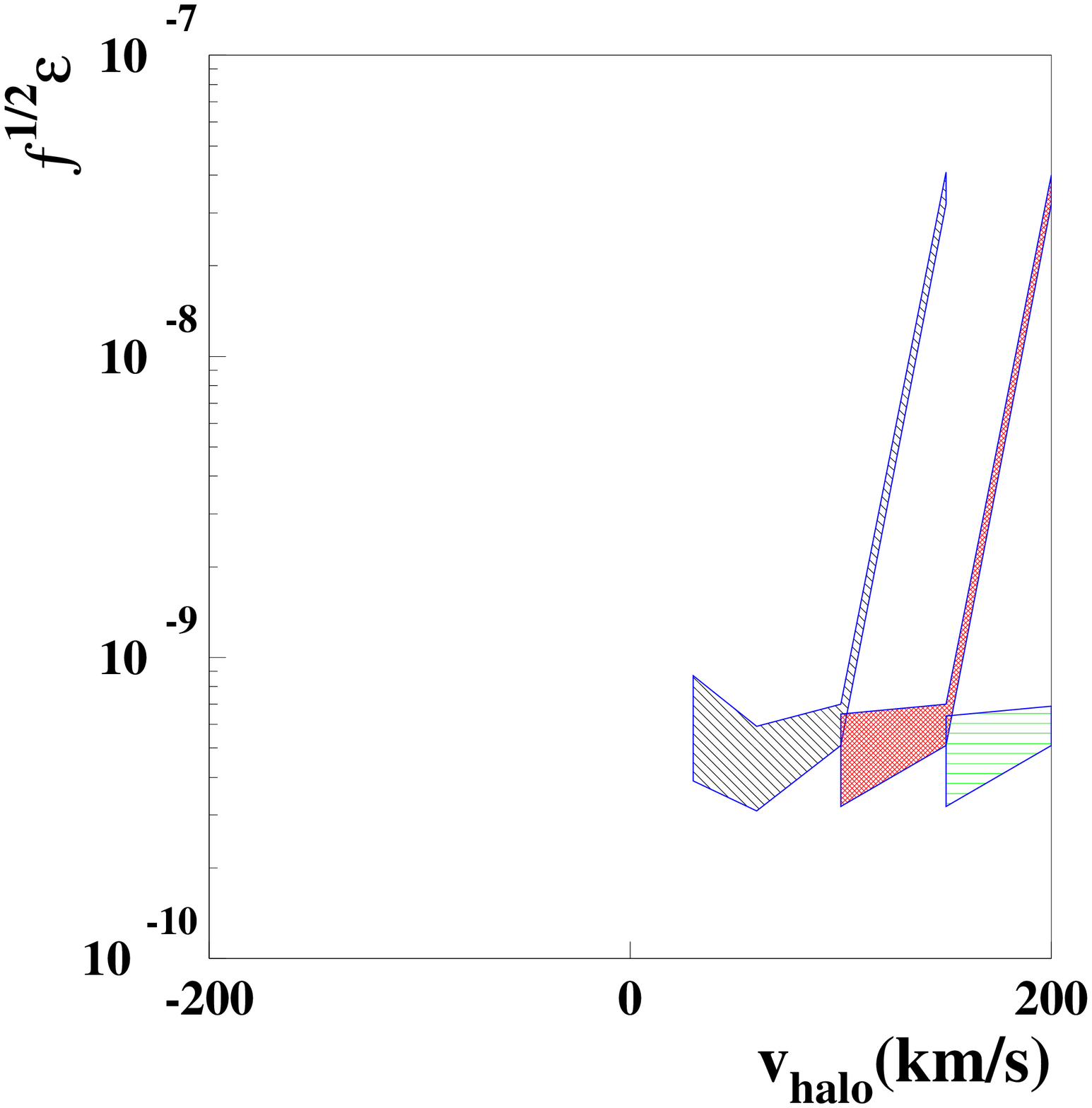}
	\end{center}
	\vspace{-0.8cm}
	\caption{
		Allowed regions for the $\sqrt{f}\epsilon$ parameter
		as function of $v_{halo}$. 
		The three graphs refer to different dark halo compositions with the same temperature $T=10^{4} \;{\textrm K}$,
		the same set A and the same scenario (d):
		{\it Left}:   composite dark halo H$'$(12.5\%), He$'$(75\%), C$'$(7\%), O$'$(5.5\%).
		{\it Center}: composite dark halo H$'$(20\%), He$'$(74\%), C$'$(0.9\%), O$'$(5\%), Fe$'$(0.1\%).
		{\it Right}:  composite dark halo H$'$(24\%), He$'$(75\%), Fe$'$(1\%).
		The three contours in each plot correspond to
		$v_{0} = 170$ km/s (area with diagonal lines) (gray area on-line), 
                $v_{0} = 220$ km/s (shaded area) (red area on-line), 
                $v_{0} = 270$ km/s (area with horizontal lines) (green area on-line), 
		respectively.
	}
	\label{fg:cfr_v02}   
\end{figure}

\begin{figure}[!h]
	\begin{center}
		\includegraphics [width=0.32\textwidth]{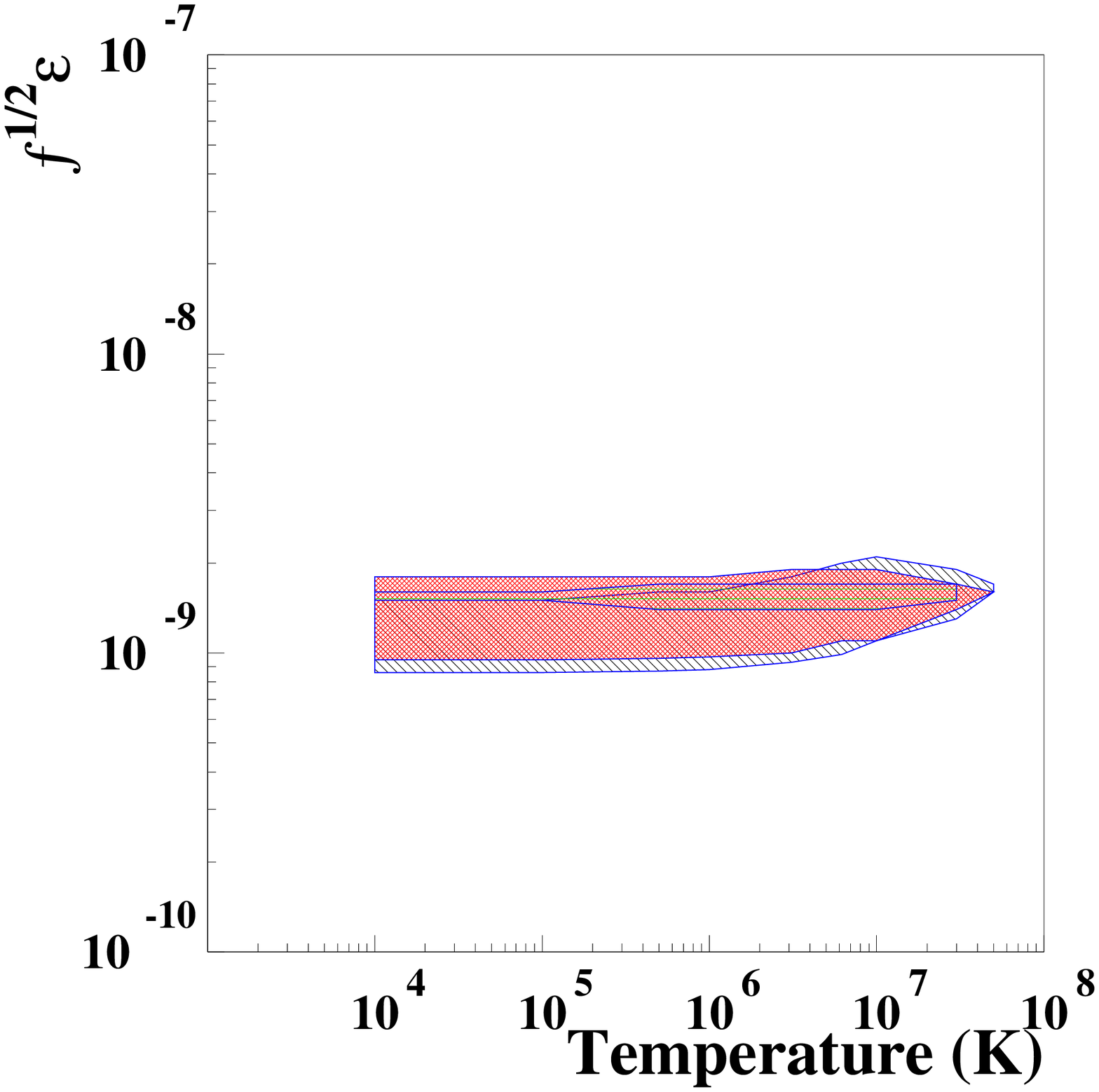}
		\includegraphics [width=0.32\textwidth]{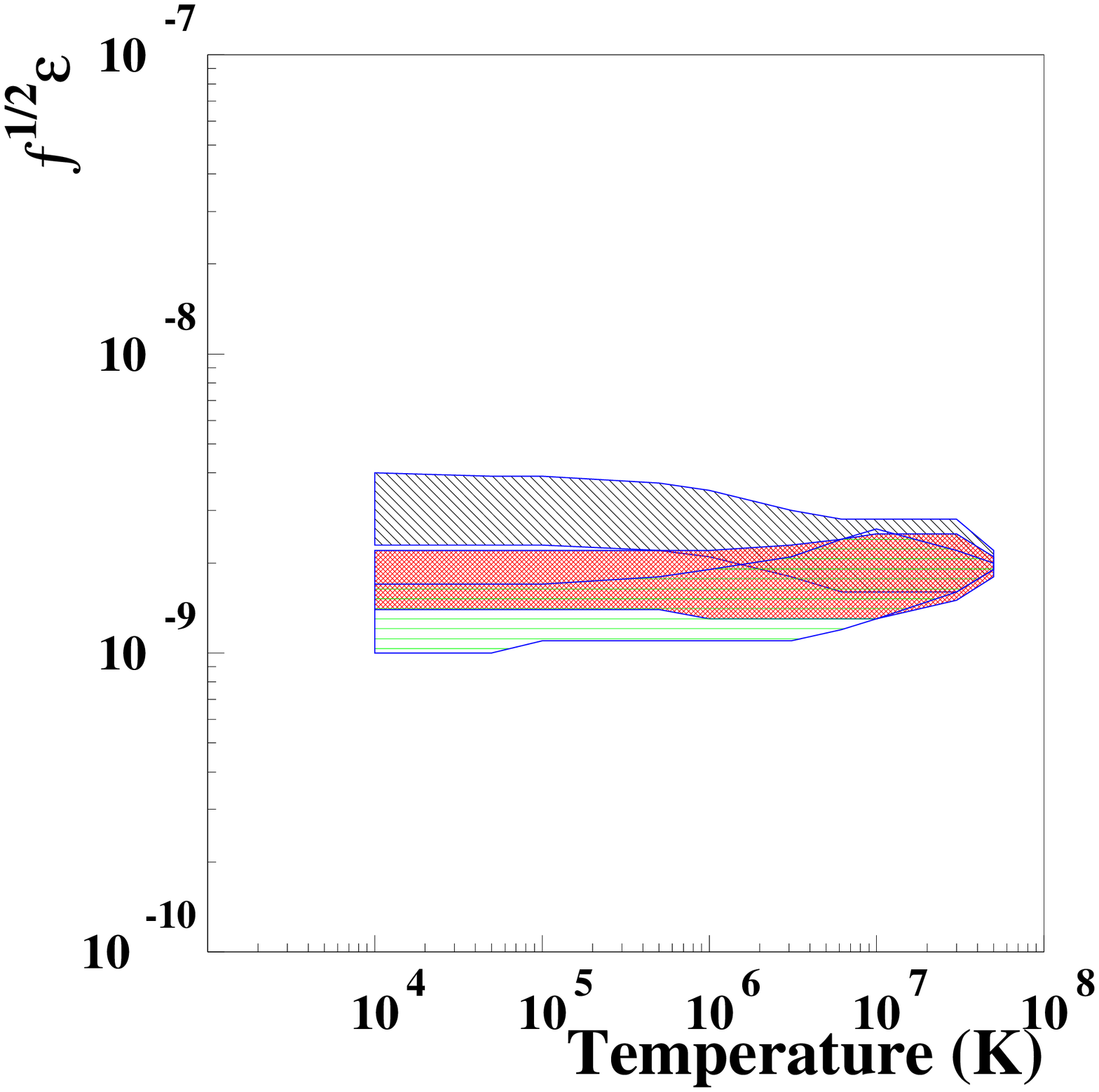}
		\includegraphics [width=0.32\textwidth]{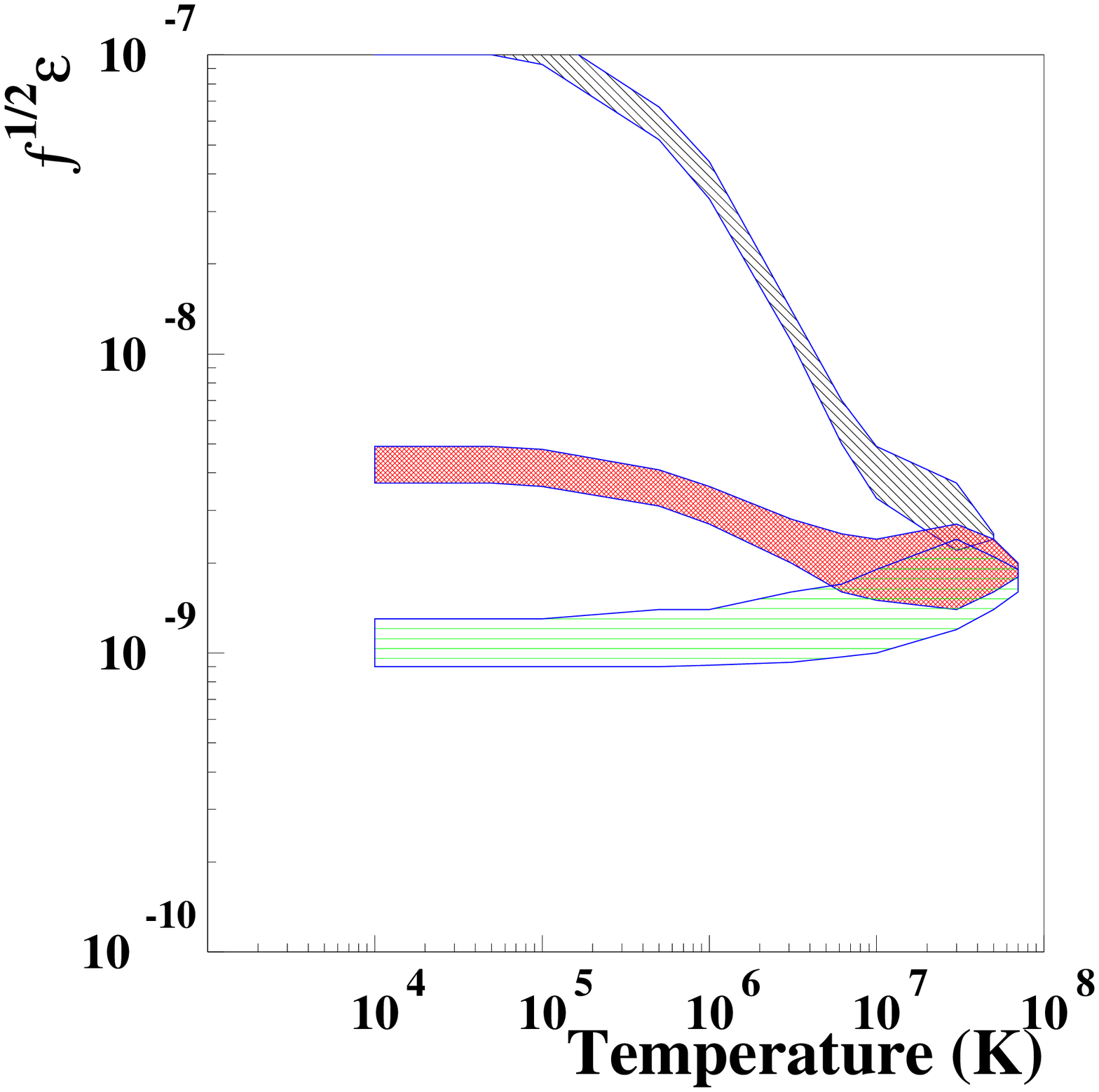}
	\end{center}
	\vspace{-0.8cm}
	\caption{
		Allowed regions for the $\sqrt{f}\epsilon$ parameter
		as function of the halo temperature. 
		The three graphs refer to different dark halo composition and allow 
		to compare the results obtained by considering different $v_{0}$:
		{\it Left}:  composite dark halo H$'$(12.5\%), He$'$(75\%), C$'$(7\%), O$'$(5.5\%), with
		$v_{halo} = -100$ km/s,  scenario (e)  and parameters in the set B. 
		{\it Center}: composite dark halo H$'$(20\%), He$'$(74\%), C$'$(0.9\%), O$'$(5\%), Fe$'$(0.1\%), with
		$v_{halo} = 30$ km/s, scenario (a) and parameters in the set B. 
		{\it Right}:  composite dark halo H$'$(24\%), He$'$(75\%), Fe$'$(1\%), with
		$v_{halo} = 150$ km/s, scenario (c)  and parameters in the set B. 
                The three contours in each plot correspond to
		$v_{0} = 170$ km/s (area with diagonal lines) (gray area on-line), 
                $v_{0} = 220$ km/s (shaded area) (red area on-line), 
                $v_{0} = 270$ km/s (area with horizontal lines) (green area on-line), 
		respectively.
	}
	\label{fg:cfr_v03}   
\end{figure}

In the Fig. \ref{fg:cfr_v02}
the allowed regions for the $\sqrt{f}\epsilon$ parameter
as a function of the halo velocity for 
the three different $v_{0}$ values: 170, 220 and 270 km/s, are reported.  The different plots in this figure
refer to different dark halo compositions with the same temperature $T=10^{4} \;{\textrm K}$,
the same set A and the same scenario (d).
From this figure it is possible to see the impact of the $v_{0}$ parameter in the
evaluation of the allowed regions. 
Fig. \ref{fg:cfr_v03} shows the allowed regions for the $\sqrt{f}\epsilon$ parameter
as a function of the halo temperature for 
the three different $v_{0}$ values by considering different dark halo. 
It is worth noting that the $v_{0}$ parameter in the considered range of variability 
has impact on the allowed regions for low temperature halo when the
halo velocity is positive and larger than 100 km/s.

\begin{figure}[!pt]
	\begin{center}
		\includegraphics [width=0.32\textwidth]{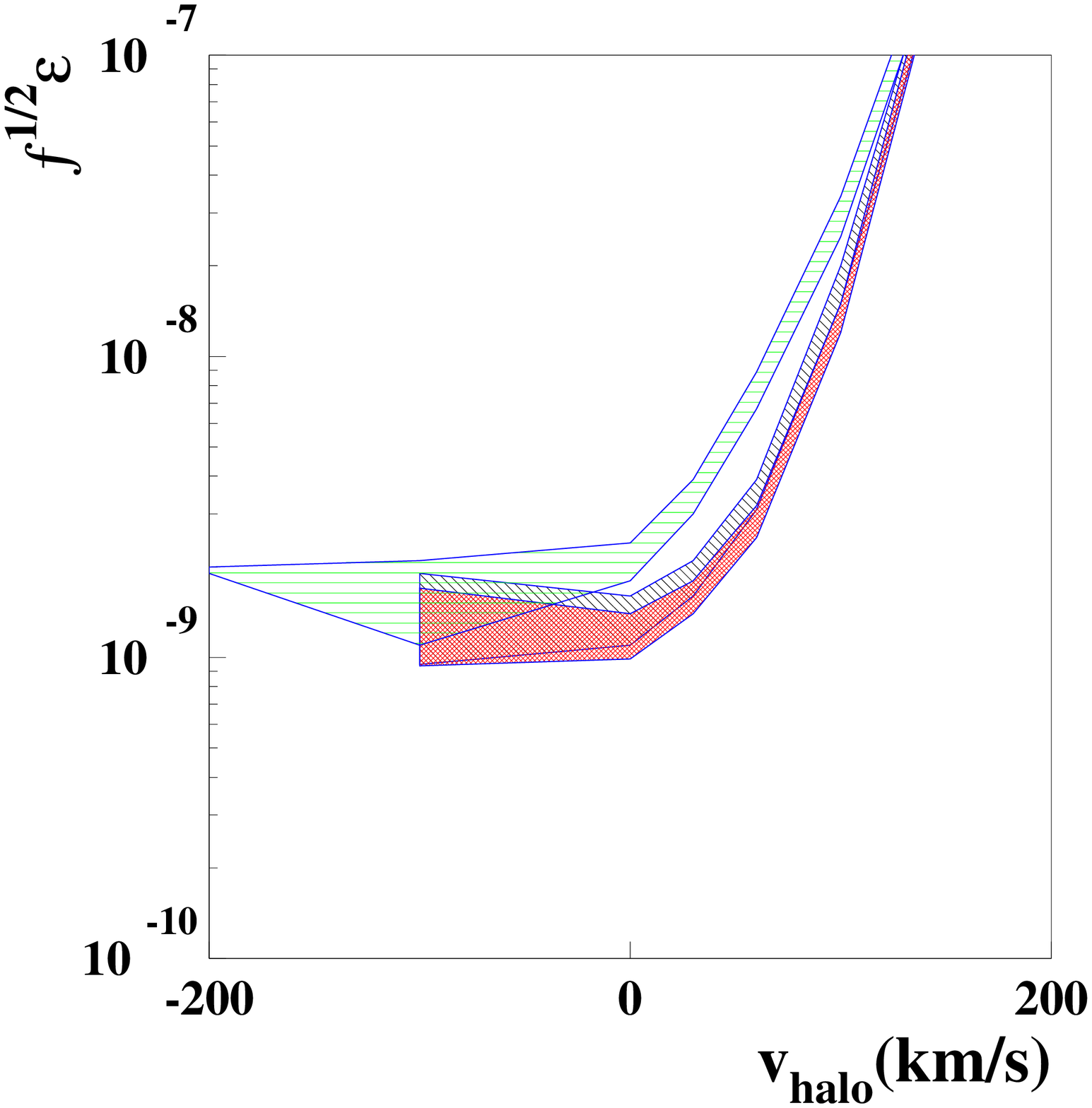}
		\includegraphics [width=0.32\textwidth]{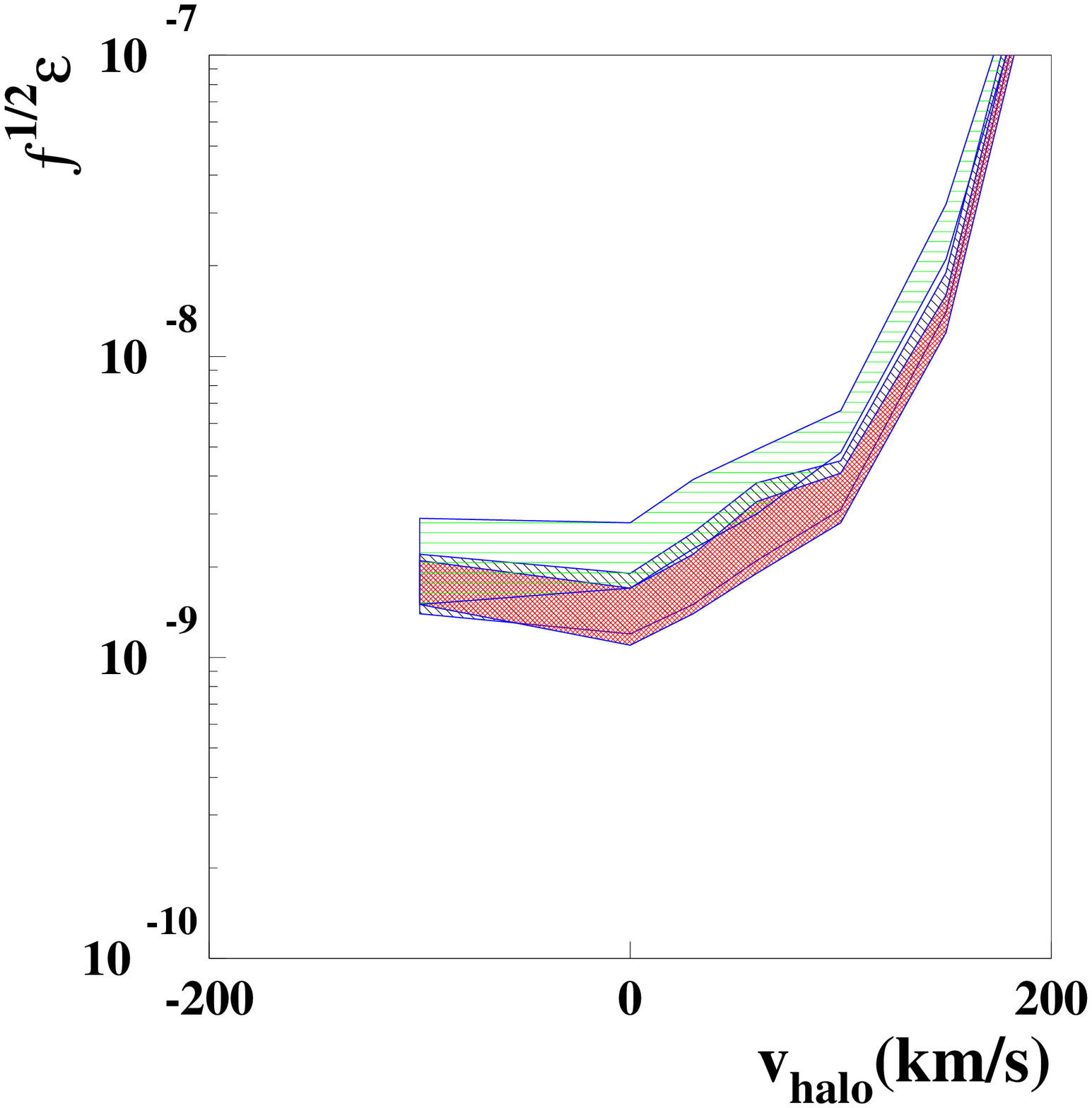}
		\includegraphics [width=0.32\textwidth]{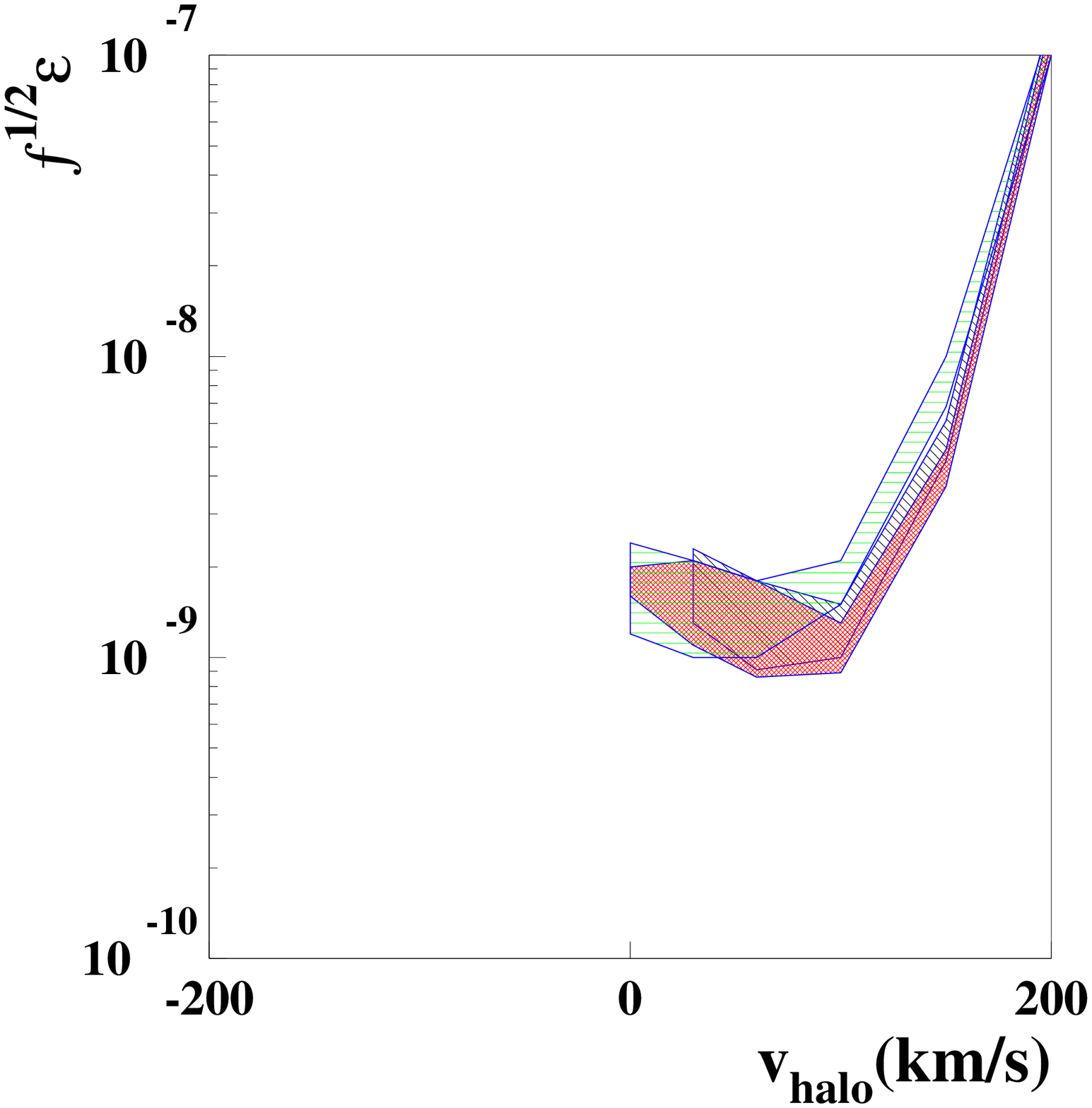}
	\end{center}
	\vspace{-0.8cm}
	\caption{Allowed regions for the $\sqrt{f}\epsilon$ parameter
		as function of $v_{halo}$. 
		The three graphs refer to different dark halo compositions in the same scenario (c),
                the same temperature $T=10^{4} \;{\textrm K}$ and $v_{0} = 220$ km/s:
		{\it Left}:  composite dark halo H$'$(12.5\%), He$'$(75\%), C$'$(7\%), O$'$(5.5\%).
		{\it Center}: composite dark halo H$'$(20\%), He$'$(74\%), C$'$(0.9\%), O$'$(5\%), Fe$'$(0.1\%).
		{\it Right}:  composite dark halo H$'$(24\%), He$'$(75\%), Fe$'$(1\%).
		The three contours in each plot correspond to:
		set C  (area with horizontal lines) (green area on-line), 
                set A (area with diagonal lines) (gray area on-line), 
                set B (shaded area) (red area on-line), respectively.
                The results obtained by considering the different sets of the parameters can be compared.
	}
	\label{fg:cfr_set1}   
\end{figure}

Finally, to point out the impact of the uncertainties in the values of some nuclear parameters,
represented by set A, B, and  C, described above,
in Fig. \ref{fg:cfr_set1} the allowed regions for the $\sqrt{f}\epsilon$ parameter
as a function of the halo velocity in the Galactic frame are reported for
three different dark halo with the same temperature $T=10^{4} \;{\textrm K}$
and $v_{0} = 220$ km/s. In each plot the three different allowed regions correspond 
to the set A, B and C.

In conclusion, Figs. \ref{fg:cfr_que1}, \ref{fg:cfr_que3}, \ref{fg:cfr_v02}, \ref{fg:cfr_v03}, \ref{fg:cfr_set1}
show that the allowed values of the $\sqrt{f}\epsilon$
parameter span over almost two orders of magnitude
depending on the halo temperature and on the halo velocity; these two
parameters have a great impact in the allowed regions.  As it can be
noted in Fig. \ref{fg:cfr_que1} and \ref{fg:cfr_que3} the allowed regions have a clear dependence
on the chosen scenario for the response of the detector (as in Table \ref{tb:scenarios}); 
scenarios with a better response at low energy, such e.g. the scenario (b), favour smaller values
of $\sqrt{f}\epsilon$. The uncertainties on the
Galactic local velocity, once the halo temperature is fixed, play a role only for positive halo
velocities larger than about 100 km/s (see for example Fig. \ref{fg:cfr_v02}).
The uncertainties on the  parameters used in the nuclear form factors
(the three different set A, B and C) have smaller impacts on the 
allowed regions.  Finally, it is worth noting that many configurations
exist that are well compatible with cosmological bounds. Obviously,
introduction of other uncertainties and modelling is expected to further
enlarge the allowed regions.

\section{Conclusions}
The mirror matter model has been considered to analyze the 
DM model-independent annual modulation effect observed by the DAMA Collaboration
with NaI(Tl) target detectors. 
In the analysis we have assumed that a fraction $f$ of the DM halo in the Galaxy
is composed by mirror atoms of various species and we have derived
allowed physical intervals for the parameters $\sqrt{f}\epsilon$, in various halo models.
We have also accounted for some of the possible existing uncertainties.
The results demonstrate that many configurations and halo models favoured by the 
annual modulation effect observed by DAMA corresponds to $\sqrt{f}\epsilon$ values 
well compatible with cosmological bounds.

Finally it is worth noting that our analysis predict in most halo models
an increase of the DM Mirror signal below 2 keV. These behaviours can be 
tested with the present DAMA/LIBRA phase2 that now is running.

\section{Acknowledgments}

The authors gratefully acknowledge the whole DAMA Collaboration for the work
which has produced the experimental results discussed here in particular framework
of mirror Dark Matter.
The work of A.A. and Z.B. was partially supported by the MIUR research grant 
"Theoretical Astroparticle Physics" PRIN 2012CPPYP7. 
The work of P.V. was supported by the 
MINECO grant "Ayudas a la movilidad predoctoral para la realizaci\'on de estancias breves 
en centros I+D 2014", EEBB-I-15-10012.


\begin{thebibliography}{99}


\bibitem{Drukier}   
  A.~K.~Drukier, K.~Freese and D.~N.~Spergel,
  Phys.\ Rev.\ D {\bf 33} (1986) 3495.

\bibitem{Freese}    
  K.~Freese, J.~A.~Frieman and A.~Gould,
  Phys.\ Rev.\ D {\bf 37} (1988) 3388.
  
  
\bibitem{prop}      P. Belli, R. Bernabei, C. Bacci, A. Incicchitti, R. Marcovaldi, D. Prosperi, DAMA proposal 
                    to INFN Scientific Committee II, April 24$^{th}$ 1990.
\bibitem{allDM1}    R. Bernabei et al., Phys. Lett. B 389 (1996) 757.
\bibitem{allDM2}    R. Bernabei et al., Phys. Lett. B 424 (1998) 195. 
\bibitem{allDM3}    R. Bernabei et al., Phys. Lett. B 450 (1999) 448. 
\bibitem{allDM4}    P. Belli et al.,    Phys. Rev. D 61 (2000) 023512.   
\bibitem{allDM5}    R. Bernabei et al., Phys. Lett. B 480 (2000) 23.
\bibitem{allDM6}    R. Bernabei et al., Phys. Lett. B 509 (2001) 197. 
\bibitem{allDM7}    R. Bernabei et al., Eur. Phys. J. C 23 (2002) 61. 
\bibitem{allDM8}    P. Belli et al.,    Phys. Rev. D 66 (2002) 043503.
\bibitem{Nim98}     R. Bernabei et al., Il Nuovo Cim. A 112 (1999) 545.
\bibitem{Sist}      R. Bernabei et al., Eur. Phys. J. C 18 (2000) 283.
\bibitem{RNC}       R. Bernabei el al., La Rivista del Nuovo Cimento 26 n.1 (2003) 1-73.
\bibitem{ijmd}      R. Bernabei et al., Int. J. Mod. Phys. D 13 (2004) 2127.
\bibitem{ijma}      R. Bernabei et al., Int. J. Mod. Phys. A 21 (2006) 1445.
\bibitem{epj06}     R. Bernabei et al., Eur. Phys. J. C 47 (2006) 263.
\bibitem{ijma07}    R. Bernabei et al., Int. J. Mod. Phys. A 22 (2007) 3155.
\bibitem{chan}      R. Bernabei et al., Eur. Phys. J. C 53 (2008) 205.
\bibitem{wimpele}   R. Bernabei et al., Phys. Rev. D 77 (2008) 023506.
\bibitem{ldm}       R. Bernabei et al., Mod. Phys. Lett. A 23 (2008) 2125.


\bibitem{allRare1}  R. Bernabei et al., Phys. Lett. B408 (1997) 439.      
\bibitem{allRare2}  P. Belli et al., Phys. Lett. B460 (1999) 236.       
\bibitem{allRare3}  R. Bernabei et al., Phys. Rev. Lett. 83 (1999) 4918.   
\bibitem{allRare4}  P. Belli et al., Phys. Rev. C60 (1999) 065501.   
\bibitem{allRare5}  R. Bernabei et al., Il Nuovo Cimento A112 (1999) 1541. 
\bibitem{allRare6}  R. Bernabei et al., Phys. Lett. B 515 (2001) 6.        
\bibitem{allRare7}  F. Cappella et al., Eur. Phys. J.-direct C14 (2002) 1. 
\bibitem{allRare8}  R. Bernabei et al., Eur. Phys. J. A 23 (2005) 7.
\bibitem{allRare9}  R. Bernabei et al., Eur. Phys. J. A 24 (2005) 51.
\bibitem{allRare10} R. Bernabei et al., Astrop. Phys. 4 (1995) 45.
\bibitem{IDM96}     R. Bernabei, in the volume {\it The Identification of Dark Matter}, World Sc. Pub. (1997) 574.

\bibitem{perflibra} R. Bernabei et al., Nucl. Instr. and Meth. A 592 (2008) 297. 
\bibitem{modlibra}  R. Bernabei et al., Eur. Phys. J. C 56 (2008) 333.
\bibitem{modlibra2} R. Bernabei et al., Eur. Phys. J. C 67 (2010) 39.
\bibitem{modlibra3} R. Bernabei et al., Eur. Phys. J. C 73 (2013) 2648.
\bibitem{bot11}     P. Belli et al., Phys. Rev. D 84 (2011) 055014.
\bibitem{pmts}      R. Bernabei et al., J. of Instr. 7 (2012) P03009.
\bibitem{mu}        R. Bernabei et al., Eur. Phys. J. C 72 (2012) 2064.
\bibitem{review}    R. Bernabei et al., Int. J. of Mod. Phys. A  28 (2013) 1330022.

\bibitem{papep}     R. Bernabei et al., Eur. Phys. J. C 62 (2009) 327. 
\bibitem{cnc-l}     R. Bernabei et al., Eur. Phys. J. C 72 (2012) 1920. 
\bibitem{IPP}       R. Bernabei et al., Eur. Phys. J. A 49 (2013) 64.
\bibitem{diu}       R. Bernabei et al., Eur. Phys. J. C 74 (2014) 2827.
\bibitem{norole}    R. Bernabei et al., Eur. Phys. J. C 74 (2014) 3196.
\bibitem{shadow}    R. Bernabei et al., Eur. Phys. J. C 75 (2015) 239.

\bibitem{mirror2015}   
  A.~Addazi et al., 
  Eur.\ Phys.\ J.\ C {\bf 75}, no. 8, 400 (2015)
  [arXiv:1507.04317 [hep-ex]].



\bibitem{Mirror1} 
  T. D. Lee and C. N. Yang, 
  Phys. Rev. 104 (1956) 254. 

\bibitem{Mirror2}
 I. Yu. Kobzarev, L. B. Okun and I. Ya. Pomeranchuk,
 Sov.\ J. Nucl.\ Phys.  3 (1966) 837. 

\bibitem{Mirror4}
  S.~I.~Blinnikov and M.~Khlopov,
  Sov.\ Astron.\  {\bf 27}, 371 (1983)
  [Astron.\ Zh.\  {\bf 60}, 632 (1983)].

\bibitem{Mirror5}
  R.~Foot, H.~Lew and R.~R.~Volkas,
 Phys.\ Lett. \ B 272 (1991) 67.

\bibitem{Khlopov:1989fj} 
  M.~Y.~Khlopov et al.,  
  Sov.\ Astron.\  {\bf 35}, 21 (1991)
  [Astron.\ Zh.\  {\bf 68}, 42 (1991)].
  
\bibitem{Hodges:1993yb} 
  H.~M.~Hodges,
  Phys.\ Rev.\ D {\bf 47}, 456 (1993).

\bibitem{Berezhiani:1995am} 
  Z.~G.~Berezhiani, A.~D.~Dolgov and R.~N.~Mohapatra,
  Phys.\ Lett.\ B {\bf 375}, 26 (1996)
  [hep-ph/9511221].

\bibitem{Berezhiani:1996sz} 
  Z.~G.~Berezhiani,
  Acta Phys.\ Polon.\ B {\bf 27}, 1503 (1996)
  [hep-ph/9602326].

\bibitem{Mohapatra:1996yy}
  R.~N.~Mohapatra and V.~L.~Teplitz,
  Astrophys.\ J.\  {\bf 478}, 29 (1997).

\bibitem{Akhmedov:1992hh} 
  E.~K.~Akhmedov, Z.~G.~Berezhiani and G.~Senjanovic,
  Phys.\ Rev.\ Lett.\  {\bf 69}, 3013 (1992)
  [hep-ph/9205230].

\bibitem{Berezhiani:1995yi} 
  Z.~G.~Berezhiani and R.~N.~Mohapatra,
  Phys.\ Rev.\ D {\bf 52}, 6607 (1995)
  [hep-ph/9505385].
  
\bibitem{Berezhiani:2000gh} 
  Z.~Berezhiani, L.~Gianfagna and M.~Giannotti,
  Phys.\ Lett.\ B {\bf 500}, 286 (2001)
  [hep-ph/0009290].
  
\bibitem{Berezhiani:1999qh} 
  Z.~Berezhiani and A.~Drago,
  Phys.\ Lett.\ B {\bf 473}, 281 (2000)
  [hep-ph/9911333].

\bibitem{Berezhiani:2000gw} 
  Z.~Berezhiani, D.~Comelli and F.~L.~Villante,
  Phys.\ Lett.\ B {\bf 503}, 362 (2001)
  [hep-ph/0008105].
  
\bibitem{Ignatiev}
  A.~Y.~Ignatiev and R.~R.~Volkas,
  Phys.\ Rev.\ D {\bf 68}, 023518 (2003)
  [hep-ph/0304260].
  
\bibitem{Berezhiani:2003xm} 
  Z.~Berezhiani,
  Int.\ J.\ Mod.\ Phys.\ A {\bf 19}, 3775 (2004)
  [hep-ph/0312335].

\bibitem{Berezhiani:2003wj} 
  Z.~Berezhiani, P.~Ciarcelluti, D.~Comelli and F.~L.~Villante,
  Int.\ J.\ Mod.\ Phys.\ D {\bf 14}, 107 (2005)
  [astro-ph/0312605].

\bibitem{Ciarcelluti:2004ip} 
  P.~Ciarcelluti,
  Int.\ J.\ Mod.\ Phys.\ D {\bf 14}, 223 (2005)
  [astro-ph/0409633].
  
 \bibitem{Berezhiani:2005ek}
  Z.~Berezhiani, ``Through the looking-glass: Alice's adventures in mirror world,''
  In  {\it From Fields to Strings: Circumnavigating Theoretical Physics}, Eds. M. Shifman et al., 
  vol. 3, pp. 2147-2195  [hep-ph/0508233].
  
\bibitem{Berezhiani:2008zza} 
  Z.~Berezhiani,
  Eur.\ Phys.\ J.\ ST {\bf 163}, 271 (2008).

\bibitem{Berezhiani:2005vv} 
  Z.~Berezhiani, S.~Cassisi, P.~Ciarcelluti and A.~Pietrinferni,
  Astropart.\ Phys.\  {\bf 24}, 495 (2006)
  [astro-ph/0507153].
 
\bibitem{Bento:2001rc} 
  L.~Bento and Z.~Berezhiani,
  Phys.\ Rev.\ Lett.\  {\bf 87}, 231304 (2001)
  [hep-ph/0107281].

\bibitem{Bento:2002sj} 
  L.~Bento and Z.~Berezhiani,
  Fortsch.\ Phys.\  {\bf 50}, 489 (2002); 
 Symposium on 100 Years Werner Heisenberg: Works and Impact, 
 Ed. D. Papenfuss et al.,  hep-ph/0111116.

\bibitem{Berezhiani:2005hv} 
  Z.~Berezhiani and L.~Bento,
  Phys.\ Rev.\ Lett.\  {\bf 96}, 081801 (2006)
  [hep-ph/0507031].

\bibitem{Holdom}
  B.~Holdom,
  Phys.\ Lett.\ B {\bf 166}, 196 (1986).

  
\bibitem{Glashow:1985ud} 
  S.~L.~Glashow,
  Phys.\ Lett.\ B {\bf 167}, 35 (1986).

\bibitem{Gninenko:1994dr} 
  S.~N.~Gninenko,
  Phys.\ Lett.\ B {\bf 326}, 317 (1994).

\bibitem{Berezhiani:1996ii} 
  Z.~Berezhiani,
  Phys.\ Lett.\ B {\bf 417}, 287 (1998)
  [hep-ph/9609342].
  
\bibitem{Addazi:2016rgo} 
  A.~Addazi, Z.~Berezhiani and Y.~Kamyshkov,
  arXiv:1607.00348 [hep-ph].

\bibitem{Foot:1995pa} 
  R.~Foot and R.~R.~Volkas,
  Phys.\ Rev.\ D {\bf 52}, 6595 (1995)
  [hep-ph/9505359].


\bibitem{Mohapatra:2005ng} 
  R.~N.~Mohapatra, S.~Nasri and S.~Nussinov,
  Phys.\ Lett.\ B {\bf 627}, 124 (2005)
  [hep-ph/0508109].

\bibitem{Berezhiani:2006je} 
  Z.~Berezhiani and L.~Bento,
Phys.\ Lett.\ B {\bf 635}, 253 (2006)
  [hep-ph/0602227].
  
\bibitem{Berezhiani:2011da} 
  Z.~Berezhiani and A.~Gazizov,
  Eur.\ Phys.\ J.\ C {\bf 72}, 2111 (2012)
  [arXiv:1109.3725 [astro-ph.HE]].

\bibitem{Berezhiani:2008bc} 
  Z.~Berezhiani,
  Eur.\ Phys.\ J.\ C {\bf 64}, 421 (2009)
  [arXiv:0804.2088 [hep-ph]].

\bibitem{Berezhiani:2012rq} 
  Z.~Berezhiani and F.~Nesti,
  Eur.\ Phys.\ J.\ C {\bf 72}, 1974 (2012)
  [arXiv:1203.1035 [hep-ph]].

\bibitem{Berezhiani:2016ong} 
  Z.~Berezhiani,
  arXiv:1602.08599 [astro-ph.CO].

  

\bibitem{Foot:2003iv} 
  R.~Foot,
  Phys.\ Rev.\ D {\bf 69}, 036001 (2004)
  [hep-ph/0308254].

\bibitem{Foot:2012cs} 
  R.~Foot,
  Phys.\ Rev.\ D {\bf 88}, no. 2, 025032 (2013)
  [arXiv:1209.5602 [hep-ph]].

\bibitem{Berezhiani:2006ac} 
  Z.~Berezhiani,
  AIP Conf.\ Proc.\  {\bf 878}, 195 (2006)
  [hep-ph/0612371].
  

\bibitem{bovy2013}   J. Bovy and H.W. Rix, Astrophys. J. 779 (2013) 115. 


  
\bibitem{Berezhiani:2007zf} 
  Z.~Berezhiani, D.~Comelli, F.~Nesti and L.~Pilo,
  Phys.\ Rev.\ Lett.\  {\bf 99}, 131101 (2007)
  [hep-th/0703264 [HEP-TH]].

\bibitem{Berezhiani:2009kx} 
  Z.~Berezhiani, L.~Pilo and N.~Rossi,
  Eur.\ Phys.\ J.\ C {\bf 70}, 305 (2010)
  [arXiv:0902.0146 [astro-ph.CO]].
  
\bibitem{Berezhiani:2009kv} 
  Z.~Berezhiani, F.~Nesti, L.~Pilo and N.~Rossi,
  JHEP {\bf 0907}, 083 (2009)
  [arXiv:0902.0144 [hep-th]].

\bibitem{Abbott:2016blz}
  B.~P.~Abbott {\it et al.} [LIGO Scientific and Virgo Collaborations],
  Phys.\ Rev.\ Lett.\  {\bf 116} (2016) no.6,  061102
  [arXiv:1602.03837 [gr-qc]]; 
  Phys.\ Rev.\ Lett.\  {\bf 116}, no. 24, 241103 (2016)
  [arXiv:1606.04855 [gr-qc]].


\bibitem{scho09}    R. Schoenrich, J. Binney and W. Dehnen, MNRAS 403, 1829-1833 (2010).
\bibitem{delh65}    J. Delhaye in ``Stars and Stellar Systems'', Univ. of Chicago Press, vol. 5 (1965) 73.
\bibitem{mcmi10}    P. McMillan, J. Binney, MNRAS 402, 934-940 (2010).

\bibitem{astr_v0}   P.J.T. Leonard and S. Tremaine, Astrophys. J. 353 (1990) 486;
                    C.S. Kochanek, Astrophys. J. 457 (1996) 228;
                    K.M. Cudworth, Astron. J. 99 (1990) 590.

\bibitem{starlink}  Starlink Project, http://starlink.jach.hawaii.edu/starlink.


\bibitem{Badertscher:2006fm}
  A.~Badertscher {\it et al.},
  Phys.\ Rev.\ D {\bf 75} (2007) 032004
  [hep-ex/0609059].

\bibitem{Carlson:1987si} 
  E.~D.~Carlson and S.~L.~Glashow,
  Phys.\ Lett.\ B {\bf 193}, 168 (1987).

\bibitem{Berezhiani:2008gi} 
  Z.~Berezhiani and A.~Lepidi,
  Phys.\ Lett.\ B {\bf 681}, 276 (2009)
  [arXiv:0810.1317 [hep-ph]].
  
\bibitem{BKK} 
Z.~Berezhiani, S. Karshenboim, A. Kobakhidze, in preparation. 

\bibitem{Berezhiani:2013dea} 
  Z.~Berezhiani, A.~D.~Dolgov and I.~I.~Tkachev,
  Eur.\ Phys.\ J.\ C {\bf 73}, 2620 (2013)
  [arXiv:1307.6953 [astro-ph.CO]].


\bibitem{Helm1}     R.~H~Helm, Phys. Rev. 104 (1956) 1466.
\bibitem{Helm2}     J.~D.~Lewin and P.~F.~Smith, Astropart. Phys. 6 (1996) 87.
\bibitem{Tretyak}   V.I. Tretyak, Astrop. Phys. 33 (2010) 40.

\bibitem{collar_qnai}  J.I. Collar, Phys. Rev. C 88 (2013) 035806.
\bibitem{Mat08}     S. I. Matyukhin, Tech. Phys. 53 (2008) 1578.
\bibitem{gelmini}   N. Bozorgnia et al., J. of Cosm. and Astropart. Phys. 11 (2010) 19.

\end{thebibliography}
\end{document}